\documentclass[]{raa}

\usepackage{graphicx,times}
\usepackage{natbib}
\usepackage{amssymb,amsmath}
\bibpunct{(}{)}{;}{a}{}{,}
\usepackage{float}

\usepackage[pagebackref=true]{hyperref}
\begin{document}

    \title{Broadband Spectral Analysis of 1ES\,2344+514: A Multi-epoch study}

 \volnopage{ {\bf 20XX} Vol.\ {\bf X} No. {\bf XX}, 000--000}
    \setcounter{page}{1}

    \author{Kishor Chaudhury %(周爱英) %% Put your Chinese name in "( )" if you like. Note to open line 11 "\usepackage[UTF8]{ctex}"
         \inst{1,2}
    \and Abhradeep Roy
         \inst{3}
    \and Varsha R. Chitnis
         \inst{3}
      \and Prajval Shastri
          \inst{4}
      \and Rajat K. Dey
            \inst{2}
    }

    \institute{Alipurduar University, Alipurduar, West Bengal 736122, India; {\it chaudhurykishor@gmail.com}\\
            \and
                   University of North Bengal, Siliguri, Darjeeling, West Bengal 734013, India\\
            \and
                   Tata Institute of Fundamental Research, Homi Bhabha Road, Colaba, Mumbai 400005, India\\
            \and Raman Research Institute, C. V. Raman Avenue, Sadashivanagar, Bengaluru 560080, India\\
\vs \no
    {\small Received 20XX Month Day; accepted 20XX Month Day}
}

\abstract{We present the results of the multi-epoch broadband spectral study of 1ES\,2344+514 and study the evolution of physical parameters. We used nearly simultaneous data obtained from 2017 June 6 to 2022 August 6 (MJD 57910 -- 59797) in optical, UV, X-ray and $\gamma$-ray wavebands from various instruments including \textit {Swift}-UVOT, \textit{Swift}-XRT, \textit{NuSTAR}, AstroSat (SXT and LAXPC), \textit{Fermi}-LAT, and TeV flux from MAGIC. During 2017 July, 1ES\,2344+514 appeared to be in the highest flaring state compared to other periods. We used the 0.5 -- 7.0 keV and 3.0 -- 20.0 keV data, respectively, from SXT and LAXPC of AstroSat and also 0.3 -- 8.0 keV and 3.0 -- 79.0 keV data, respectively, from \textit{Swift}-XRT and \textit{NuSTAR}. A joint fit between SXT and LAXPC, and between \textit{Swift}-XRT and \textit{NuSTAR} has been done for constraining the synchrotron peak. A clear shift in the synchrotron peak has been observed, which included 1ES\,2344+514 in the HSP BL Lac family. A `harder-when-brighter' trend is observed in X-rays, and the opposite trend, i.e., `softer-when-brighter', is seen in the $\gamma$-rays. The multi-epoch broad-band spectral energy distributions (SEDs) of this source were built and studied to get an idea of the radiation processes. The SEDs were fitted using a steady-state leptonic one-zone synchrotron+SSC model, and the fitted parameters of the emission region are consistent with those of other TeV BL Lacs. In this study, we found a weak correlation tendency between bolometric luminosity and magnetic field (B), as well as between bolometric luminosity and the break Lorentz factor ($\gamma_{break}$).
}

    \authorrunning{K. Chaudhury et al. }                  %author_head in even pages
    \titlerunning{Multi-epoch broadband study of 1ES\,2344+514}   % title_head in odd pages
    \maketitle
%________________________________________________ sections below
% 
\section{Introduction}                %% first-level sections will be auto-capitalized
\label{sect:intro}

Blazars are a subclass of accreting supermassive black holes (SMBHs) with relativistic jets that show high polarization and extreme variability at multiple frequencies on a range of time scales. This extreme behaviour is understood to be a consequence of their relativistic jet being oriented at very close angles to the line of sight \citep[e.g.][]{Blandford_1979}, and the non-thermal emission from the jet therefore dominating the total emission due to Doppler beaming. Blazars include Flat Spectrum Radio Quasars (FSRQs) that show prominent broad optical emission lines originated due to Keplerian gas clouds around the SMBH, and BL Lacertae objects (BL Lacs), which typically exhibit weak or no emission lines \citep{Urry_1995}. \citet{Abdollahi_2020} introduced the fourth \textit{Fermi}--LAT AGN catalogue (4FGL catalogue), identifying 3137 blazars, 42 radio galaxies, and 28 other active galactic nuclei (AGNs) with detection significance above $4\sigma$ in the 50~MeV to 1~TeV energy range. Among this sample of blazars, 1131 were identified or associated with BL Lacs, 694 were identified or associated with FSRQs, and 1312 were blazar candidates of uncertain type (BCUs). Analysis of \textit{Fermi}-LAT observations reveals distinct differences between FSRQs and BL Lacs. Specifically, FSRQs exhibit softer gamma-ray spectra (photon spectral index $(\Gamma)> 2.2$, corresponding to energy spectral index $> 1.2$) and higher gamma-ray luminosities ($> 10^{46} \, \text{erg} \, \text{s}^{-1}$). In contrast, BL Lacs display harder gamma-ray spectra ($\Gamma< 2.2$ or energy spectral index $< 1.2$) and lower luminosities \citep{Ghisellini_2009, Ghisellini_2010, Ackermann_2011, Ackermann_2015}. Additionally, FSRQs demonstrate a `harder-when-brighter' tendency, whereas BL Lacs exhibit more diverse trends in $\gamma$-ray band. LSP and ISP BL Lacs often display a `harder-when-brighter' trend, while HSP BL Lacs tend to show `softer-when-brighter' patterns \citep{Abdo_2010a}.

The spectral energy distributions (SEDs) in blazars exhibit a double-humped structure. 
The low-energy hump is located in the radio to X-ray bands, and the high-energy one is located in the $\gamma$-ray band. The production mechanism of broadband continuum spectra is commonly explained with two types of models: the leptonic model and the hadronic model. In both cases (FSRQs and BL Lacs), the low-energy hump of the SEDs is widely believed to be arising from the synchrotron radiation by relativistic electrons in the jet, occasionally having significant contributions from the host galaxy thermal emission in the optical to infrared (IR) band \citep{Urry_1995}. In the leptonic model, the high-energy hump is believed to be produced by inverse-Compton (IC) scattering of either synchrotron photons (known as synchrotron self-Compton \citep[SSC;][]{Maraschi_1992, Krawczynski_2004, Albert_2007, Tavecchio_2010, Aleksic_2013}) or of low-energy photons from external regions like the accretion disk, broad-line region, torus (known as external Compton (EC) or external inverse-Compton (EIC)), or a combination of both \citep[see; e.g.,][]{Dermer_1993, Sikora_1994, Fan_2006, Böttcher_2007, Abdo_2010c, Ghisellini_2010, Costamante_2018}. The spectral and variability differences between FSRQs and BL Lacs are attributed to distinct emission mechanisms, with FSRQs dominated by EC scattering and BL Lacs by SSC processes \citep{Ghisellini_2009, Ghisellini_2010}. Alternatively, the origin of high-energy components of SEDs is also described using hadronic models via photo-meson production, neutral pion decay, and proton synchrotron radiation or synchrotron-pair cascading (see; e.g., \citep{Mannheim_1993, Aharonian_2000, Mucker_2001, Böttcher_2013, Cerruti_2015, Goswami_2024}). Based on the synchrotron peak frequency ($\nu_{\rm sp}$), blazars are subdivided into low-frequency peaked blazars (LSPs; $\nu_{\rm sp} < 10^{14}$ Hz), intermediate-frequency peaked blazars (ISPs; $10^{14} < \nu_{\rm sp} < 10^{15}$ Hz), and high-frequency peaked blazars (HSPs; $\nu_{\rm sp} \geq 10^{15}$ Hz) \citep{Abdo_2010b, Padovani_2017}. Extreme high-frequency peaked blazars (EHSPs), a subset of HSPs, have synchrotron peaks at $\nu_{\text{sp}} \geq 10^{17} \, \text{Hz}$ and IC peaks typically at energies $\geq 1 \, \text{TeV}$ \citep{Costamante_2001, Aharonian_2007, Acciari_2010}. Regarding jet composition of FSRQs and BL Lacs, FSRQ jets may be more magnetized, potentially Poynting-flux dominated, while BL Lac jets are likely less magnetized and particle dominated \citep{Zhang_2014,Tavecchio_2016}.

1ES\,2344+514 (RA = ${23}^{h}$${47}^{m}$${04.837}^{s}$, Dec = {+51$^{\circ}$}${42}^{'}$${17.878}^{''}$, J2000) located at redshift z = 0.044 \citep{Perlman_1996} is a BL Lac. The estimated black hole mass of this object is $10^{(8.80\pm0.16)}$M$_\odot$ \citep{Barth_2003}. It was first observed during the Einstein Slew Survey \citep{Elvis_1992} in the X-ray energy range (0.2 -- 4 keV). It is also one of the extra-galactic objects detected in the very-high-energy (VHE, $>$ 100 GeV) band. It was detected by the Whipple 10 m telescope at energies above 350 GeV during a bright flare of 1-day duration on 1995 December 20, with the flux level of $\sim$ 60\% of the Crab Nebula \citep{Catanese_1998, Schroedter_2005}. Strong variability, with a factor of $\sim$ 2 variations in 2-10 keV flux, was observed with BeppoSAX in December 1996 on a timescale of about 5 ks during the flaring state \citep{Giommi_2000}. Later, 1ES\,2344+514 was also observed by \textit{Swift} in the X-ray band in 2005 on various occasions, with its 2 -- 10 keV flux steady around $1\times10^{-11}$ erg cm$^{-2}$ s$^{-1}$, which is lower than the flux measured in 1996 December \citep{Tramacere_2007}. \citet{Kapanadze_2017} reported that during the monitoring campaign with \textit{Swift} during 2005 -- 2015, the source showed high variability on longer time scales (weeks to months) with the 0.3 -- 10 keV flux varying by a factor of 13.3. 1ES\,2344+514 has been studied through \textit{Fermi}-LAT observations, revealing its gamma-ray characteristics and variability, such as a low flux state in 2008 \citep{Aleksic_2013} and an enhanced state in 2016 \citep{Acciari_2020} with a flux of approximately $1.2\times 10^{-8}~\text{cm}^{-2} \text{s}^{-1}$ in the 0.3-300 GeV range, as noted in multi-wavelength campaigns. No significant flaring activity was found in the VHE $\gamma$-ray band with the Very Energetic Radiation Imaging Telescope Array System (VERITAS) between 2007 and 2015, which corresponds to the quiescent state of 1ES\,2344+514 in VHE \citep{Allen_2017}. Multiwavelength studies of the source have been carried out using simultaneous and quasi-simultaneous data, and the SEDs are modelled on a few occasions \citep{Albert_2007, Godambe_2007, Acciari_2011, Aleksic_2013, Acciari_2020}. \citet{Acciari_2020} reported that 1ES 2344+514 was detected in a high flaring state in VHE $\gamma$-rays during 2016 August by ground-based $\gamma$-ray telescopes, including the First G-APD Cherenkov Telescope (FACT) and the Major Atmospheric Gamma Imaging Cherenkov Telescopes (MAGIC) and observed that the synchrotron peak shifted to frequencies exceeding $10^{18} \, \text{Hz}$. Also, the peak of the synchrotron emission was found near frequencies of the order of $\approx$ $10^{16}$ -- $10^{17}$ Hz during the quiescent state \citep{Giommi_2000, Aleksic_2013, Nilsson_2018, Abe_2024}.

In this paper, we report the simultaneous and quasi-simultaneous observations of 1ES\,2344+514 from 2017 to 2022, in optical, UV, X-ray and $\gamma$-ray bands. We present the results of individual and joint spectral fittings covering soft and hard X-rays. We used the data in optical, UV, X-ray and $\gamma$-ray bands for a multiwavelength study to understand the broadband properties of the source. The SED in different epochs has also been constructed and fitted with the two-component and one-zone synchrotron+SSC model. The paper is structured as follows: In section~\ref{sec:Multi_obs} of this paper, we describe the details of the instruments used for multiwavelength observations, procedures for data reduction, and analysis. In section~\ref{sec:ana_res}, we present the results of this analysis. Details of the SED modelling along with results are discussed in section~\ref{sec:multi_sed}. We present the discussion in section~\ref{sec:discus}, and finally, the conclusions are listed in section~\ref{sec:conl}.

\section{Multiwavelength Observations, data reduction and data analysis}

\label{sec:Multi_obs}
1ES\,2344+514 was observed on several occasions in the X-ray band with the Soft X-ray Telescope (SXT) and Large Area X-ray Proportional Counters (LAXPC) onboard AstroSat\footnote{\url{https://www.isro.gov.in/AstroSat.html}} \citep{Singh_2014}, X-Ray Telescope (XRT) onboard Neil Gehrels \textit{Swift} Observatory \citep{Burrows_2004} and Nuclear Spectroscopic Telescope Array (\textit{NuSTAR}) \citep{Harrison_2013}. In addition to these, optical and UV observations are available from the Ultra-violet Optical Telescope (UVOT) onboard \textit{Swift}. In addition , it was observed in the gamma-ray band by the Large Area Telescope (LAT) onboard \textit{Fermi} \citep{Atwood_2009}. In this section, we discuss the details of the observations, data reduction, and data analysis procedures of various instruments. A log of the observations, along with the effective exposure times and the mean count rates in different energy bands, is given in Table~\ref{tab:table1}. 
The TeV flux measurements, published by the MAGIC team from their July 2020 observations, were utilised in the associated SED modelling.

\begin{table*}[!ht]%[h!]
	\centering
	\caption{ X-ray, UV, and Optical Observations of 1ES\,2344+514}
	\label{tab:table1}
             \resizebox{\textwidth}{!}{
	\begin{tabular}{lcccccc} 
		\hline
                  \hline
                  \noalign{\smallskip}
		Instrument & Energy band & Obs. Id & Start time & Stop time & Exposure time &Mean count rate\\
                  &(keV)&&YYYY-MM-DD HH:MM:SS&YYYY-MM-DD HH:MM:SS&(sec)&(counts/s)\\
		\hline
		&& 9000001276&2017-06-06 15:36:34& 2017-06-06 21:06:10 & 2803&0.357$\pm$0.012
\\
		&& 9000001368 &2017-07-09 08:33:18 &2017-07-09 16:32:50 & 9565&0.718$\pm$0.009
\\
		&& 9000001438 &2017-08-07 05:41:13&2017-08-07 13:32:51 & 8837&0.428$\pm$0.007
\\
	      && 9000001524&2017-09-07 18:46:48&2017-09-08 03:35:55 & 5316&0.247$\pm$0.007
\\
            && 9000001632&2017-10-21 22:03:13&2017-10-22 07:52:31 & 4591&0.277$\pm$0.008
\\
            {\bf{SXT}}&0.5 -- 7.0&9000001710&2017-11-22 01:44:24&2017-11-22 05:53:31 & 3031&0.212$\pm$0.009
\\
             && 9000001754&2017-12-07 22:11:13&2017-12-08 06:50:15 & 3595&0.247$\pm$0.009
\\
            && 9000002266&2018-08-01 17:31:24&2018-08-01 23:03:30 & 7063&0.514$\pm$0.009
\\
            && 9000002356&2018-09-11 18:01:33&2018-09-12 02:05:12 & 8345&0.452$\pm$0.008
\\
            && 9000004598 &2021-07-28 01:52:16&2021-07-31 01:54:59 & 61920&0.257$\pm$0.002
\\
            && 9000004626&2021-08-05 15:09:14&2021-08-07 01:21:42 & 41190&0.156$\pm$0.002
\\
            \hline
            && 9000001276& 2017-06-06 15:36:34& 2017-06-06 21:06:09& 10590&2.633$\pm$0.045
\\
                  && 9000001368 & 2017-07-09 08:50:51& 2017-07-09 16:32:50& 11940&5.169$\pm$0.046
\\
                  && 9000001438 & 2017-08-07 09:39:31& 2017-08-07 13:32:51& 10730&2.629$\pm$0.045
\\
                  && 9000001524& 2017-09-07 19:49:41& 2017-09-08 03:35:56& 11300&5.255$\pm$0.048
\\
                  && 9000001632& 2017-10-21 22:50:42& 2017-10-22 07:52:31& 9616&1.404$\pm$0.048
\\
{\bf{LAXPC}}& 3.0 -- 20.0&9000001710& 2017-11-22 00:27:27& 2017-11-22 05:53:33& 11680&2.492$\pm$0.045
\\
                  && 9000001754& 2017-12-07 23:34:44& 2017-12-08 06:50:13& 12960&1.661$\pm$0.043
\\
                  && 9000002266& 2018-08-01 19:08:21& 2018-08-01 23:03:30& 12670&3.490$\pm$0.044
\\
                  && 9000002356& 2018-09-11 19:13:45& 2018-09-12 02:05:12& 12060&2.077$\pm$0.046
\\
                  && 9000004598 & 2021-07-28 01:46:56& 2021-07-31 01:55:00& 102600&1.496$\pm$0.022
\\
                  && 9000004626& 2021-08-05 15:09:13& 2021-08-07 01:21:42& 55860&1.127$\pm$0.025
\\
                  \hline
 & &00035031184 &2020-07-22 02:01:19&2020-07-23 23:02:52 & 9660&0.379$\pm$0.006\\
 {\bf{\textit{Swift}-XRT}}&0.5 -- 7.0 & \& 00081310002& & & &\\
                  && 00035031241& 2022-08-06 03:58:51& 2022-08-06 04:24:53& 1556&0.356$\pm$0.015
\\
                  \hline
                  
 & &00035031184 &2020-07-22 02:01:19&2020-07-23 23:02:52 & 8002&   --\\
 {\bf{\textit{Swift}-UVOT}}& --   & \& 00081310002& & & &\\
                  && 00035031241& 2022-08-06 03:58:51& 2022-08-06 04:24:53& 1535& --
\\

                  \hline
 {\bf{\textit{NuSTAR}}}&3.0 -- 20.0&60160836002 &2020-07-22 19:26:09 &2020-07-23 07:11:09 &20880&0.127$\pm$0.003
\\
                  &&80801650002&2022-08-05 18:26:09 &2022-08-06 07:11:09 &22910&0.075$\pm$0.002\\

 \hline

	\end{tabular}
            }
\end{table*}

\subsection{AstroSat observations}
\label{sec:astrosat} 
AstroSat, the multi-wavelength space observatory launched on 2015 September 28 \citep{Rao_2016} has five scientific payloads which can simultaneously observe objects from optical-UV to hard X-ray energies. These are the Scanning Sky Monitor \citep[SSM;][]{Ramadevi_2018}, the Ultraviolet Imaging Telescope \citep[UVIT;][]{Tandon_2017a,Tandon_2017}, the Soft X-ray Telescope \citep[SXT;][]{Singh_2016,Signh_2017}, the Large Area X-ray Proportional Counters \citep[LAXPCs;][]{Yadav_2016,Agrawal_2017}, and the Cadmium Zinc Telluride Imager \citep[CZTI;][]{Rao_2017}. 1ES\,2344+514 was observed by AstroSat on several occasions from 2017 June to 2021 August, details of which are given in Table~\ref{tab:table1}. We have downloaded publicly available data from SXT and LAXPC for these observations from the Indian Space Science Data Center (ISSDC)\footnote{\url{https://astrobrowse.issdc.gov.in/astro_archive/archive/Home.jsp}} and used it in the present work. 

\subsubsection{SXT observations}
\label{sec:sxt} % used for referring to this section from elsewhere

The SXT is a focusing X-ray telescope based on the grazing incidence principle with a focal length of 2 metres. It has a thermoelectrically cooled CCD in the focal plane, which collects the X-ray photons. Its operational energy range is 0.3 -- 8.0 keV with an energy resolution of 5 -- 6\% at 1.5 keV. 1ES\,2344+514 was observed in the photon counting mode (PC). The raw Level1 (L1) data of individual orbits were received from the satellite to the data acquisition center at the Indian Space Science Data Center and transferred to the SXT Payload Operation Center (POC). Later, X-ray data from individual orbits were reduced to level2 (L2) data files through \textsc{SXTPIPELINE} (Version:1.4b)\footnote{\url{https://www.tifr.res.in/~astrosat_sxt/index.html}} at the POC. The task of event generation, time tagging of events above a pre-set threshold and the electronic noise level, the coordinate transformation from the detector to sky coordinates, bias subtraction, flagging of spurious pixels, and calibration of source events were carried out in the pipeline. Event grading \citep[similar to \textit{Swift}-XRT; see][]{Romano_2005} and removal of the events with grades $>$ 12, the event pulse height amplitude (PHA) construction, the conversion from the event PHA to the X-ray energy (Pulse Invariant; PI), a search and removal of hot and flickering pixels were also involved. Further, the events with a bright Earth-avoidance angle of $\ge$ 110$^{\circ}$ were screened and those data during the passage through the South Atlantic Anomaly (SAA) were removed using the criteria that the Charged Particle Monitor (CPM) rate is below 12 counts s$^{-1}$. Good Time Intervals (GTIs) were constructed and filtered, generating the level2 (L2) events files. The science products were generated from L2 event files. The orbit-wise L2 SXT data files were downloaded from the AstroSat archive. The individual orbits L2 data contained cleaned event files. These individual cleaned event files were merged using the Julia package {SXTMerger.jl}\footnote{\url{https://github.com/gulabd/SXTMerger.jl}}, also available at the SXT website\footnote{\url{http://astrosat-ssc.iucaa.in/uploads/threadsPageNew_SXT.html}}, which takes care of exposure overlapping within GTIs. The merged cleaned event files were compatible with HEASoft software. An image of the cleaned events file was created using DS9 (version 8.3) and from this image, a circular region of 10 arcmin radius and an annular region with a radius extending from 13 to 19 arcmin centered on the source were chosen as the source and the background region, respectively, based on a radial profile similar to \cite{Chau_2018}. \textsc{XSELECT} (version 2.4) tool of \textsc{HEASoft} (version 6.28) was used to extract the source and background filtered light curves and spectra from the SXT merged cleaned events file. The XSPEC (version: 12.11.1) \citep{ARNAUD_1996} compatible ancillary response file (ARF; sxt\_pc\_excl01\_v03.arf), the response matrix file (RMF; sxt\_pc\_mat\_g0to12.rmf) and the blank-sky background spectrum file (SkyBkg\_comb\_EL3p5\_Cl\_Rd16p0\_v01.pha) provided by SXT team\footnote{\url{ https://www.tifr.res.in/~astrosat_sxt/dataanalysis.html}} were used for spectral analysis. The spectra from SXT were grouped using GRPPHA (v3.1.0) in such a way that a minimum of 20 counts were contained in each energy bin. The spectra over the energy range 0.5 -- 7.0 keV were fitted using XSPEC (v12.11.1) with an absorbed power-law model: 

\begin{equation}
    dE/dN = k{(E/E_0)}^{-\Gamma}
    \label{mod:powlaw}
\end{equation} 
where $\Gamma$ is the photon spectral index; ${E_0} = 1$ keV is the pivot energy and k is the normalization. The soft X-ray spectrum is affected by the line-of-sight absorption in the interstellar gas, and this is modelled using the Tuebingen-Boulder Inter-Stellar Medium absorption model \citep[tbabs,][]{Wilms_2000} with neutral hydrogen column density ($N_H$) fixed at 1.41 $\times$ $10^{21}$ cm$^{-2}$ \citep{HI4PI_2016}. 

\subsubsection{LAXPC observations}
\label{sec:laxpc} 

The LAXPC onboard AstroSat has three identical co-aligned independently operating proportional counter units (LAXPC10, LAXPC20, and LAXPC30) with a broad energy range of 3 -- 80 keV, energy resolution in the 22 -- 60 keV range of about 10 -- 12\%, and a collimator field of view of $55'\times 55'$. The LAXPC30 was suspected to have undergone a gas leakage resulting in a continuous gain shift \citep{Antia_2017}. The LAXPC10 was reported to have problems in background normalization for faint sources such as RE J1034+396 \citep{Chau_2018}. Therefore, in the present work, only LAXPC20 data has been used. In Event Analysis (EA) mode, each photon arrival time is recorded with a time resolution of 10 $\mu$s and a dead-time of about 42 $\mu$s \citep{Yadav_2016, Agrawal_2017, Antia_2017}. The effective area of LAXPC20 is $\sim$ 2000 cm$^2$ in 5 -- 20 keV. The SXT and LAXPC instruments are misaligned by $\sim$6 arcmin\footnote{\url{http://astrosat-ssc.iucaa.in/uploads/APPS/NoteOnRelativeAngleBetwwenPayloads_Astrosat_15072016.pdf}}, much smaller than the field of view (FOV) of either instrument. We used event analysis mode (EA) data from LAXPC20 for the timing and spectral analyses of 1ES\,2344+514. The laxpcl1 tool of the LAXPC software (LAXPCsoft: version 3.4.4)\footnote{\url{https://www.tifr.res.in/~astrosat_laxpc/LaxpcSoft.html}} was used for extracting the light curve and the spectrum of the source and background from level1 (L1) event analysis mode (EA) data files. The backshiftv3 tool was used to apply a gain shift to the background spectrum, accounting for gain mismatch between source and background runs and to identify required response files \citep[for details, see][]{Antia_2017}. Data from the top layer (L1, L2) of the LAXPC20 detector have been used for all analyses to minimise the detector background effect. The spectra from LAXPC20 were grouped using GRPPHA (v3.1.0) in such a way that a minimum of 20 counts were contained in each energy bin. The spectra were fitted over the energy range 3 -- 20 keV only for LAXPC20 because of poor statistics beyond 20 keV. LAXPC spectra were fitted with the power-law model using XSPEC (v12.11.1). In addition to individual fits, the SXT and LAXPC20 spectra were also jointly fitted to get a better handle on the model parameters by covering a larger dynamic range of about 0.5 to 20 keV with a log-parabola model \citep{Massaro_2004} including relative normalization between the SXT and LAXPC20 instruments with a fixed value for neutral hydrogen column density ($N_{H}$) to model line of sight absorption. Relative normalization accounts for cross-calibration between two detectors as well as for the fact that data from these two detectors are not strictly-simultaneous coupled with the variable nature of the source. The mathematical form of the log-parabola model is as follows:

\begin{equation}
{\frac{dN}{dE}} = k\left(\frac{E}{E_b} \right)^{-\alpha - \beta \log(E/E_b)}
\label{mod:logpar}
\end{equation}
where $\alpha$ is the photon spectral index at the energy ${E_b}$, $\beta$ is the curvature parameter, $k$ is the normalization, and ${E_b}$ is the reference energy that was kept fixed at 1 keV during the fitting. The synchrotron peak energy is estimated using the following equation \citep{Massaro_2006}: 

\begin{equation}
{E_p} = {E_b} 10^{(2-\alpha)/2\beta}
\label{eq:logpar_sp}
\end{equation}
The synchrotron peak frequency is obtained in XSPEC using the eplogpar model.

\subsection{\textit{Swift}-XRT observations}
\label{sec:xrt} % used for referring to this section from elsewhere

1ES\,2344+514 has been observed on various occasions with \textit{Swift}-XRT \citep{Burrows_2005}, the grazing incidence focusing telescope operating in the energy range 0.3 -- 8.0 keV. In this work, we used data that were simultaneous with \textit{NuSTAR} observations. These data were recorded in Photon Counting mode (PC), was obtained from the \textit{Swift} public archive\footnote{\url{https://swift.gsfc.nasa.gov/archive/}} (see Table~\ref{tab:table1} for details). All XRT data were processed using the \textsc{XRTDAS} software package (version 3.6.0) available under \textsc{HEASoft} (version 6.28). The \textit{Swift}-XRT calibration files (25 July 2023 files) were used within standard \textit{Swift} \textsc{XRTPIPELINE} (version 0.13.5) for calibrating and cleaning the event files. A circular region of radius 20 pixels centered at RA and Dec of the source was chosen as the source region, and a circular region of 40-pixel radius away from the source was chosen as the background region. \textsc{XRTPRODUCTS} (v0.4.2) was used to extract the source and background spectra. The ancillary response files were generated via the task 'xrtmkarf', taking care of corrections for point spread function (PSF) losses and CCD defects using the exposure map. XRT spectra obtained from observation IDs 00035031184 and 0081310002 were combined using \textsc{ADDSPEC} v1.4.0) as they are quasi-simultaneous with \textit{NuSTAR} observation ID 60160836002. Spectra were grouped using \textsc{GRPPHA} (v3.1.0) with a minimum of 20 counts in each energy bin. Individual spectra in the energy range 0.3 – 8 keV from \textit{Swift}-XRT, with proper response, and ancillary file were loaded into \textsc{XSPEC} (v12.11.1) and fitted with the power-law model with fixed hydrogen column density, $N_{H}$ = 1.41 $\times$ $10^{21}$ cm$^{-2}$. The unabsorbed flux was computed using the cflux convolution model in the energy range of 0.5 – 7 keV. 

\subsection{\textit{Swift}-UVOT observations}
\label{sec:swift_uvot}
\textit{Swift} has the capability of getting simultaneous Optical-UV and X-ray observational data. Optical-UV data were obtained from \textit{Swift}-UVOT \citep{Roming_2005}, simultaneous with July 2020 and August 2022 X-ray observations. It has three filters UW1, UM2, and UW2, in the ultraviolet band and three filters V, B, and U in the optical band. \textit{Swift}-UVOT data were analysed using the tools within \textsc{HEASoft} (v6.28) package. UVOTIMSUM tool was used to integrate all observational data of selected epochs. To estimate flux values from integrated image files for various filters, circular regions with a radius of 5 arcseconds and 8 arcseconds, centered around the source position, were used respectively for visible and UV filters, whereas background flux values were obtained using circular regions with a radius of 25 arcseconds located away from the source. The flux magnitudes were extracted using the \textsc{UVOTSOURCE} tool and then corrected for the Galactic extinction of E(B-V) = 0.1819 mag \citep{Schlegel1998} for all filters as obtained using a web tool\footnote{\url{https://ned.ipac.caltech.edu/forms/calculator.html}}.

\subsection{\textit{NuSTAR} observations}
\label{sec:nustar} % used for referring to this section from elsewhere

\textit{NuSTAR} \citep{Harrison_2013} is the first focusing high energy X-ray telescope covering the energy range 3 -- 79 keV, launched on 2012 June 13. 1ES\,2344+514 was observed on a few occasions with its two co-aligned X-ray telescopes i.e. Focal Plane Module A (FPMA) and B (FPMB). We used publicly available data from the HEASARC archive{\footnote{\url{https://heasarc.gsfc.nasa.gov/docs/archive.html}}} for two observations carried out in July 2020 and August 2022, details of which are given in Table~\ref{tab:table1}. The level 1 \textit{NuSTAR} data were processed using the \textsc{NuSTARDAS} software package (version 2.0.0) available within the \textsc{HEASoft} package. The latest \textit{NuSTAR} calibration files (17 October 2023 files) were used within \textsc{NUPIPELINE} (v.0.4.8) for calibrating and cleaning the event files. The spectra of the source for different observations were extracted using \textsc{NUPRODUCTS} (v.0.3.2) with a circle of radius 12 pixels centered on the RA and Dec of the source, and the background was chosen from a region away from the source to extract the background spectra. Spectra were grouped using \textsc{GRPPHA} (v3.1.0) with a minimum of 30 counts in each energy bin for FPMA/B onboard \textit{NuSTAR}. Individual spectra of the FPMA/B in the energy range 3 – 79 keV, with appropriate response, and ancillary file were loaded into XSPEC (v12.11.1) and fitted with the power-law model. The corresponding flux was computed in the energy range 3–20 keV.Similarly, the combined \textit{Swift}-XRT and \textit{NuSTAR} FPMA/B spectra were fitted with constant*tbabs*logpar in the broadband X-ray energy range 0.3 -- 79 keV. The synchrotron peak frequency is obtained in XSPEC using the eplogpar model.

\subsection{\textit{Fermi}-LAT observations}
\label{sec:fermi_lat} % used for referring to this section from elsewhere
The Large Area Telescope (LAT) instrument is a pair-production telescope onboard the \textit{Fermi} satellite launched in 2008 \citep{Atwood_2009, Ackermann_2012}. It covers the 20 MeV to 1 TeV energy band with $\sim$2.3 sr field of view (FOV) and covers the entire sky every three hours with an orbital period of $\sim$96 mins. \textit{Fermi}-LAT observations of 1ES\,2344+514 have been available since 2008. For the multi-waveband study, we have obtained the data from \textit{Fermi}-LAT data archive\footnote{\url{https://fermi.gsfc.nasa.gov/ssc/data/access/}} in the energy range 30 MeV to 1 TeV from a region of radius 30$^{\circ}$ centered on the source. The data from 6th June 2017 15:36:34 to 6th August 2022 07:11:09 over the energy range of 0.1 to 500 GeV were analysed to get $\gamma$-ray spectra simultaneous with X-ray observations. 

The analysis is performed using the standard software package \textsc{FERMITOOLS}(v1.2.1) supplied by \textit{Fermi}-LAT collaboration and user-contributed ENRICO python script \citep{Sanchez_2013}. A circle of radius 10$^{\circ}$ centered at the source 1ES\,2344+514 was considered as the region of interest (ROI) for photon events extraction with evclass=128 and evtype=3. Good time intervals were restricted using the filter ``DATA QUAL$>$0 \&\& LAT CONFIG==1" as recommended by the \textit{Fermi}-LAT collaboration\footnote{\url{https://fermi.gsfc.nasa.gov/ssc/data/analysis/documentation/}}. Also, events with zenith angles smaller than 90$^{\circ}$ were chosen to avoid the contribution from the Earth limb's $\gamma$-rays. The galactic interstellar emission model ``gll\_iem\_v07.fits" and isotropic spectral template ``iso\_P8R3\_SOURCE\_V3\_v1.txt" were used as the diffused background models\footnote{\url{http://fermi.gsfc.nasa.gov/ssc/data/access/lat/BackgroundModels.html}} to take care of background $\gamma$-ray emission. The source spectrum is obtained using the binned and unbinned maximum likelihood approach with the instrumental response function P8R3\_SOURCE\_V3.

The $\gamma$-ray light curve and SEDs were extracted from \textit{Fermi}-LAT data. For this purpose, an XML file was created that contains the spectral information of all the $\gamma$-ray sources lying within the radius of 20$^{\circ}$ (ROI + 10$^{\circ}$) centered at 1ES\,2344+514. The spectral parameters of the sources were taken from the fourth \textit{Fermi}-LAT source catalogue Data Release 2 \citep[4FGL-DR2,][]{Abdollahi_2020}. All the spectral parameters of the sources within 5$^{\circ}$ radius of 1ES\,2344+514 were left free during the likelihood fitting process. For sources within 5$^{\circ}$ to 10$^{\circ}$, all spectral parameters except for normalization were fixed to 4FGL catalogue values, whereas for sources within 10$^{\circ}$ to 20$^{\circ}$, all parameters were fixed to 4FGL catalogue values. The determination of the photon flux and spectral parameters for the source was carried out using a binned and unbinned likelihood analysis method \textsc{GTLIKE}. The likelihood analysis was performed in an iterative manner. Sources with test statistics less than 1 were eliminated during each fitting iteration. Log-Parabola and PowerLaw2 was used to model the source spectra for different time epochs. The LogParabola model is defined in equation~\ref{mod:logpar}, and the PowerLaw2 model is given as follows:

\begin{equation}
{\frac{dN}{dE}} = \frac{N(\gamma - 1)E^{\gamma}}{E_{max}^{\gamma+1} - E_{min}^{\gamma+1} }
\label{mod:powlaw2}
\end{equation}
where $\gamma$ is the spectral index, $E_{min}$ and $E_{max}$ are the energy bounds and $N$ is the integral flux over the energy range considered. The LAT model fits, and the butterfly contours were extended up to the maximum energy of 500 GeV, and the correction for extragalactic background light (EBL) absorption was applied to $\gamma$-ray spectra using the model of \citet{Franceschini_2017}.

The flux points in the SED unit are estimated by dividing the entire energy range of the source into a few bins. For bins with test statistics (TS) $<$ 9 (i.e. $\lesssim$3$\sigma$ significance), the flux upper limit is calculated at 95\% confidence level using the profile likelihood method.

\section{Results}
\label{sec:ana_res}
\subsection{AstroSat}
\label{sec:sxt_laxpc} 
\begin{figure}[!htb]%[h!]
      \centering
      \includegraphics[width=0.65 \linewidth, angle=270]{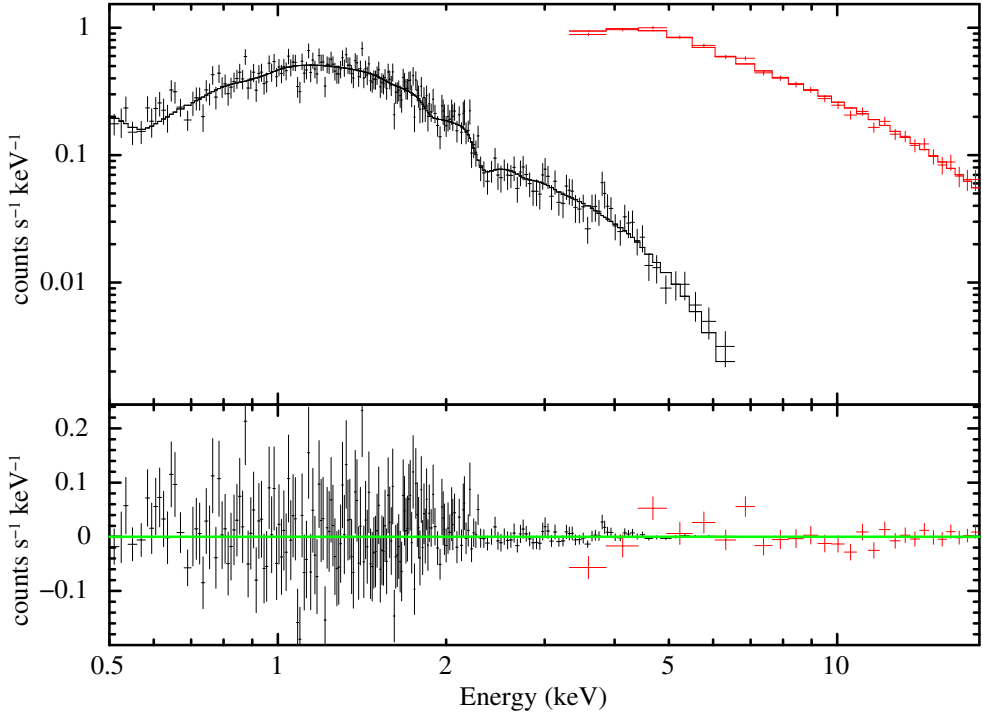}
      \caption{
      Joint fit for SXT (black) and LAXPC20 (red) spectra for 2017 July 9 observations with the log-parabola model along with the line of sight absorption. Data over energy ranges 0.5-7.0 keV and 3.0-20.0 keV are used for SXT and LAXPC20, respectively. The top panel shows data and model fit, whereas the bottom panel shows the residuals.}
      \label{fig:figure1}
\end{figure}

1ES\,2344+514 was observed simultaneously by SXT and LAXPC onboard AstroSat on several occasions, the details of which are given in Table~\ref{tab:table1}. The light curves and spectra were generated for all these observations following the procedure outlined in section~\ref{sec:sxt} and section~\ref{sec:laxpc}. In addition, the individual and combined spectra of SXT and LAXPC were fitted using the procedure mentioned in sections~\ref{sec:sxt} and ~\ref{sec:laxpc}. The best-fit parameters for individual SXT and LAXPC spectra along with unabsorbed flux are presented in Table~\ref{tab:table2} for 11 observation epochs during 2017-2021. Figure~\ref{fig:figure1} shows the combined spectral plot for the SXT (0.5 -- 7.0 keV) and LAXPC20 (3 -- 20 keV) observations carried out on 2017 July 9 with the best-fitted log-parabola model. The values of the best-fit parameters along with the peak energy ($E_{p}$) and $\chi^2$/dof for the log-parabola model are given in Table~\ref{tab:table4} for 11 epochs.

\begin{table*}[!ht]%[h!]
      \centering
      \caption{Best-fit absorbed power-law results for SXT and LAXPC spectral analysis }
      \resizebox{\textwidth}{!}{
      \begin{tabular}{lccccccc}
      \hline
      \hline
      \noalign{\smallskip}
 & \multicolumn{3}{c}{SXT (0.5 -- 7.0 keV)} & &\multicolumn{3}{c}{LAXPC (3.0 -- 20.0 keV)}\\
             Obs. Date& $\Gamma$&   ${Flux}^{\mathrm{a}}$ & $\chi^{2}/dof$   &&  $\Gamma$& ${Flux}^{\mathrm{b}}$ &$\chi^{2}/dof$   \\
\hline
06-06-2017&2.014$\pm$0.073&2.887$\pm$0.106&45.67/46&&2.114$\pm$0.041&2.276$\pm$0.041&29.59/25\\
09-07-2017&1.873$\pm$0.027&6.013$\pm$0.087&199.44/211&&2.037$\pm$0.021&4.451$\pm$0.043&48.13/25\\
07-08-2017&1.930$\pm$0.037&3.502$\pm$0.068&158.21/148&&2.064$\pm$0.042&2.243$\pm$0.042&28.48/25\\
07-09-2017&2.030$\pm$0.068&1.945$\pm$0.064&71.47/63&&2.022$\pm$0.022&4.524$\pm$0.045&48.67/25\\
21-10-2017&1.897$\pm$0.064&2.322$\pm$0.080&53.93/60&&2.285$\pm$0.092&1.243$\pm$0.047&15.09/14\\
21-11-2017&2.100$\pm$0.100&1.688$\pm$0.078&29.31/32&&2.205$\pm$0.047&2.108$\pm$0.042&36.76/19\\
07-12-2017&1.858$\pm$0.084&2.065$\pm$0.089&54.31/43&&2.221$\pm$0.068&1.415$\pm$0.040&33.9/17\\
01-08-2018&1.849$\pm$0.037&4.311$\pm$0.087& 157.02/146&&2.171$\pm$0.031&2.969$\pm$0.040&30.05/24\\
11-09-2018&1.939$\pm$0.038&3.699$\pm$0.072& 140.95/152&&2.564$\pm$0.055&1.875$\pm$0.039&26.55/17\\
28-07-2021&2.131$\pm$0.019&2.058$\pm$0.019&342.3/267 &&2.366$\pm$0.030&1.258$\pm$0.019&40.59/25\\
05-08-2021&2.130$\pm$0.034&1.234$\pm$0.020&248.37/214&&2.280$\pm$0.058&0.936$\pm$0.022&37.88/23\\
\hline
\end{tabular}
}
\label{tab:table2}
\begin{list}{}{}
\item[$^{\mathrm{a}}$] Unabsorbed $Flux$ in units of $10^{-11}$ erg cm$^{-2}$ s$^{-1}$ in 0.5--7 keV energy range.
\item[$^{\mathrm{b}}$] $Flux$ in units of $10^{-11}$ erg cm$^{-2}$ s$^{-1}$ in 3--20 keV energy range.
\end{list}
\end{table*}

\citet{Abe_2024} have carried out a detailed multiwaveband analysis of 1ES\,2344+514. 
They used AstroSat-SXT data observed on 2021 August 6 and fitted the spectrum using a log-parabola model. We also used these data for our spectral analysis. The measured values of the spectral parameters presented in this paper are found to be consistent with those obtained by \citet{Abe_2024} (see their Table 3). 

\subsection{\textit{Swift}-XRT and \textit{NuSTAR}}
\label{sec:xrt_nus} % used for referring to this section from elsewhere
The near-simultaneous X-ray data from \textit{Swift}-XRT and \textit{NuSTAR} observations were chosen for this work. The details of the observations are listed in Table~\ref{tab:table1}. The combined and individual spectral fitting for both instruments was discussed in sections ~\ref{sec:xrt} and ~\ref{sec:nustar}. 

\begin{table*}[!ht]%[h!]
      \centering
      \caption{ Best-fit absorbed power-law results for \textit{Swift}-XRT and \textit{NuSTAR} spectral analysis }
       \resizebox{\textwidth}{!}{
      \begin{tabular}{cccccccccc}
      \hline
      \hline
      \noalign{\smallskip}
             &   \multicolumn{3}{c}{\textit{Swift}-XRT (0.3 -- 8 keV)}&   \multicolumn{3}{c}{\textit{NuSTAR}-FPMA (3 -- 79 keV)}&   \multicolumn{3}{c}{\textit{NuSTAR}-FPMB (3 -- 79 keV)}\\
             \hline
               Obs. Date&    $\Gamma$&   ${Flux}^{\mathrm{a}}$ &   $\chi^{2}/dof$   &    $\Gamma$&   ${Flux}^{\mathrm{b}}$ &   $Chi^{2}$/DOF   &    $\Gamma$&   ${Flux}^{\mathrm{b}}$ & $\chi^{2}/dof$   \\
             \hline
             22-07-2020&   2.025$\pm$0.029&1.834$\pm$0.032&129.66/140&2.318$\pm$0.045&1.254$\pm$0.025&57.48/72&2.293$\pm$0.046&1.291$\pm$0.026&80.17/72\\
             05-08-2022&   2.096$\pm$0.082&1.682$\pm$0.075&29.13/24&2.469$\pm$0.059&0.769$\pm$0.019&50.86/49&2.422$\pm$0.058&0.865$\pm$0.021&48.8/51\\
            \hline
      \end{tabular}
      }
      \label{tab:table3}
      \begin{list}{}{}
      \item[$^{\mathrm{a}}$] Unabsorbed $Flux$ in units of $10^{-11}$ erg cm$^{-2}$ s$^{-1}$ in 0.5--7 keV energy range.
      \item[$^{\mathrm{b}}$] $Flux$ in units of $10^{-11}$ erg cm$^{-2}$ s$^{-1}$ in 3--20 keV energy range.
      \end{list}
      \end{table*}

\begin{figure}[!htb]%[h!]
      \centering
      \includegraphics[width=0.65\linewidth, angle=270]{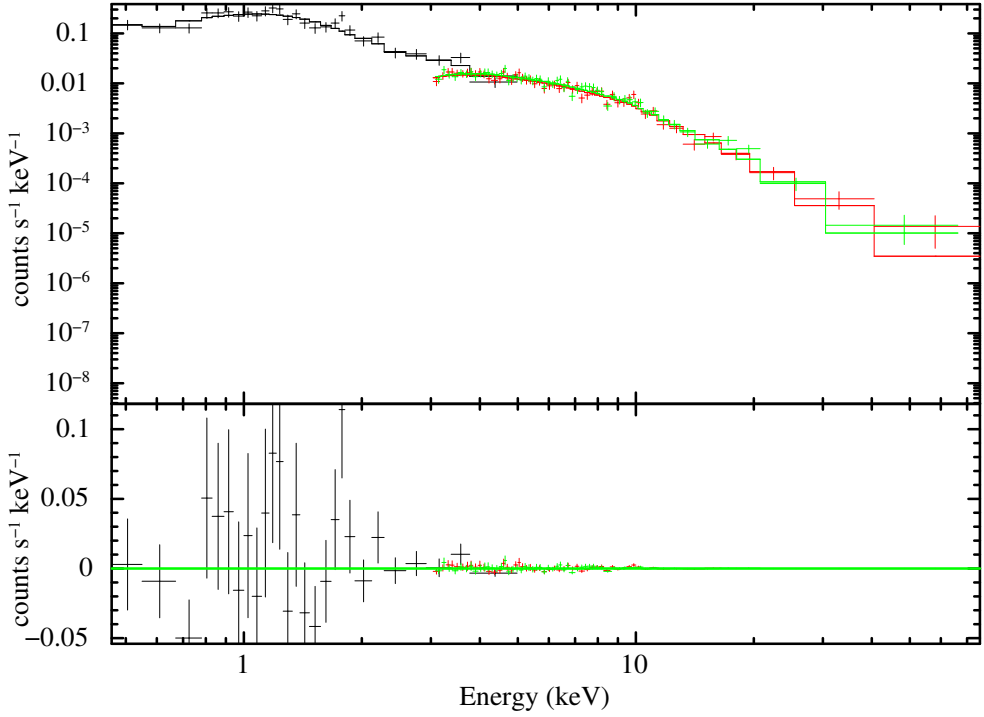}
      \caption{ 
      Joint fit for \textit{Swift}-XRT (black) and \textit{NuSTAR} FPMA/FPMB (red/green) spectra for 2020 July 22 observations with log-parabola model along with the line of sight absorption. Data over energy ranges 0.3-8 keV and 3-79 keV are used for \textit{Swift}-XRT and \textit{NuSTAR}, respectively. The top panel shows data and model fit, whereas the bottom panel shows the residuals. }
      \label{fig:figure2}
\end{figure}

The best-fitted spectral parameters along with unabsorbed flux are tabulated in Table~\ref{tab:table3} for two observation epochs, 2020 July 22 and 2022 August 5. The results of the combined fit parameters are listed in Table~\ref{tab:table4}. Figure~\ref{fig:figure2} shows the combined best-fitted spectral plot of for the \textit{Swift}-XRT (0.3 -- 8.0 keV) and \textit{NuSTAR}-FPMA/B (3 -- 79 keV) for 2020 July 22 observations. The constant component in the fitted model corresponds to the relative normalization between \textit{Swfit}-XRT and \textit{NuSTAR} FPMA/B, and the corresponding best-fitted peak energy ($E_{p}$) is listed in table~\ref{tab:table4}.

\begin{table*}[!htbp]%[h!]										
      \centering											
      \caption{Results of joint SXT-LAXPC and XRT-\textit{NuSTAR} spectral fit with constant*tbabs*logpar model. }	

       \resizebox{\textwidth}{!}{
      \begin{tabular}{llcccccc}					 																		
      \hline																							
      \hline																																									
          Instruments&Obs. Date	&	$\alpha$	&	$\beta$	&	Peak Energy($E_{p}$)	&	$\text{Constant}^{c}$ &	$Flux^{d}$	&	$\chi^{2}/dof$	\\
            \hline
          &06-06-2017	&	1.905	$\pm$	0.097	&	0.142	$\pm$	0.067	&	2.163	$\pm$	1.021	&	1.187	$\pm$	0.086	&	1.730	$\pm$	0.107	&	71.9/75	\\
          &09-07-2017	&	1.780	$\pm$	0.037	&	0.155	$\pm$	0.029	&	4.450	$\pm$	0.642	&	0.998	$\pm$	0.032	&	3.851	$\pm$	0.097	&	224.9/236	\\
          &07-08-2017	&	1.876	$\pm$	0.051	&	0.119	$\pm$	0.043	&	3.320	$\pm$	0.904	&	0.909	$\pm$	0.041	&	2.162	$\pm$	0.076	&	186.45/176	\\
          &07-09-2017	&	1.925	$\pm$	0.083	&	0.069	$\pm$	0.052	&	3.484	$\pm$	1.846	&	3.206	$\pm$	0.195	&	1.218	$\pm$	0.065	&	 116.47/88	\\
          &21-10-2017	&	1.637	$\pm$	0.100	&	0.487	$\pm$	0.103	&	2.361	$\pm$	0.318	&	0.903	$\pm$	0.074	&	1.343	$\pm$	0.084	&	54.64/74	\\
          SXT \& LAXPC&21-11-2017	&	2.010	$\pm$	0.134	&	0.130	$\pm$	0.091	&	0.913	$\pm$	1.138	&	2.145	$\pm$	0.211	&	0.929	$\pm$	0.078	&	64.86/50	\\
          &07-12-2017	&	1.640	$\pm$	0.120	&	0.398	$\pm$	0.097	&	2.834	$\pm$	0.488	&	1.059	$\pm$	0.093	&	1.257	$\pm$	0.089	&	78.24/60	\\
          &01-08-2018	&	1.732	$\pm$	0.052	&	0.272	$\pm$	0.040	&	3.106	$\pm$	0.335	&	1.020	$\pm$	0.041	&	2.658	$\pm$	0.089	&	184.23/170	\\
          &11-09-2018	&	1.737	$\pm$	0.058	&	0.517	$\pm$	0.058	&	1.796	$\pm$	0.152	&	1.079	$\pm$	0.052	&	1.885	$\pm$	0.070	&	169.26/169	\\
          &28-07-2021	&	2.016	$\pm$	0.027	&	0.234	$\pm$	0.032	&	0.924	$\pm$	0.130	&	1.221	$\pm$	0.036	&	1.039	$\pm$	0.022	&	303.94/264	\\
          &05-08-2021	&	2.034	$\pm$	0.047	&	0.206	$\pm$	0.052	&	0.826	$\pm$	0.247	&	1.485	$\pm$	0.072	&	0.626	$\pm$	0.022	&	249.6/204	\\
	\hline																							
            XRT \& \textit{NuSTAR}&22-07-2020&	1.850	$\pm$	0.036	&	0.275	$\pm$	0.034	&	1.854	$\pm$	0.185	&	1.241	$\pm$	0.123	&	1.049	$\pm$	0.031	&	266.43/297	\\
            &05-08-2022&	1.988	$\pm$	0.095	&	0.270	$\pm$	0.061	&	1.052	$\pm$	0.416	&	0.948	$\pm$	0.082	&	0.850	$\pm$	0.060	&	131.69/132	\\
																
      \hline															
      \end{tabular}
      }
      \label{tab:table4}	
\begin{list}{}{}
\item[$^{\mathrm{c}}$] The relative normalization between instruments. 
\item[$^{\mathrm{d}}$] Intrinsic/Unabsorbed $Flux$ in units of $10^{-11}$ erg cm$^{-2}$ s$^{-1}$ calculated using $Cflux$ multiplicative model in energy range 2 -- 10 keV.
\end{list}
\end{table*}

\begin{figure}[ht]%[h!]
      \centering
      \includegraphics[width = \linewidth]{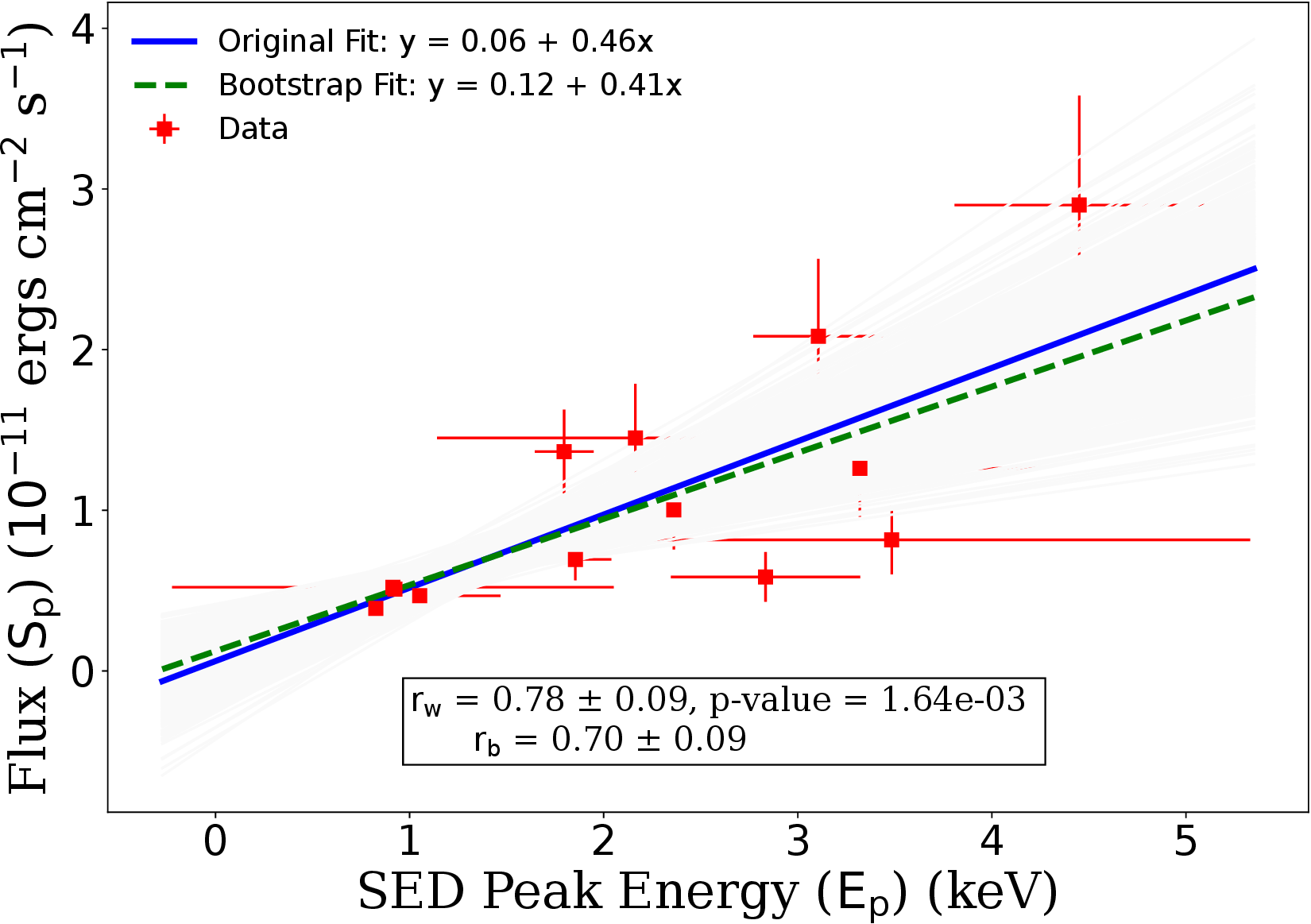}
      \caption{The scatter plot of estimated flux data at the synchrotron peak ($S_{\rm p}$) vs peak energy ($E_{\rm p}$) for SXT-LAXPC and XRT-\textit{NuSTAR} fitted with weighted linear regression and bootstrap linear regression. The grey lines are bootstrapped regression lines, the green dashed line is the mean of them, and the blue line is the weighted linear regression line.}      \label{fig:figure3}
\end{figure}

We investigated a correlation between the peak energy of SED ($E_{\rm p}$) and corresponding peak flux ($S_{\rm p}$) obtained using the log-parabola model. Figure~\ref{fig:figure3} shows the scatter plot of the peak energy of SED vs. the corresponding peak flux along with the fitted weighted and mean bootstrapped regression lines. A positive correlation with the weighted Pearson's correlation coefficient~\citep{fuller_1987,cheng_1999}, $r_{w}$ = $0.78\pm0.09$, the corresponding p-value = $1.64\times10^{-3}$, and the bootstrap mean correlation coefficient~\citep{efron_1979,carroll_1988}, $r_{b}$ = $0.70\pm0.09$ is seen between these parameters, indicating that the spectral index hardens and the peak of SED shifts towards higher energies during the flares. This is similar to the behaviour of other blazars such as Mrk 501, Mrk 421, and others \citep{Massaro_2004, Massaro_2006, Massaro_2008, Tramacere_2009} reported earlier.

\subsection{X-ray flux -- photon spectral index correlation}
\label{sec:x_corr}
\begin{figure}[ht]%[h!]
      \centering
      \includegraphics[width=1\linewidth]{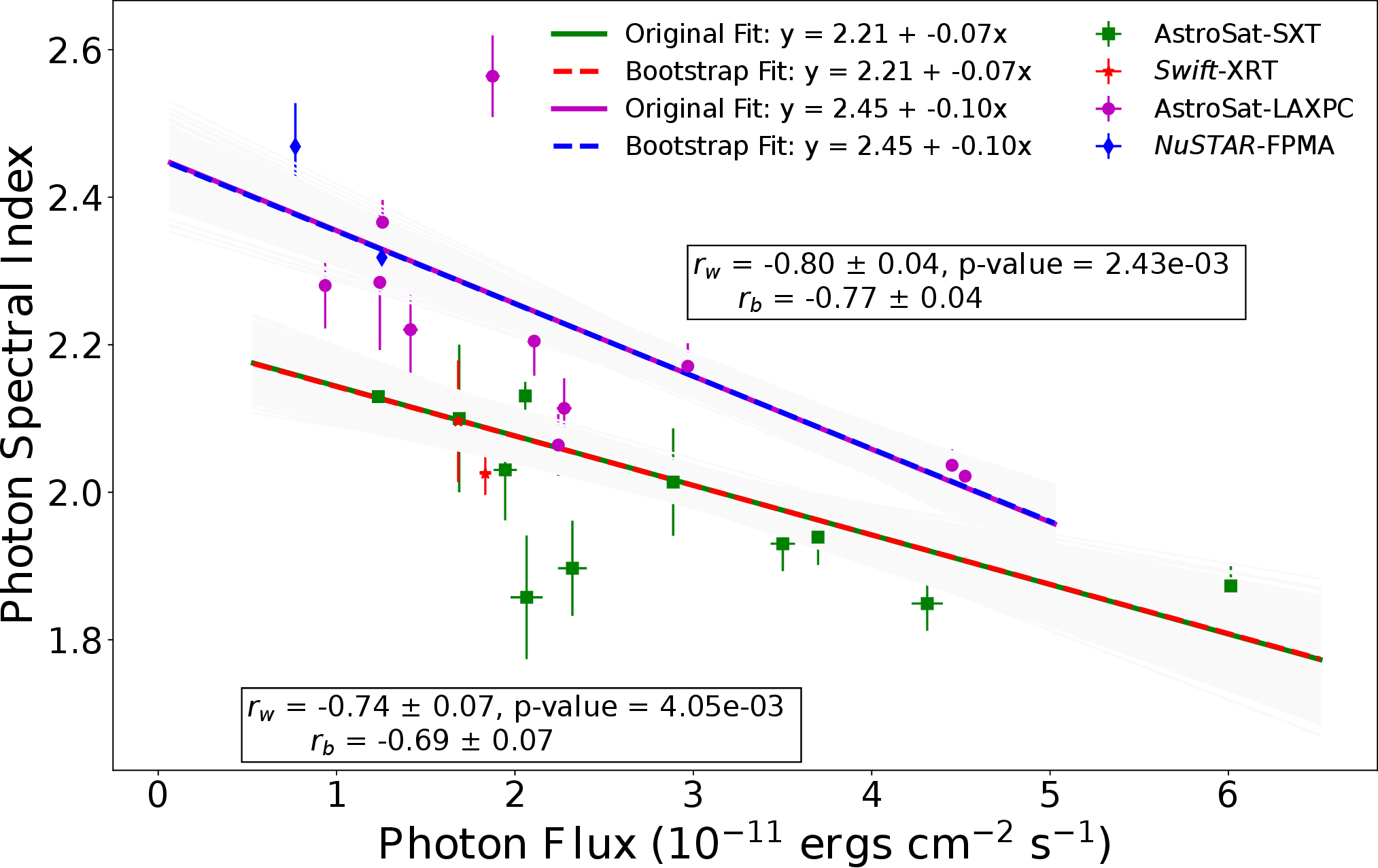}
      \caption{ The scatter plot illustrates the relationship between the power-law  photon spectral index and X-ray photon flux. Flux values for \textit{Swift}-XRT and SXT are calculated over the 0.5–7 keV energy range, while LAXPC and \textit{NuSTAR} cover the 3–20 keV range. These plots are analyzed using weighted linear regression and bootstrap linear regression. Gray lines represent bootstrapped regression lines for both plots. For the SXT-XRT data (bottom), the red dashed line indicates the mean of bootstrapped regression lines, and the green line shows the weighted linear regression. For the LAXPC-FPMA data (top), the blue dashed line represents the mean of bootstrapped regression lines, and the purple line depicts the weighted linear regression.
}

    \label{fig:figure4}
\end{figure}

We used simultaneous or quasi-simultaneous X-ray data between SXT and LAXPC onboard AstroSat and also between \textit{Swift}-XRT and \textit{NuSTAR} FPMA/B to study the index hardening/softening from 2017 June 6 to 2022 August 6 observations. We computed the X-ray unabsorbed flux values and the corresponding photon spectral indices using the power-law model, and the results are summarised in Tables~\ref{tab:table2} and \ref{tab:table3}. Figure~\ref{fig:figure4} shows the plot of the power-law photon spectral index ($\Gamma$) as a function of the X-ray integral flux for AstroSat-SXT and \textit{Swift}-XRT (0.5-7 keV) and AstroSat-LAXPC and \textit{NuSTAR}-FPMA/B (3-20 keV). The lower and higher energy data points are fitted separately. We observed a negative correlation between the flux and the $\Gamma$ for both plots. In this plot, the SXT-XRT data show a negative correlation with the weighted Pearson's correlation coefficient, $r_{w}$ = $-0.74\pm0.07$, the corresponding p-value = $4.05\times10^{-3}$, and the bootstrapped correlation coefficient, $r_{b}$ = $-0.69\pm0.07$. It was fitted with weighted linear regression and bootstrap linear regression with a slope of -0.10. In the same plot, the LAXPC-FMPA/B data also show a negative correlation with the weighted Pearson's correlation coefficient, $r_{w}$ = $-0.80\pm0.04$, the corresponding p-value = $2.43\times10^{-3}$, and the bootstrapped correlation coefficient, $r_{b}$ = $-0.77\pm0.04$. It was fitted with weighted linear regression and bootstrap linear regression with a slope of -0.07. This plot shows the `harder--when-brighter' trend in the X-ray energy region. The measured AstroSat-SXT and \textit{Swift}-XRT $\Gamma$ values range from 1.85 to 2.13 and the flux varies by a factor of 4.9 over the energy range of 0.5 -- 7 keV. Whereas for AstroSat-LAXPC and \textit{NuSTAR}-FPMA/B in the energy range 3 -- 20 keV, the $\Gamma$ varies from 2.02 to 2.56 and the flux varies by a factor of 5.1.

\subsection{\textit{Swift}-UVOT}
The corrected observed flux magnitudes obtained for various UVOT filters, following the procedure mentioned in section ~\ref{sec:swift_uvot}, were converted into the fluxes in the SED unit using zero-point magnitudes \citep{Poole2008}. The flux values obtained for the July 2020 and August 2022 observations for various filters are listed in Table~\ref{table:uvot_flux}.
   
 \begin{table*}[!ht]%[h!]
       \centering
         \caption{Summary of the {\textit{Swift}-UVOT} data analysis. The fluxes of six UVOT filters are reported.}

       \resizebox{\textwidth}{!}{
       \begin{tabular}{lccccccc}
       \hline
            Obs Date/Filters   &   & V & B   & U   & W1 & M2 & W2 \\
               & Units &   &   &   &   &   & \\
               \hline
               & &   &   &   &   &   & \\
             22-07-2020    & $10^{-11}$ $\text{erg}$ $\text{cm}^{-2}$ $\text{s}^{-1}$ & 2.322$\pm$0.076 & 1.476$\pm$0.044 & 0.853$\pm$0.031 & 0.717$\pm$0.031 & 0.700$\pm$0.038 & 0.602$\pm$0.025\\
             & &   &   &   &   &   & \\
             05-08-2022 & $10^{-11}$ $\text{erg}$ $\text{cm}^{-2}$ $\text{s}^{-1}$ &   2.155$\pm$0.088 & 1.432$\pm$0.053 & 0.768$\pm$0.037 & 0.620$\pm$0.037 & 0.697$\pm$0.051 & 0.559$\pm$0.031\\
             & &   &   &   &   &   & \\
               \hline
       \end{tabular}
       \label{table:uvot_flux}
       }
      \end{table*}

\subsection{\textit{Fermi}-LAT}
\label{sec:fermi_lat_an} % used for referring to this section from elsewhere
 
\begin{figure}[!htp]%[h!]
      \centering
      \includegraphics[width = \linewidth]{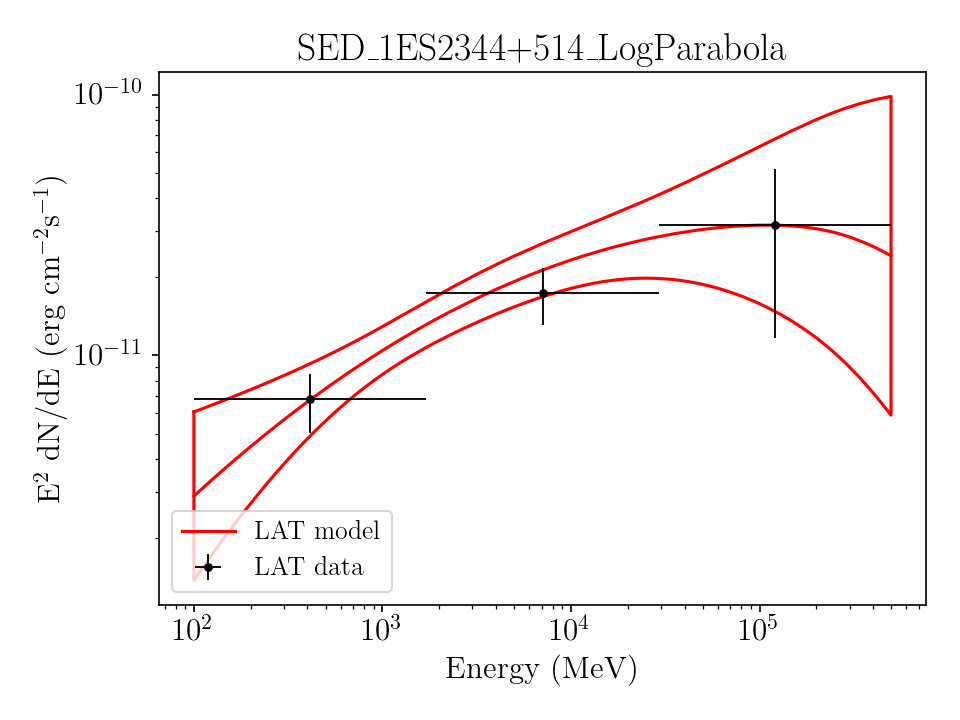}
      \caption{The $\gamma$-ray SED, extracted from the \textit{Fermi}-LAT data observed on 2017 July 9 and fitted with a log parabola model using the binned likelihood fit method with a bin size of 30 days.} 
      \label{fig:figure5}
\end{figure}

The $\gamma$-ray \textit{Fermi}-LAT SED of 2017 July fitted with a log-parabola model is shown in Figure~\ref{fig:figure5}. Here \textit{Fermi}-LAT observation stretches were selected centering on X-ray observation epochs. The observation duration was considered to vary from 30 days to 153 days, depending on the $\gamma$-ray flux level, to achieve good statistics.

\begin{figure}[ht]%[h!]
      \centering
      \includegraphics[width=1\linewidth]{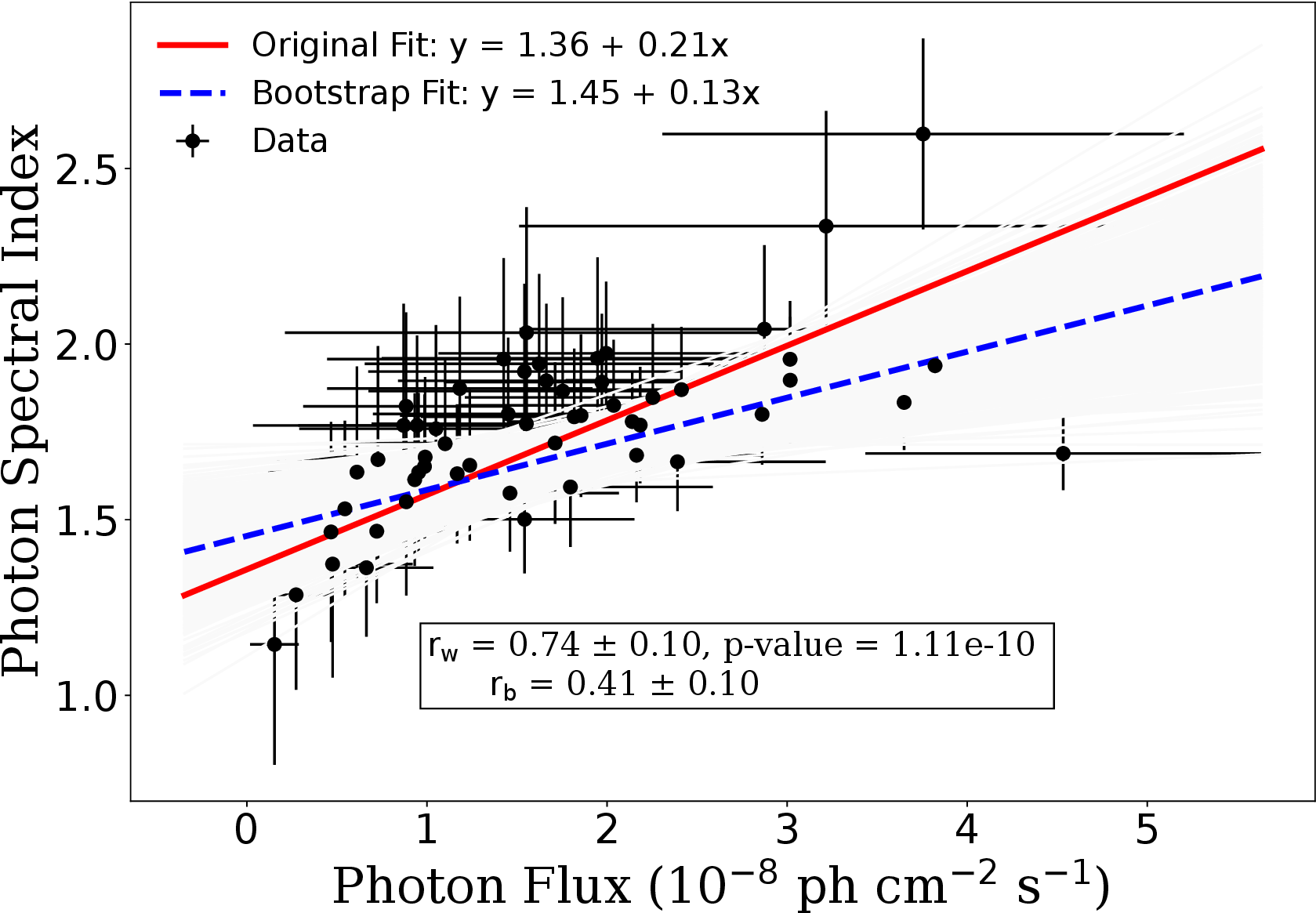}
      \caption{Plot for the correlation between $\gamma$-ray photon spectral index vs flux for 30-day binned \textit{Fermi}-LAT data. The data were fitted with weighted linear regression and bootstrap linear regression. The gray lines are bootstrapped regression lines, the blue dashed line is the mean of them, and the red line is the weighted linear regression line.
}
      \label{fig:figure6}
\end{figure}

The $\gamma$-ray light curve is extracted using the aforementioned likelihood analysis method on the LAT data divided into monthly time bins using the power-law model (see the bottom panel of Figure~\ref{fig:figure7}). Also, the flux-$\Gamma$ correlation for $\gamma$-ray data has been studied. Figure~\ref{fig:figure6} shows the photon flux vs $\Gamma$ plot fitted with weighted linear regression and bootstrap linear regression.   
A positive correlation between photon flux and $\Gamma$ is observed with weighted Pearson's correlation coefficient, $r_{w}$ = $0.74\pm0.10$, the corresponding p-value = $1.11\times10^{-10}$, and the bootstrapped correlation coefficient, $r_{b}$ = $0.41\pm0.10$ in $\gamma$-rays. The trend follows the weighted linear regression line with a slope of 0.21, and the bootstrap mean linear regression line with a slope of 0.13, indicating a `softer-when-brighter' trend.

\subsection{Multiwavelength light curves} 
\label{sec:multi_lc} % used for referring to this section from elsewhere

\begin{figure*}[h!]
      \centering
      \includegraphics[width=0.9 \textwidth]{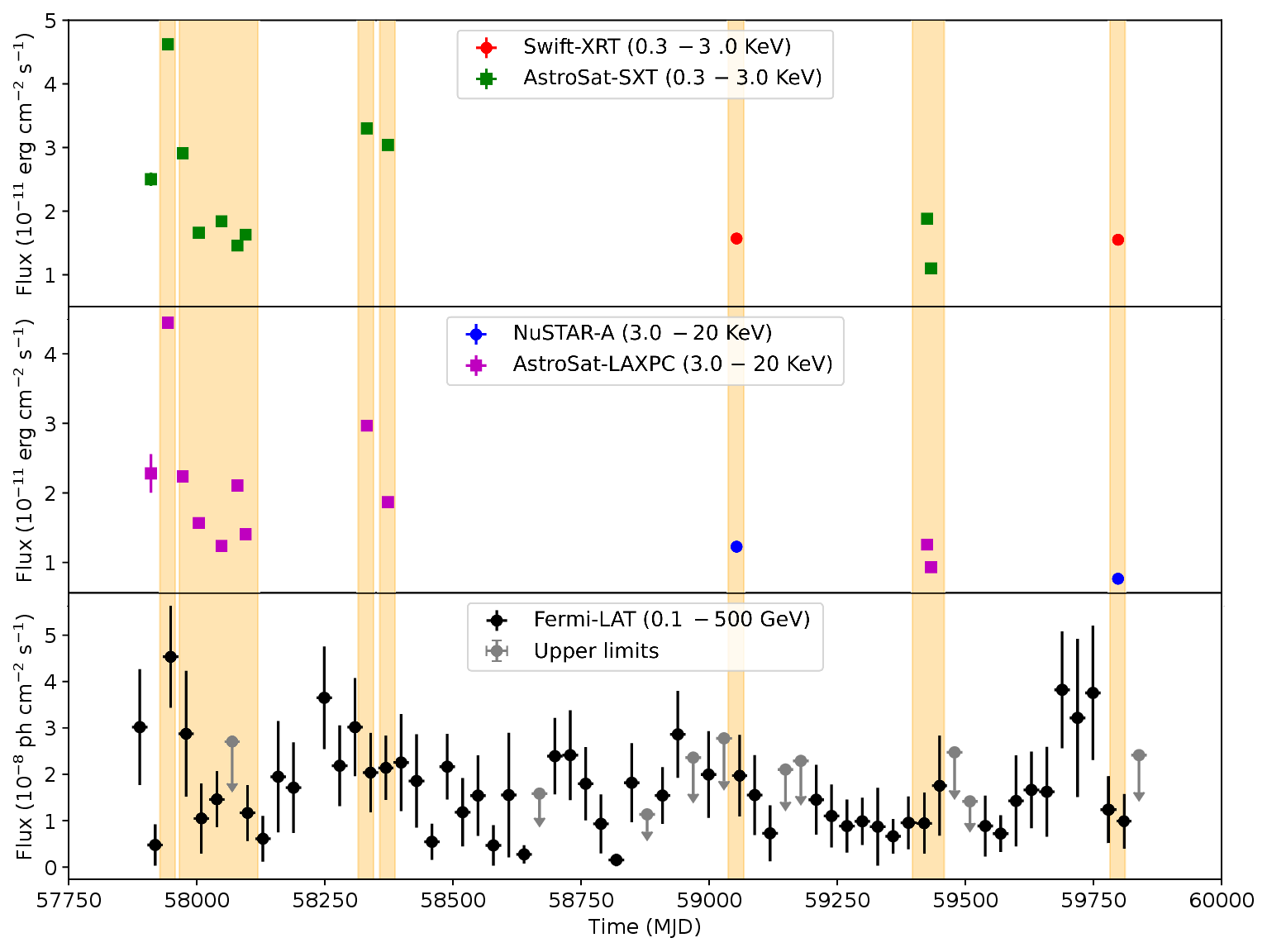}
      \caption{Multiwavelength light curves of 1ES 2344+514 between 2017 June 6 to 2022 August 6. The top panel shows the \textit{Swift}-XRT (red) and AstroSat-SXT (green) fluxes in 0.3 -- 3.0 keV. The second panel from the top displays the \textit{NuSTAR}-FPMA (blue) and AstroSat-LAXPC (cyan) fluxes in the energy range of 3.0 -- 20 keV. The third panel from the top represents the \textit{Fermi}-LAT (black) fluxes in the 0.1 -- 500 GeV energy band with 30 days binning and the upper limits. Yellow shaded areas correspond to epochs for which SEDs are generated.}
      \label{fig:figure7}
\end{figure*}

The multiwavelength light curves from 2017 June 6 to 2022 August 6 spanning the X-ray and $\gamma$-ray bands are shown in Figure~\ref{fig:figure7}. The top panel of the figure displays observation-wise averaged \textit{Swift}-XRT and AstroSat-SXT fluxes in the 0.3 -- 3.0 keV energy range.    
The middle panel shows the \textit{NuSTAR} and LAXPC fluxes in the energy range of 3.0 -- 20 keV for various observations. In the lowermost panel, the \textit{Fermi}-LAT fluxes in the 0.1--500 GeV energy band are plotted with monthly time binning, as shown by black markers, and the upper limits are shown using gray markers. A bright flare was observed by SXT and LAXPC onboard AstroSat on 2017 July 9 with a simultaneous increase in \textit{Fermi}-LAT flux. The source was reported to be in a flaring state in very high-energy $\gamma$-rays by MAGIC during July 2020 and August 2021. The epochs correspond to 30-day or longer duration of \textit{Fermi}-LAT observational windows centered around X-ray observations. The epochs are marked as a vertical yellow shaded area in Figure~\ref{fig:figure7}. Variability is seen in all X-ray and $\gamma-$ray light curves.

\section{Broadband Spectral Energy Distributions}
\label{sec:multi_sed}

To study the evolution of multiwaveband SEDs with time and to understand underlying emission mechanisms, SEDs were generated for various epochs, where X-ray observations are available. The epochs with quasi-simultaneous X-ray and gamma-ray observations are marked with the yellow shaded area (see Figure~\ref{fig:figure7}). Multiwaveband SEDs were generated and studied for these epochs, typically using the \textit{Fermi}-LAT data with a duration of 30 days. In the case of low-flux $\gamma$-rays , longer duration data, for example, 62-days or 153-days, have been used while generating SEDs. 

\begin{figure}[ht]%[h!]
      \centering
      \includegraphics[width = \linewidth]{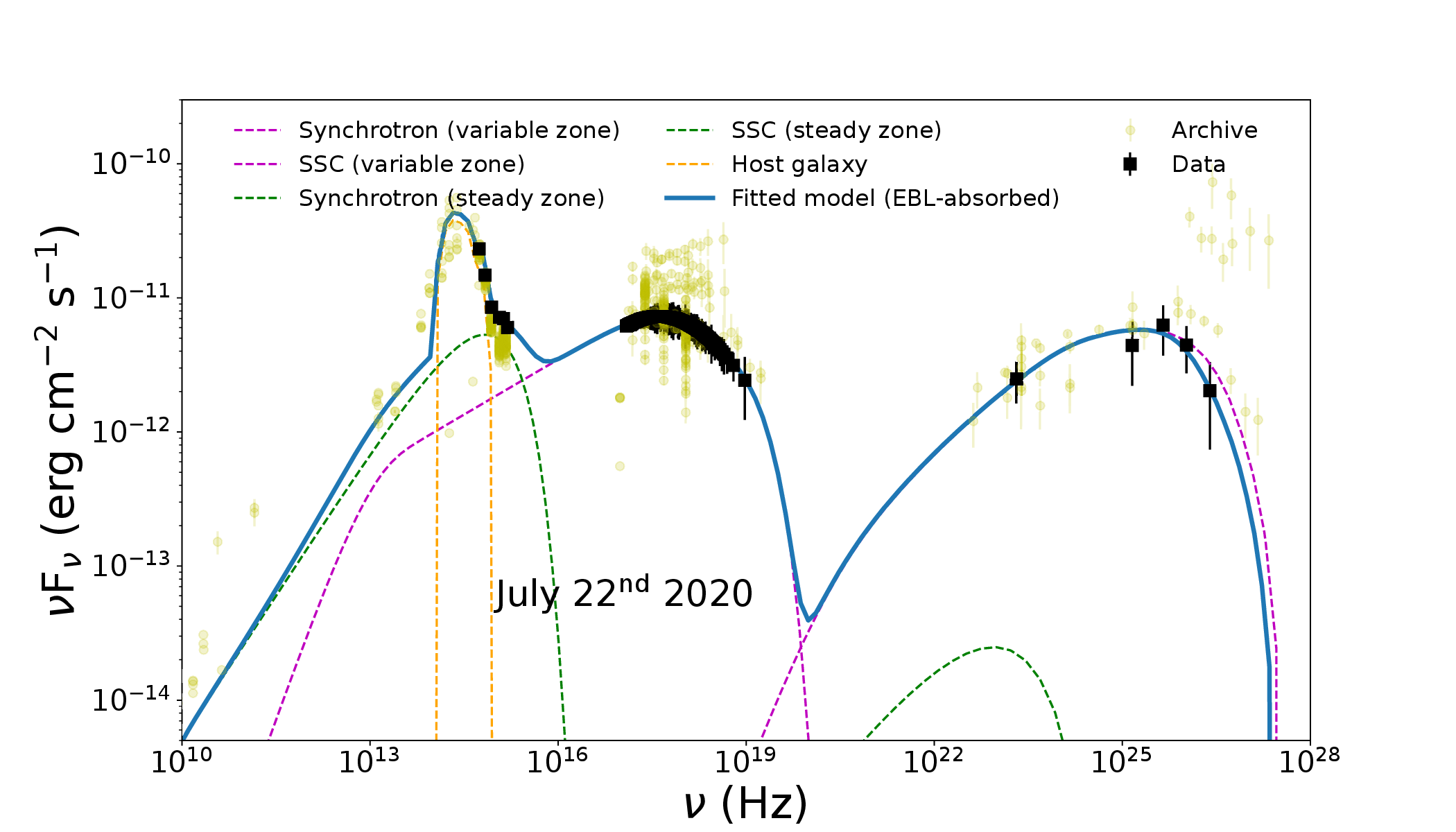}
      \caption{Broad-band SED of \textit{Swift}-UVOT, \textit{Swift}-XRT, \textit{NuSTAR}, \textit{Fermi}-LAT, and MAGIC observations during 2020 July (fainter state) fitted with a two-zone synchrotron+SSC model along with host galaxy component, and archival data shown as yellow points.}
      \label{fig:figure8}
\end{figure}

\begin{figure}[ht]%[h!]
      \centering
      \includegraphics[width = \linewidth]{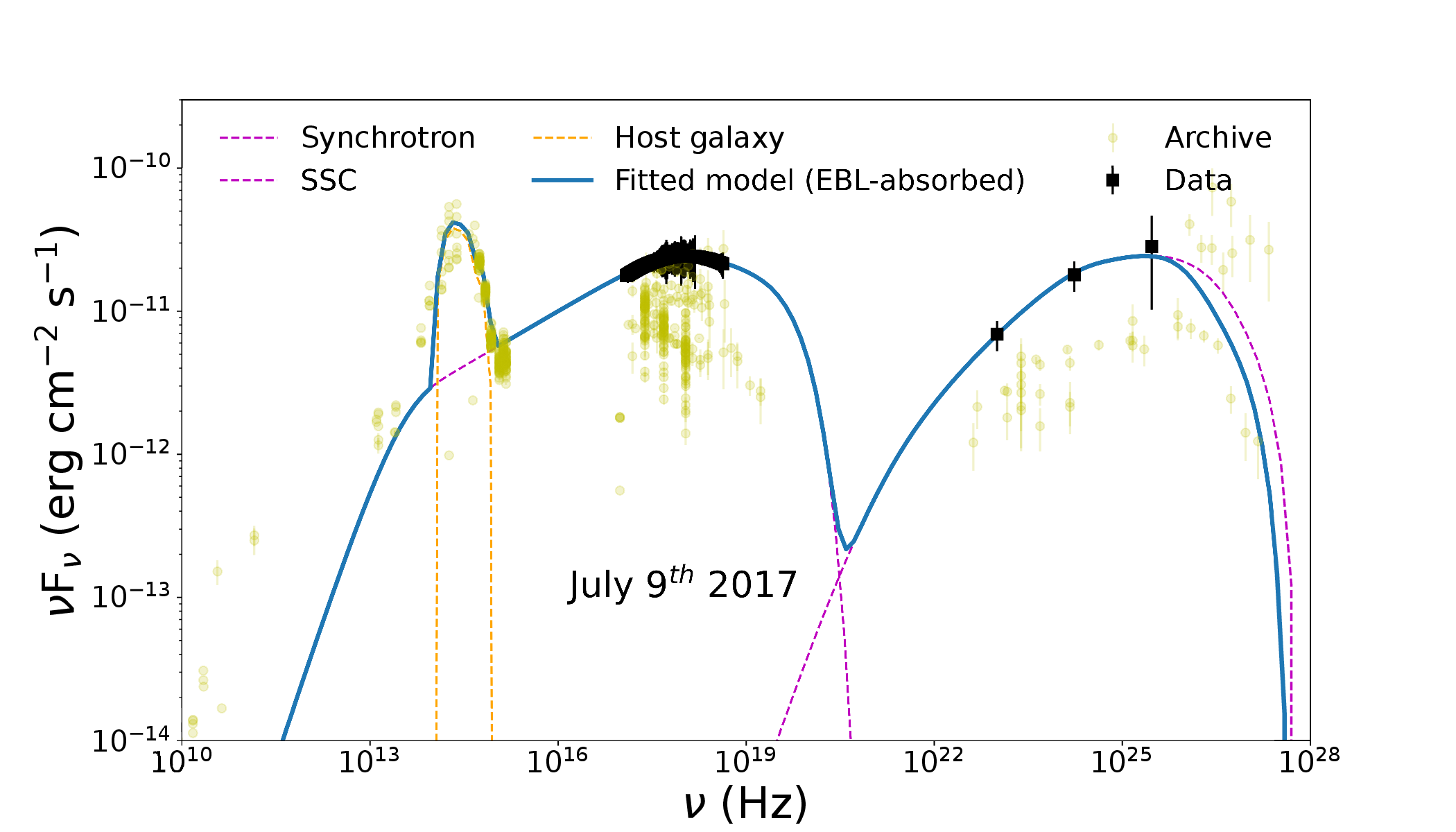}
      \caption{Broad-band SED of SXT and LAXPC onboard AstroSat and \textit{Fermi}-LAT observations during 2017 July (brightest state) fitted with a one-zone synchrotron+SSC model along with host galaxy component. Black points represent the analysed flux measurement for the selected epochs, and yellow points represent the archived data from the earlier studies.}
      \label{fig:figure9}
\end{figure}

Two examples of broadband SED are shown in Figures ~\ref{fig:figure8} and \ref{fig:figure9}. SED shown in Figure~\ref{fig:figure8} corresponds to the 2020 July epoch and is based on data from \textit{Swift}-UVOT, \textit{Swift}-XRT, \textit{NuSTAR}, \textit{Fermi}-LAT, and MAGIC. This corresponds to the quiescent state of the source. We have included TeV flux measurements published by the MAGIC collaboration (2024) in Figure~\ref{fig:figure8}. In the present analysis, we have used \textit{Swift}-XRT data instead of XMM-Newton data used by the MAGIC collaboration (2024). Also, we do not have VHE data for all of our SEDs. SED shown in Figure~\ref{fig:figure9} corresponds to the 2017 July epoch and is based on data from AstroSat-SXT, AstroSat-LAXPC, and \textit{Fermi}-LAT. This epoch corresponds to one of the brightest states of 1ES\,2344+514 in the X-ray band. Similar SEDs have also been generated for other epochs. Each SED is fitted with an one zone synchrotron+SSC model including a gamma-ray EBL absorption by the model of \citep{Franceschini_2017}. The re-absorption of the synchrotron radiation by the synchrotron electrons themselves (synchrotron self-absorption) at low frequencies below $\approx$ $10^{12}$ Hz is considered in this model. Archival SED data points (faded yellow) are retrieved from SSDC\footnote{\url{https://tools.ssdc.asi.it/SED/}} and used for comparison (Figures~\ref{fig:figure8} and \ref{fig:figure9}). A comparative study between archive data and SEDs from different epochs of the source show that the source was in its flaring state close to 2017 July 9, as can be seen from the X-ray lightcurves in Figure~\ref{fig:figure7}. To constrain the physical properties of the source 1ES\,2344+514, the single-zone leptonic scenario is used to model the emission with the publicly available numerical code \textsc{JetSe (V1.3.0)} provided by Andrea Tramacere \citep{Massaro_2006, Tramacere_2009, Tramacere_2011, Tramacere_2020}. In synchrotron+SSC models, the synchrotron photons produced in the magnetic field (B) get up-scattered by the relativistic electrons. The model assumes that the broadband emission originated from a spherical zone (blob) of radius (R), filled with relativistic electrons accelerated by an isotropic magnetic field. 
The population of electrons is modelled with a broken power-law energy distribution,

\begin{equation}
N(\gamma)d{\gamma} \propto \begin{cases} K{\gamma}^{-p_1}d{\gamma} & \ {\gamma}_{min} \leq {\gamma} \leq {\gamma}_{break} \\
K{\gamma_{break}}^{(p_2-p_1)}{\gamma}^{-p_2}d{\gamma} & \ {\gamma}_{break} < {\gamma} \leq {\gamma}_{max} \\
\end{cases}
\end{equation}
where $K$ is the normalization constant which corresponds to the electron energy density, denoted as $U_{\rm e}$. Here, $p_1$ and $p_2$ are the particle spectral indices below and above the break at energy $\gamma_{break}$. Also, $\gamma{\rm_{min}}$, $\gamma{\rm_{break}}$ and $\gamma{\rm_{max}}$ are the minimum, break, and maximum Lorentz factor of the injected electron population, respectively. The size of the emission region (R) and the position of the region ($R_{H}$) are set to 2 $\times$ $10^{\rm 16}$ cm and $10^{17}$ cm, respectively. The fitted SED parameters of the one-zone synchrotron+SCC model for all the epochs are listed in Table~\ref{tab:table6}.   

\begin{table*}[!htb]%[!htb]
\centering
%\begin{minipage}[]{100mm}
\caption{Best-fit parameters of the one-zone synchrotron+SSC model for observed SEDs The size of the emission region (R) is set to 2 $\times$ $10^{16}$ cm and $\delta$ to 10}

    \resizebox{\textwidth}{!}{
      \begin{tabular}{ccccccccccc}
      \hline
      \hline
      & & & & & Inner zone& & &&&\\
      \hline
      \hline
      %& & & &Model&Parameters& & &&&\\
            Obs date&   B &   $N_e$ &   $p_1$&   $p_2$&   $\gamma_{min}$ &   $\gamma_{max}$ &   $\gamma_{break}$ & $U_e$& $U_B$&$\eta$ = $U_B/U_e$ \\
      %      & & & & & & & &\\
            &   $10^{-2}$ G&   $cm^{-3}$&   &   & $10^{3}$&   $10^{6}$&$10^{5}$& erg cm$^{-3}$& erg cm$^{-3}$&   \\
% & & & & & & & &&&\\
 \hline
 \hline
% & & & & & & & &&&\\
             09-07-2017&   $5.10^{+1.92}_{-1.15}$&   $1.03^{+0.24}_{-0.14}$&   $2.51^{+0.01}_{-0.01}$&   $3.31^{+0.87}_{-0.25}$&   $3.66^{+0.52}_{-0.57}$&   $6.47^{+5.90}_{-2.62}$&   $6.88^{+3.52}_{-1.43}$& 9.12$\times$$10^{-3}$& 9.93$\times$$10^{-5}$& 1.09$\times$$10^{-2}$ \\
             07-08-2017&   $5.40^{+1.23}_{-0.80}$&   $0.17^{+0.05}_{-0.04}$&   $2.66^{+0.14}_{-0.10}$&   $3.26^{+0.74}_{-0.17}$&   $11.4^{+3.42}_{-2.24}$&   $5.67^{+3.87}_{-1.60}$&   $6.28^{+6.16}_{-1.76}$& 3.91$\times$$10^{-3}$& 1.14$\times$$10^{-4}$& 2.92$\times$$10^{-2}$ \\
             21-10-2017&   $2.38^{+0.70}_{-0.37}$&   $1.57^{+0.32}_{-0.35}$&   $2.30^{+0.08}_{-0.08}$&   $4.28^{+0.48}_{-0.44}$&   $1.88^{+0.56}_{-0.35}$&   $5.92^{+3.88}_{-2.43}$&   $9.90^{+1.20}_{-1.26}$& 9.16$\times$$10^{-3}$& 2.18$\times$$10^{-5}$& 2.39$\times$$10^{-3}$ \\
             07-12-2017&   $1.97^{+0.54}_{-0.39}$&   $1.33^{+0.27}_{-0.32}$&   $2.31^{+0.09}_{-0.08}$&   $4.10^{+0.58}_{-0.43}$&   $2.55^{+0.89}_{-0.55}$&   $8.17^{+6.63}_{-2.99}$&   $11.9^{+1.83}_{-2.21}$ &1.01$\times$$10^{-2}$& 1.65$\times$$10^{-5}$ &1.64$\times$$10^{-3}$ \\
             01-08-2018&   $3.87^{+0.82}_{-0.86}$&   $0.95^{+0.22}_{-0.23}$&   $2.36^{+0.09}_{-0.09}$&   $3.68^{+0.85}_{-0.95}$&   $2.92^{+0.76}_{-0.57}$&   $2.79^{+2.63}_{-0.99}$&   $9.08^{+1.79}_{-2.46}$& 7.78$\times$$10^{-3}$& 5.97$\times$$10^{-5}$ &7.67$\times$$10^{-3}$\\
             11-09-2018&   $4.07^{+0.10}_{-0.07}$&   $5.42^{+1.16}_{-1.08}$&   $2.30^{+0.07}_{-0.06}$&   $4.56^{+0.33}_{-0.33}$&   $0.59^{+0.15}_{-0.10}$&   $4.41^{+2.07}_{-1.43}$&   $6.59^{+0.66}_{-0.43}$& 1.75$\times$$10^{-2}$& 1.12$\times$$10^{-5}$& 6.41$\times$$10^{-4}$ \\
             %&   & && && && & &\\
             22-07-2020&$3.52^{+0.67}_{-0.56}$&   $1.11^{+0.29}_{-0.18}$&$2.46^{+0.10}_{-0.08}$&   $3.85^{+0.23}_{-0.17}$&$2.53^{+0.48}_{-0.48}$&   $7.39^{+6.19}_{-3.25}$&$6.67^{+0.85}_{-0.62}$& 6.73$\times$$10^{-3}$&5.02$\times$$10^{-5}$& 7.46$\times$$10^{-3}$ \\
             28-07-2021&   $4.11^{+0.84}_{-0.72}$&   $1.04^{+0.26}_{-0.19}$&   $2.18^{+0.11}_{-0.12}$&   $3.55^{+0.19}_{-0.22}$&   $1.11^{+0.24}_{-0.18}$&   $2.52^{+0.10}_{-0.46}$&   $3.61^{+0.50}_{-0.64}$& 4.09$\times$$10^{-3}$& 6.63$\times$$10^{-5}$& 1.56$\times$$10^{-2}$ \\
             05-08-2021& $2.15^{+0.36}_{-0.25}$& $0.53^{+0.14}_{-0.10}$&   $2.08^{+0.13}_{-0.06}$&   $3.34^{+0.18}_{-0.19}$&   $2.76^{+0.64}_{-0.46}$&   $2.62^{+0.61}_{-0.37}$&   $3.85^{+0.81}_{-0.76}$& 5.99$\times$$10^{-3}$& 1.46$\times$$10^{-5}$ &2.44$\times$$10^{-3}$ \\
             05-08-2022& $1.89^{+0.63}_{-0.42}$& $0.25^{+0.08}_{-0.05}$&   $2.43^{+0.22}_{-0.14}$&   $3.92^{+0.45}_{-0.36}$&   $14.5^{+4.32}_{-4.21}$&   $5.46^{+7.03}_{-2.04}$&   $7.72^{+1.60}_{-1.78}$& 8.65$\times$$10^{-3}$& 1.39$\times$$10^{-5}$& 1.61$\times$$10^{-3}$ \\
 %            &   &   &   &   &   &   &   & &&\\
             \hline
             \hline
             &   & &&& Host Galaxy &&    & &&\\
             \hline
             \hline
                &nuFnu\_p\_host & &   &erg $cm^{-2}$$s^{-1}$   &   &   &-10.4   &&& \\
               &nu\_scale &&    & Hz &   &    & -4.2$\times$$10^{-3}$ & &&\\
            \hline
            \hline
            %&   & &&& Outer zone &&    & &&\\
            %\hline
            %\hline
            %&   & &&& &&    & &&\\
            %\hline
            %\hline
             
      \end{tabular}
      }
      \label{tab:table6}
    % \label{tab:my_label}
\end{table*}

The SED, consisting of X-ray and gamma-ray data, could be fitted with a single-zone synchrotron+SSC model. However, SED with optical-UV data could not be fitted satisfactorily just with a combination of single zone synchrotron+SSC and host galaxy, as seen by us and as seen by the MAGIC collaboration (2024) earlier, due to excess emission in UV. To account for this UV excess, following the MAGIC collaboration (2024), we have added a second zone to the model so that the entire emission is the sum of contributions from two distinct spatially separated particle populations, along with the host galaxy emission. The second zone, referred to as the `core' zone by the MAGIC collaboration (2024), contributes mostly in radio with some contribution in the IR/optical/UV part of the spectrum. Whereas, the first zone, referred to as the `variable zone', dominates emission in the optical to the gamma-ray band.
The electron distribution in the second-zone component is modelled with the simple power-law function,

\begin{equation}
N(\gamma)d{\gamma} = K{\gamma}^{-p}, \,\,{\gamma}_{min} \leq {\gamma}\leq {\gamma}_{max}
\end{equation}
where $K$ is the normalization constant and $p$ is the particle spectral index. This simplified model was adopted due to the sparseness of the radio and optical/UV data.   
We use the Doppler factor of $\delta$ = 10 for both components, which are assumed to be streaming down the jet with the same speed \citep{Hovatta_2009,Abe_2024}. The size of the second zone is set to R = $10^{17}$ cm, similar to the size of the radio core reported by \citep{Aleksic_2013} at 15.4 GHz.

\begin{figure}
      \centering
      \begin{minipage}{0.49\textwidth}
            \includegraphics[width=\textwidth]{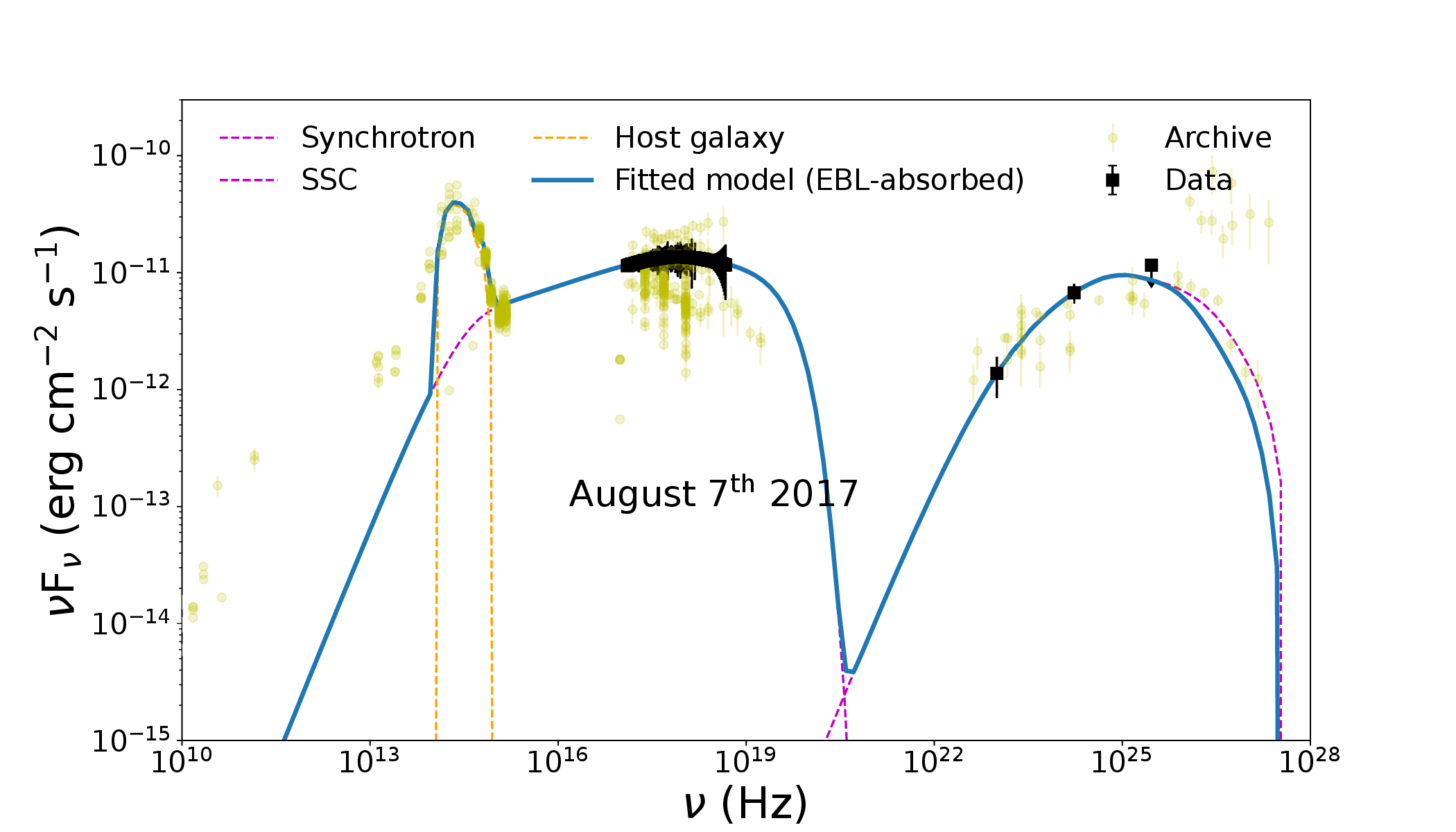}
            %\caption*{(a) Figure 1}
      \end{minipage}
      \begin{minipage}{0.49\textwidth}
            \includegraphics[width=\textwidth]{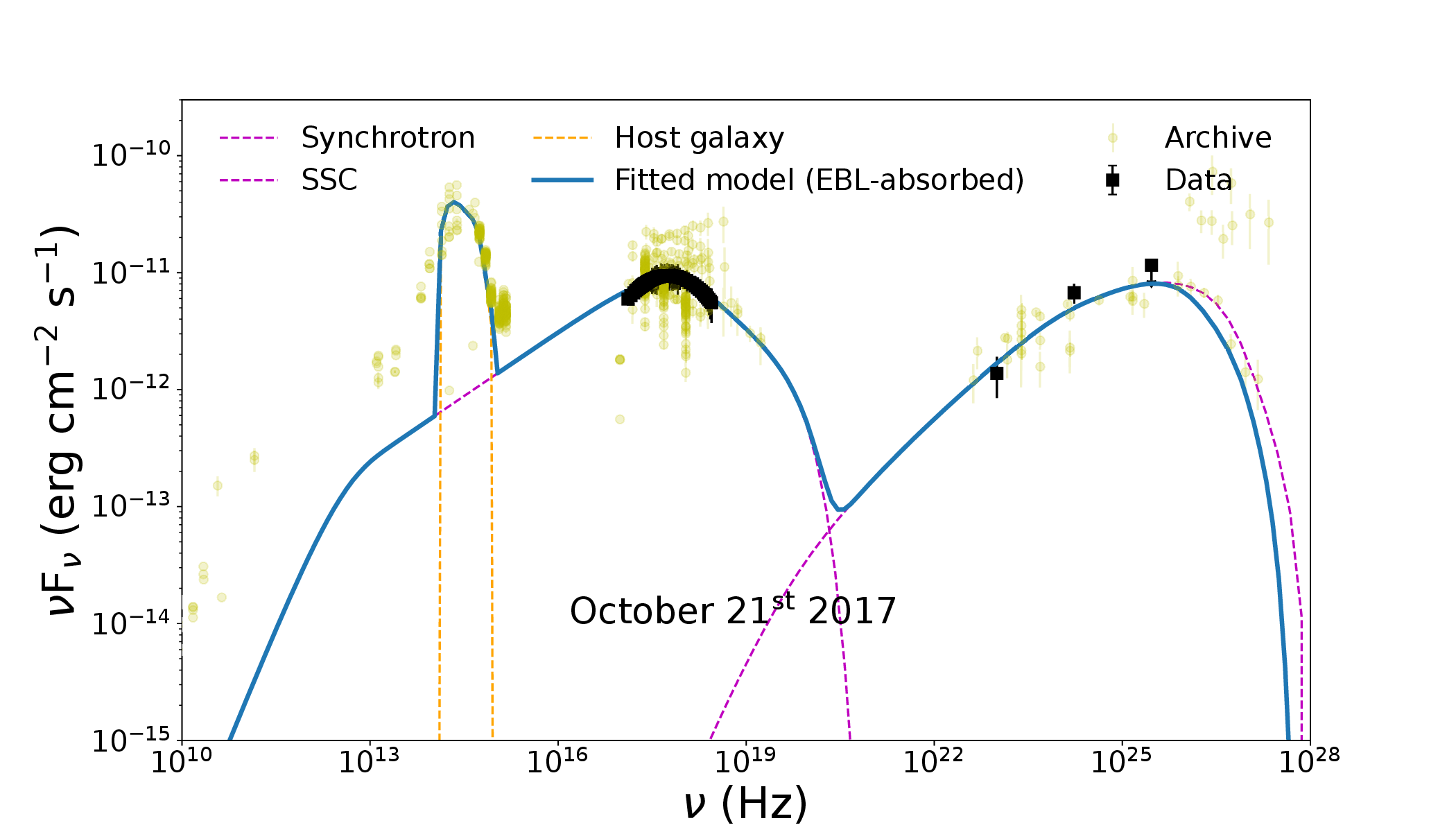}
            %\caption*{(b) Figure 2}
      \end{minipage}
      \vspace{0.2cm} % Add some vertical space
      
      \begin{minipage}{0.49\textwidth}
            \includegraphics[width=\textwidth]{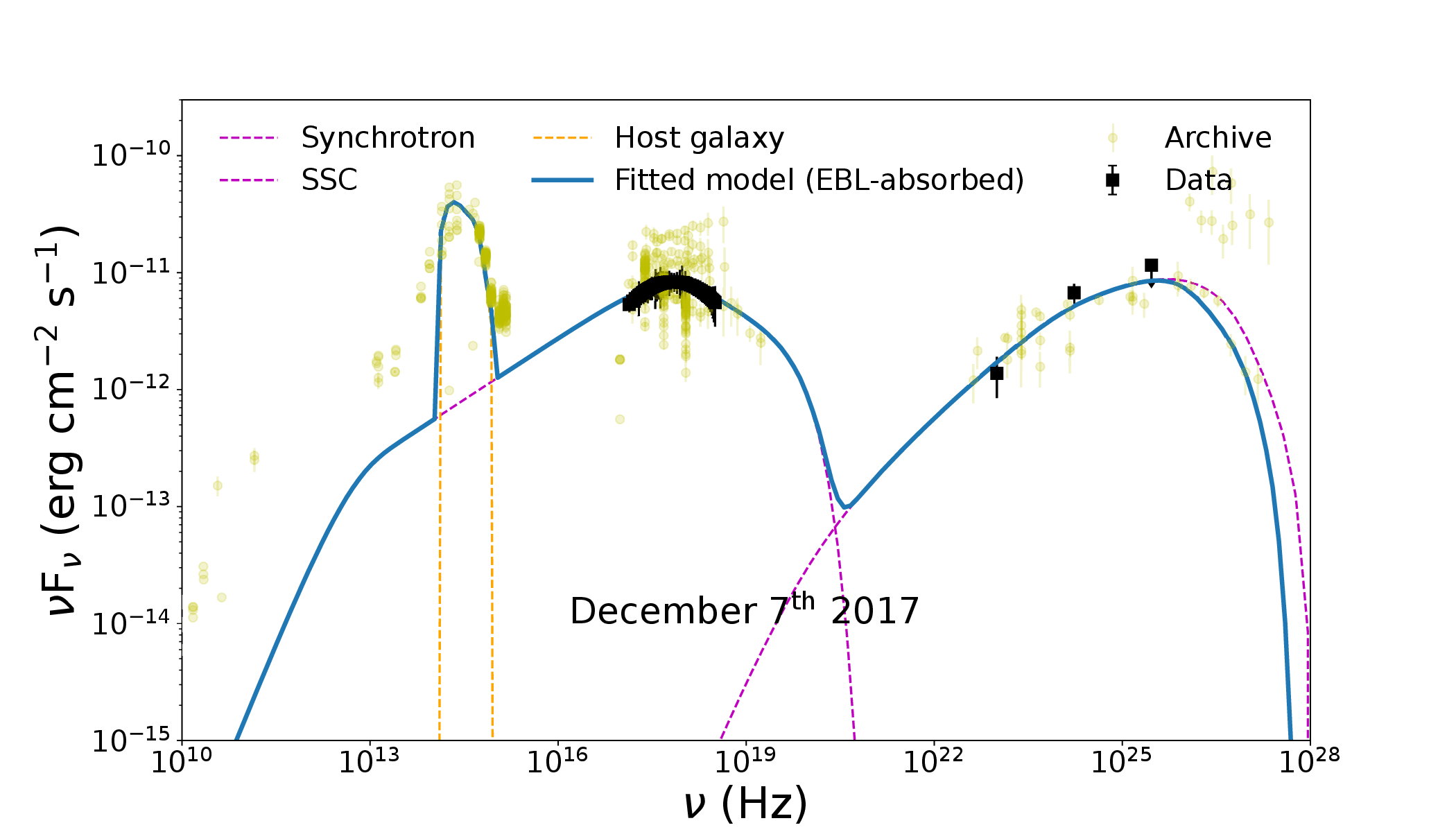}
            %\caption*{(c) Figure 3}
      \end{minipage}
      \begin{minipage}{0.49\textwidth}
            \includegraphics[width=\textwidth]{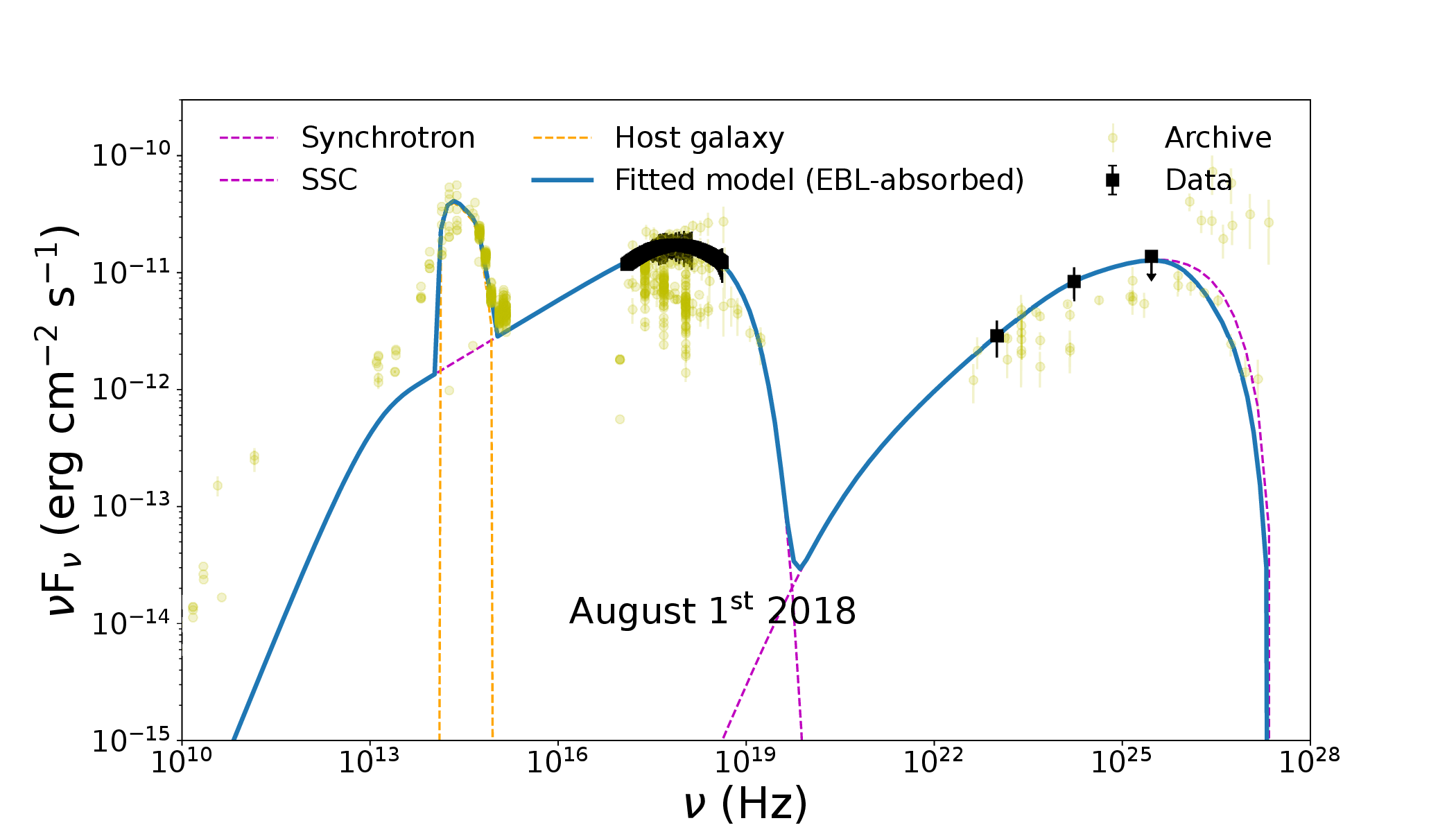}
            %\caption*{(d) Figure 4}
      \end{minipage}

      \vspace{0.2cm} % Add some vertical space
      \begin{minipage}{0.49\textwidth}
            \includegraphics[width=\textwidth]{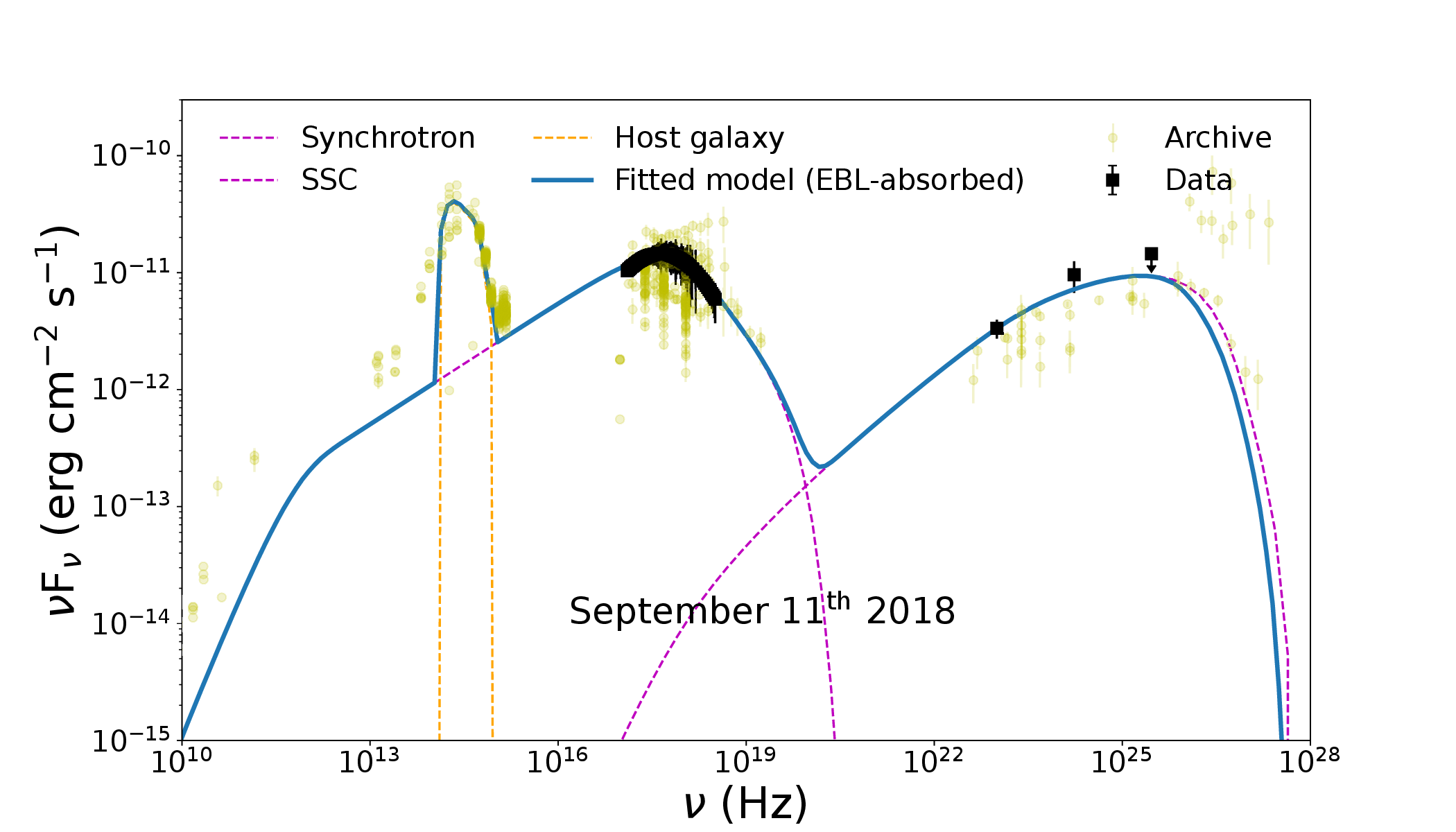}
            %\caption*{(e) Figure 5}
      \end{minipage}
      \begin{minipage}{0.49\textwidth}
            \includegraphics[width=\textwidth]{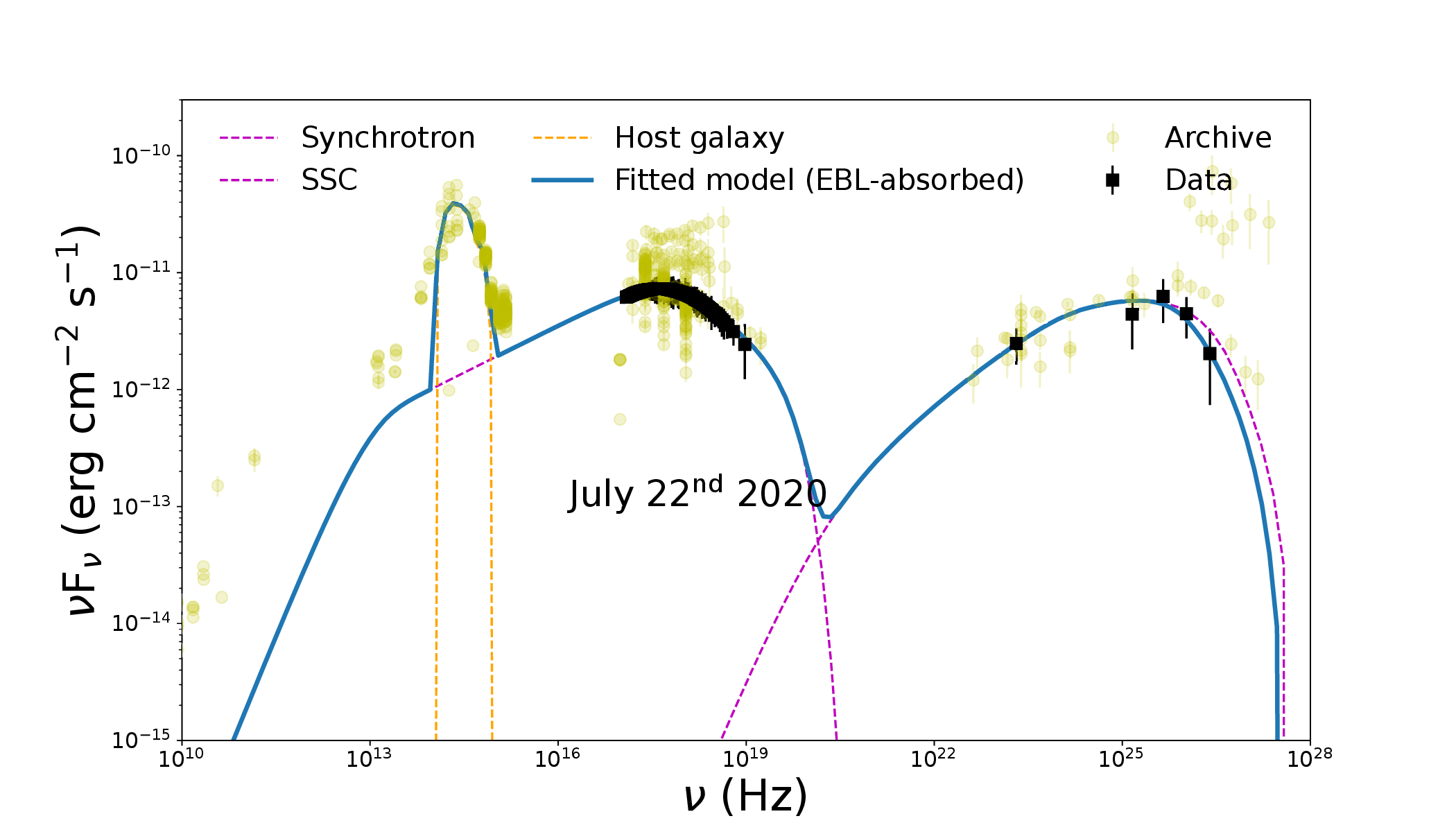}
            %\caption*{(f) Figure 6}
            \end{minipage}
      \vspace{0.2cm}
      \begin{minipage}{0.49\textwidth}
            \includegraphics[width=\textwidth]{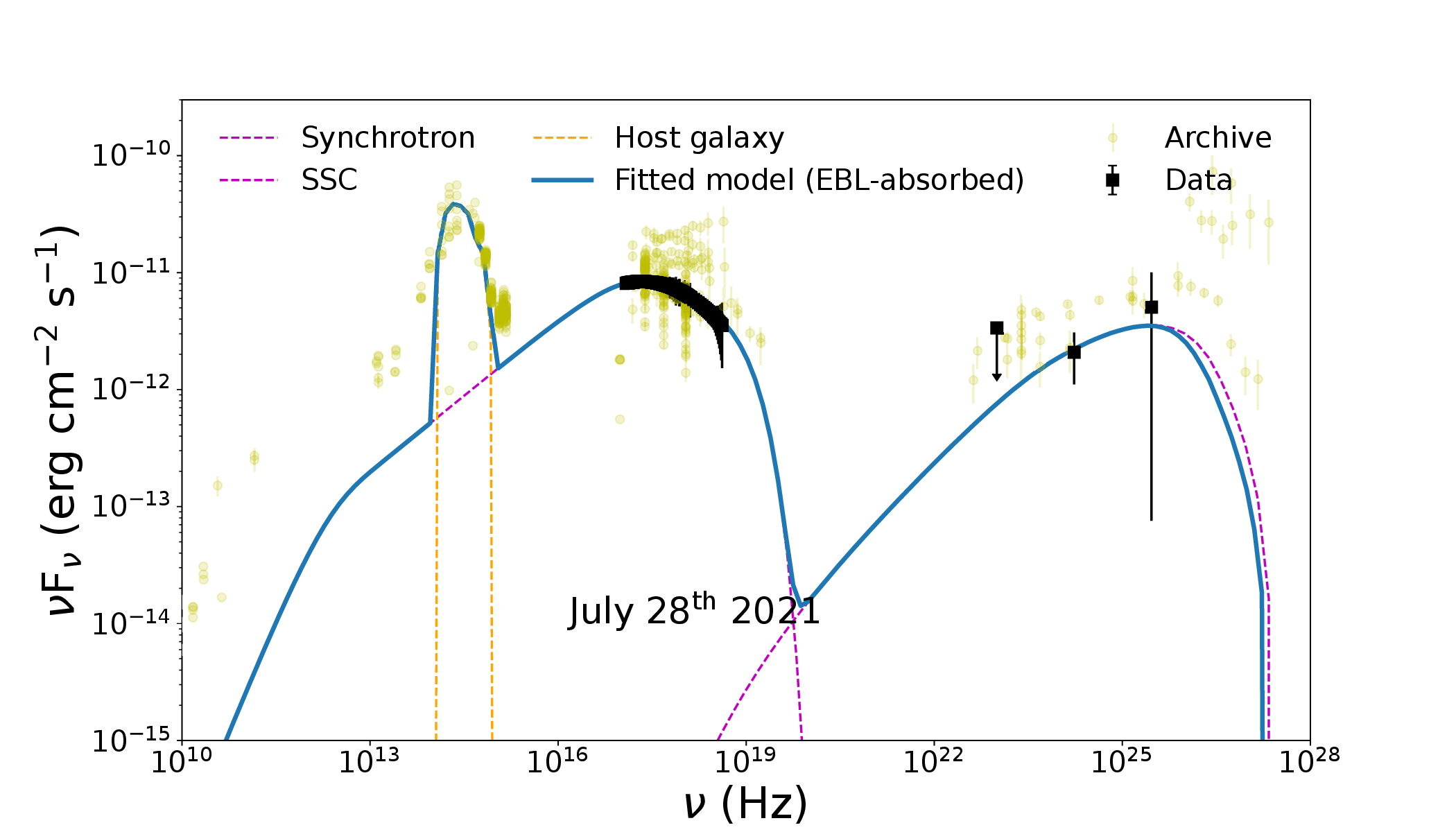}
          % \caption*{(g) Figure 7}
      \end{minipage}
      \begin{minipage}{0.49\textwidth}
            \includegraphics[width=\textwidth]{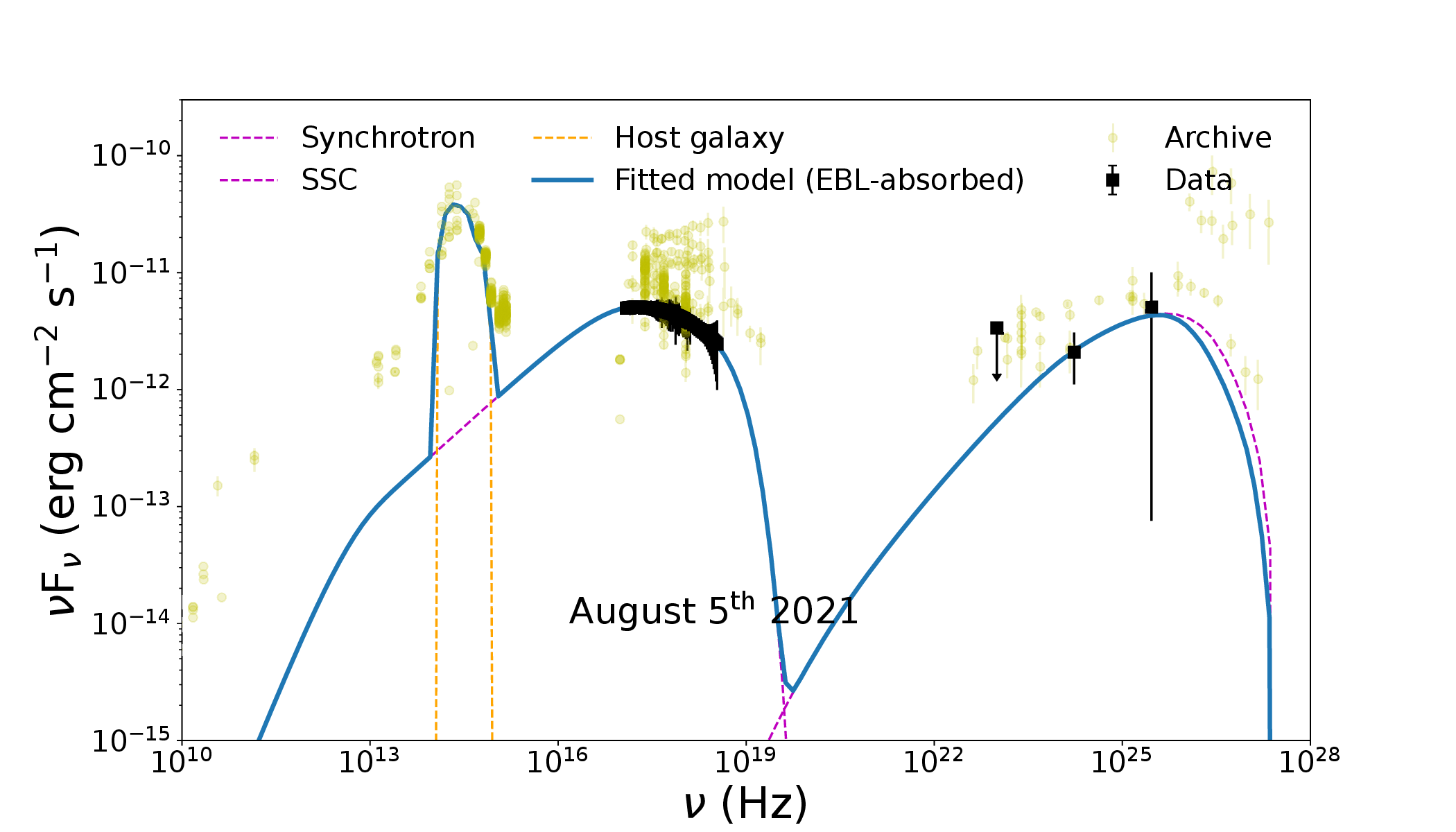}
            %\caption*{(h) Figure 8}
      \end{minipage}
      \vspace{0.2cm} % Add some vertical space
      
      \begin{minipage}{0.49\textwidth}
            \includegraphics[width = \textwidth]{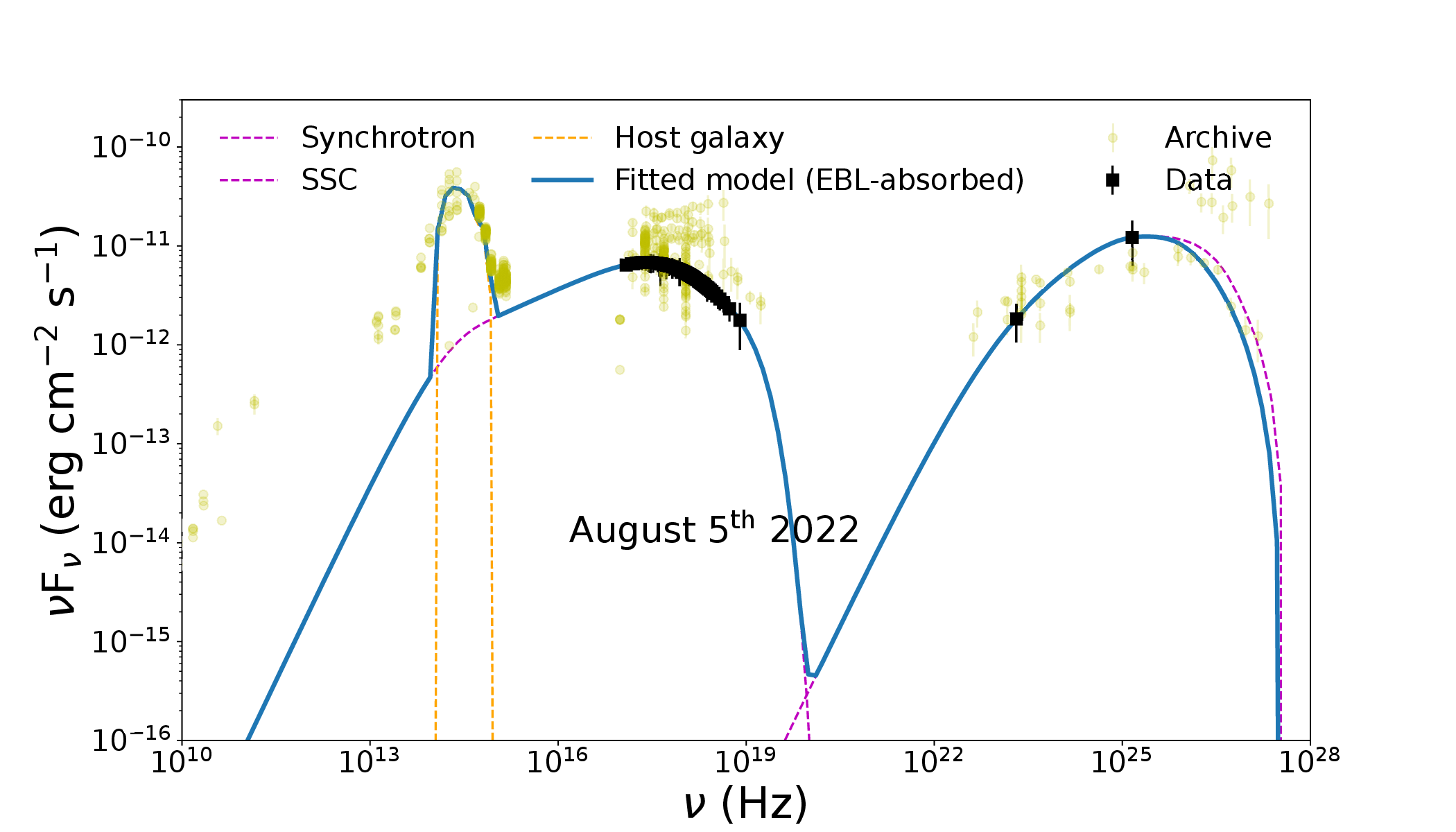} 
            %\caption*{(i) Figure 9}
      \end{minipage}
      \begin{minipage}{0.49\textwidth}
            \includegraphics[width=\textwidth]{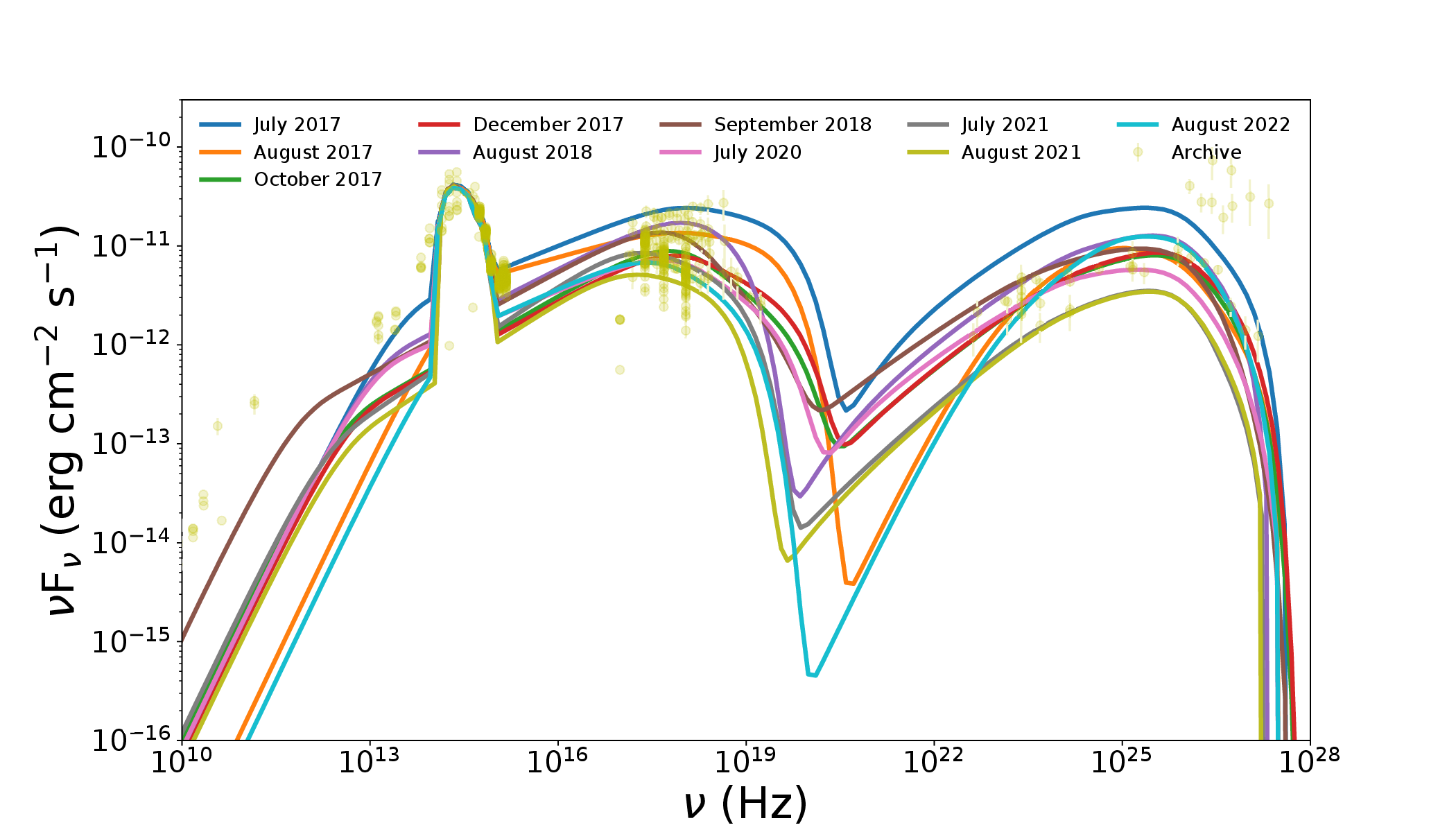}
            %\caption*{(j) Figure 10}
      \end{minipage}
      
      \caption{The nine SEDs fitted with one-zone synchrotron+SSC model for each epoch along with data from SXT and LAXPC onboard AstroSat, \textit{Swift}-XRT, \textit{NuSTAR}, \textit{Fermi}-LAT, and MAGIC during 2017 June 6 to 2022 August 6 and host galaxy component are shown in separate panels. Bottom-right panel shows the plot of fitted broadband SEDs model of the ten different epochs in a single panel for comparison, and here, 1ES\,2344+514 is observed in its brightest state during 2017 July.}
      \label{fig:figure10}
\end{figure}

The physical parameters of the second zone, as well as of the first zone, were obtained by fitting the SED for the epoch of 2020 July. The value of the parameters of the second zone is the particle density ($N_e$) = $0.23^{+0.20}_{-0.09}$ $cm^{-3}$, the magnetic field (B) = $0.22^{+0.09}_{-0.08}$ G, the low-energy spectral slope (p) = $1.76^{+0.12}_{-0.18}$, the minimum Lorentz factor ($\gamma_{min}$) = $1.5^{+1.2}_{-0.4}$, and the maximum Lorentz factor ($\gamma_{max}$) = $1.76^{+0.77}_{-0.46}$ $\times$ $10^{4}$. Whereas the value of parameters of the first zone is the particle density ($N_e$) = $1.04^{+0.23}_{-0.15}$ $cm^{-3}$, the magnetic field (B) = $3.36^{+0.0.64}_{-0.51}$ $\times$ $10^{-2}$ G, low energy spectral slope ($p_1$) = $2.44^{+0.09}_{-0.04}$, low energy spectral slope ($p_2$) = $3.85^{+0.29}_{-0.32}$, the minimum Lorentz factor ($\gamma_{min}$) = $2.56^{+0.49}_{-0.49}$ $\times$ $10^{3}$, the beak Lorentz factor ($\gamma_{break}$) = $7.70^{+1.15}_{-1.08}$ $\times$ $10^{5}$, and the maximum Lorentz factor ($\gamma_{max}$) = $4.49^{+3.13}_{-1.57}$ $\times$ $10^{6}$. The inner zone parameters are quite consistent with the results of \citet{Abe_2024} discussed in their Table 5. It is observed that the contribution of the second zone in X-ray and $\gamma$-ray is negligible. In addition, it is difficult to constrain parameters with a two-zone fit with the sparse nature of the data. So, in the rest of the work, we have not considered the second zone, which we refer to as the outer zone, and given results considering the single-zone synchrotron+ SSC model. In Figure ~\ref{fig:figure8}, the 2-component synchrotron+SSC modelling is shown using green dotted lines for emission from the ``zone-2" or ``outer zone" region and the violet dashed line for emission from the ``zone-1" or ``inner" region, whereas the blue solid line is the sum of both components along with the host galaxy contribution. 

\begin{figure}[!htp]%[h!]
      \centering
      \includegraphics[width = \linewidth]{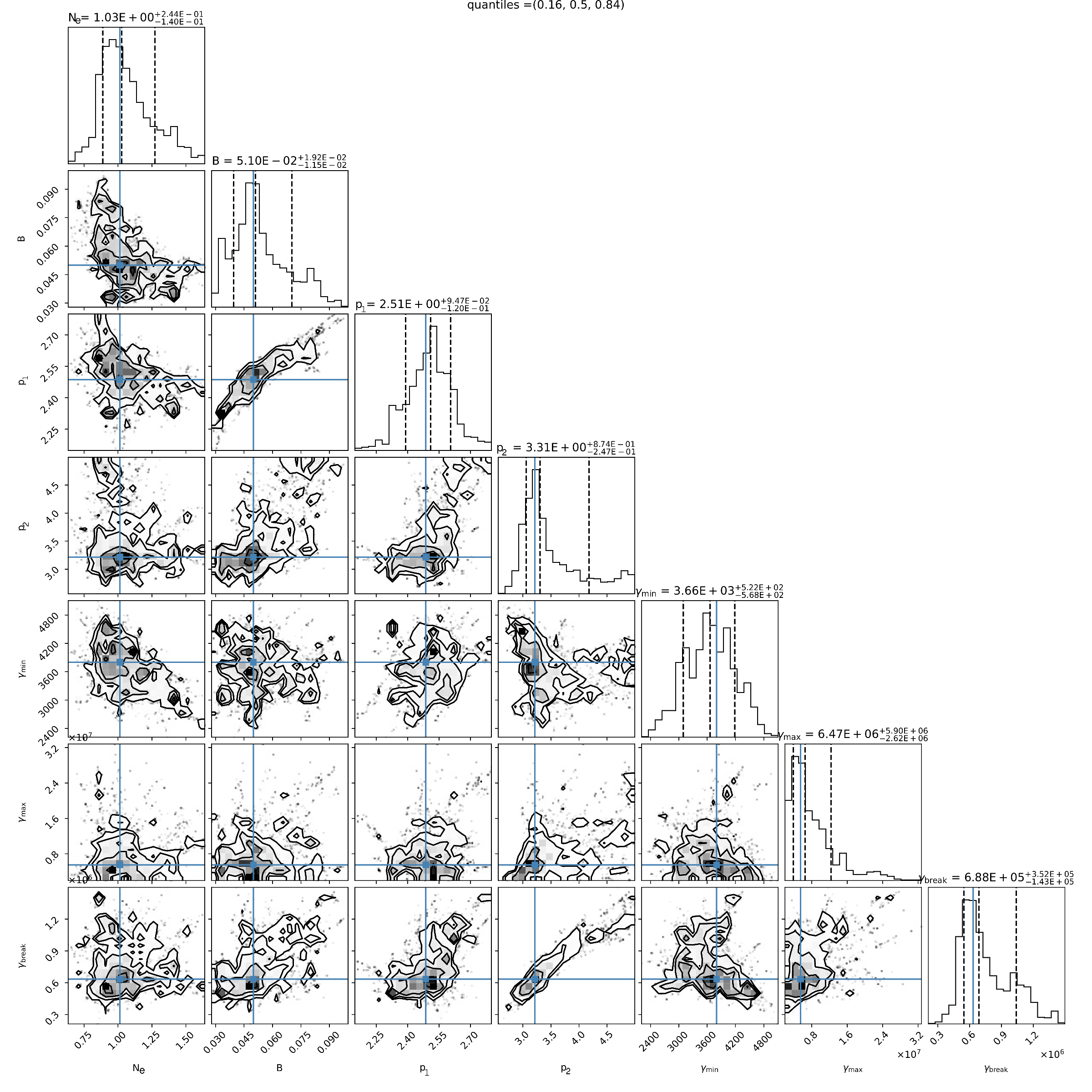}
      \caption{Corner plot of the posterior probability distribution of the free parameters from the SED fit of SXT and LAXPC onboard AstroSat and \textit{Fermi}-LAT observations during 2017 July fitted with single-zone synchrotron+SSC model. Vertical lines in 1D distributions show 16\%, 50\%, and 68\% quantiles.}
      \label{fig:figure11}
\end{figure}

A large part of the flux seen in the optical-UV region ($\sim$ $10^{14}$ -- $10^{15}$ Hz) from the archival data of long-term observation of 1ES\,2344+514 from SSDC, is due to the stellar emission from the host galaxy of the source. These fluxes are much higher than the non-thermal radiation from the jet. For modelling host galaxy contribution, we have used the archival data points from SSDC, and also the host galaxy component has been added while modelling the SED using \textsc{JetSeT}. The host galaxy contribution is shown by the orange dashed line in Figures ~\ref{fig:figure8} and \ref{fig:figure9}. The fits for all broadband SEDs fitted with a one-zone model, for various epochs during 2017 June 6 to 2022 August 6 based on data from SXT and LAXPC onboard AstroSat, \textit{Swift}-UVOT, \textit{Swift}-XRT, \textit{NuSTAR}, \textit{Fermi}-LAT, and MAGIC are shown in Fig.~\ref{fig:figure10}.

The Markov Chain Monte Carlo (MCMC) sampler is performed for all the SEDs to explore all the parameter space of various model components, such as the electron population parameter, magnetic field, the Doppler factor ($\delta$), the electron distribution indices ($p_1$ and $p_2$), and the break energy ($\gamma_{break}$) etc. using \textsc{JetSeT}. We used the corner plot to visualise the marginal posterior distributions of all the parameters, which highlights the uncertainties and parameter degeneracies. The marginal posterior distributions were also used to determine the correlation between the parameters due to degeneracies within the SED spectral fitting. We used 20 walkers with burning initial steps of 50 and a chain length of 500 for MCMC sampling of the one-zone synchrotron+SSC model. The one- and two-dimensional (1D and 2D) marginal posterior distributions for SED of 2017 July are shown in Fig.~\ref{fig:figure11}. The median values of the parameters with 68\% credible interval are shown above each 1D histogram.

\subsection{Jet Power Estimation from SED Modelling}
{The SED modelling of 1ES 2344+514 provides insights into the jet power, defined in general as the total power carried by the relativistic jet, \(P_{\text{jet}}\), following \citealt{Ghisellini_2001} and \citealt{ Celotti_2008} is
\begin{equation}
P_{\text{jet}} = L_e + L_p + L_B + L_{\text{rad}} \quad \text{(erg s$^{-1}$)},
\end{equation}
where \(L_B\), \(L_p\), \(L_e\), and \(L_{\text{rad}}\) represent the power carried by the magnetic field, cold protons, relativistic electrons, and produced radiation, respectively. The power of each component is calculated as
\begin{equation}
L_i \simeq \pi R^2 \Gamma^2 \beta c U_i,
\end{equation}
where \(U_i\) is the energy density of the \(i\)-th component in the comoving jet frame, \(\beta c\) is the blob’s bulk velocity, \(\Gamma\) is the bulk Lorentz factor, and \(R\) is the blob radius. For a jet at angle \(\theta\) to the line of sight, the Doppler factor is \(\delta = [\Gamma (1 - \beta \cos \theta)]^{-1}\).
}

{The power carried by relativistic electrons is
\begin{equation}
L_e \simeq \pi R^2 \Gamma^2 \beta c \langle \gamma \rangle N_e m_e c^2,
\end{equation}
where \(\langle \gamma \rangle\) is the average Lorentz factor of electrons in the comoving frame, \(N_e\) is the electron number density (an input parameter), and \(m_e\) is the electron rest mass. The proton power carried by 'cold' protons estimated assuming one proton per relativistic electron, is
\begin{equation}
L_p \simeq \pi R^2 \Gamma^2 \beta c N_p m_p c^2,
\end{equation}
where \(N_p \simeq N_e \) is the proton number density and \(m_p\) is the proton rest mass.}

{The magnetic field power, carried as Poynting flux, is
\begin{equation}
L_B \simeq \frac{1}{8} \pi R^2 \Gamma^2 \beta c B^2,
\end{equation}
where \(B\) is the magnetic field strength in the comoving frame.}

\begin{table*}[!ht]%[!htb]
\centering
%\begin{minipage}[]{100mm}
\caption{Jet power components and total jet power derived from synchrotron+SSC modelling of 1ES 2344+514’s SED for multiple observation dates.}

\begin{tabular}{cccccc}

\hline
\hline
&&Inner zone&&&\\
\hline
\hline
Obs. Date & $L_e$  &$L_p$  
& $L_B$ &$L_{\text{rad}}$ 
 &$P_{jet}$
\\
& $\text{erg}\,\text{s}^{-1}$&$\text{erg}\,\text{s}^{-1}$&$\text{erg}\,\text{s}^{-1}$&$\text{erg}\,\text{s}^{-1}$&$\text{erg}\,\text{s}^{-1}$\\
\hline
09-07-2017 & $3.42\times10^{43}$  &$5.71\times10^{42}$  
& $3.72\times10^{41}$ &$4.61\times10^{42}$ 
 &$4.49\times10^{43}$ 
\\
07-08-2017 & $1.47\times10^{43}$  &$9.87\times10^{41}$  
& $4.28\times10^{41}$ &$2.14\times10^{42}$ 
 &$1.82\times10^{43}$ 
\\
21-10-2017 & $3.43\times10^{43}$  &$8.46\times10^{42}$  
& $8.19\times10^{40}$ &$1.36\times10^{42}$ 
 &$4.42\times10^{43}$ 
\\
07-12-2017 & $3.78\times10^{43}$  &$7.73\times10^{42}$  
& $6.19\times10^{40}$ &$1.40\times10^{42}$ 
 &$4.70\times10^{43}$ 
\\
01-08-2018 & $2.92\times10^{43}$  &$5.96\times10^{42}$  
& $2.24\times10^{41}$ &$2.32\times10^{42}$ 
 &$3.77\times10^{43}$ 
\\
11-09-2018 & $6.55\times10^{43}$  &$2.58\times10^{43}$  
& $4.20\times10^{40}$ &$2.52\times10^{42}$ 
 &$9.39\times10^{43}$ 
\\
22-07-2020 & $2.21\times10^{43}$  
& $5.82\times10^{42}$&$1.79\times10^{41}$
 &$1.13\times10^{42}$ 
 &$2.92\times10^{43}$ 
\\
28-07-2021 & $1.53\times10^{43}$  &$5.21\times10^{42}$  
& $2.39\times10^{41}$ &$9.08\times10^{41}$ 
 &$2.17\times10^{43}$ 
\\
05-08-2021 & $2.25\times10^{43}$  &$3.06\times10^{42}$  
& $5.48\times10^{40}$ &$6.87\times10^{41}$ 
 &$2.63\times10^{43}$ 
\\
05-08-2022 & $3.24\times10^{43}$  &$1.46\times10^{42}$  & $5.21\times10^{40}$ &$1.42\times10^{42}$  &$3.54\times10^{43}$ \\
\hline
\hline

\end{tabular}
\label{tab:table7}

\end{table*}

{The radiative power, associated with non-thermal emission, is
\begin{equation}
L_{\text{rad}} = \pi R^2 \Gamma^2 \beta c U_{\text{rad}} = L' \frac{\Gamma^2}{4} = L \frac{\Gamma^2}{4 \delta^4},
\end{equation}}
{where \(U_{\text{rad}} = \frac{L'}{4 \pi R^2 c}\) is the radiation energy density in the comoving frame, \(L'\) is the comoving non-thermal luminosity, and \(L\) is the observed total luminosity. 
The total jet power, $P_{\text{jet}}$, and its individual components are calculated within the \textsc{JetSeT} framework and the results are presented in Table~\ref{tab:table7}.}

\section{Discussion}
\label{sec:discus}

In this work, we report the results of the multi-epoch broadband spectral study along with the light curve of 1ES\,2344+514 using the quasi-simultaneous multiwavelength data from 2017 June 6 to 2022 August 6 (MJD 57910 -- 59797). We have carried out a spectral analysis in the energy range of 0.5 -- 7.0 keV for AstroSat-SXT and \textit{Swift}-XRT, 3 -- 20 keV for AstroSat-LAXPC, 3 -- 79 keV for \textit{NuSTAR} and optical/UV band for \textit{Swift}-UVOT observations (Tables~\ref{tab:table2}, ~\ref{tab:table3} and ~\ref{table:uvot_flux}). Our study reveals the anti-correlation between the power-law photon spectral index and the integral flux at X-ray energies, as seen in Figure~\ref{fig:figure4}. These plots show that the spectrum hardens when the source brightens. This trend is expected in EHSP BL Lacs and is also similar to the results reported earlier \citep{Giommi_2000, Acciari_2011, Aleksic_2013, Kapanadze_2017}.

A positive correlation between the $\gamma$-ray spectral index and flux has been observed in Figure~\ref{fig:figure6}, revealing a `softer-when-brighter' trend in the $\gamma$-ray energy range. This trend is expected in HSP BL Lacs. The different trends observed in the X-ray and $\gamma$-ray bands are due to the position of the X-ray band at the falling edge of the first hump (synchrotron) and that of the $\gamma$-ray band at the rising edge of the second hump (inverse Compton) of SED. 

A strong correlation is observed between the peak energy ($E_p$) and the corresponding peak flux ($S_p$), shown in Figure~\ref{fig:figure3}. This trend, where the synchrotron peak shifts to higher energies as the source brightens, is aligned with previous observations by \citet{Giommi_2000} and \citet{Acciari_2020}. Such a displacement of the synchrotron peak between these epochs implies a significant increase in the energy of emitting electrons.
 
\begin{figure}[ht]%[h!]
      \centering
      \includegraphics[width = \linewidth]{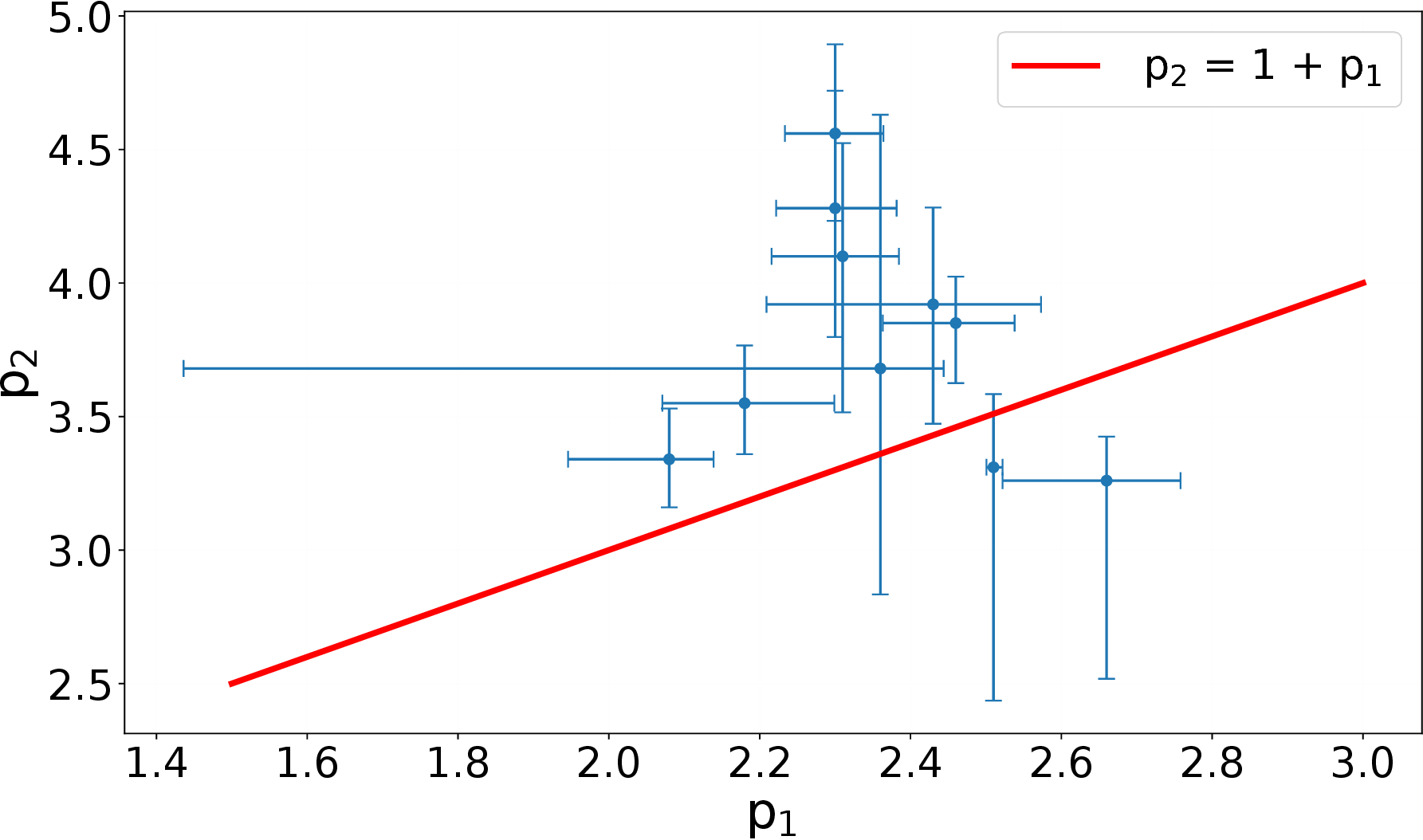}
      \caption{Plot of power-law index of broken power-law
               used for electron distribution above break energy ($p_2$) vs the index
               below break energy ($p_1$). The red line corresponds to the relation, $p_2 = 1+p_1$ }
      \label{fig:figure12}
\end{figure}

We analysed the SEDs for epochs where the Astrosat or \textit{Swift}-UVOT, \textit{Swift}-XRT, \textit{NuSTAR}, and MAGIC observations are available, alongside quasi-simultaneous \textit{Fermi}-LAT data. Initially, a steady-state, one-zone synchrotron+SSC model with a broken power-law electron energy distribution was used to fit the broadband SEDs. The fitted parameters such as particle density (N), magnetic field (B), spectral indices ($p_1$,~$p_2$), and the Lorentz factors of the minimum ($\gamma_{\text{min}}$), break ($\gamma_{\text{break}}$), and maximum ($\gamma_{\text{max}}$) Lorentz factors, as detailed in Table~\ref{tab:table6}. However, \citet{Abe_2024} reported strong evidence of two separate emitting components contributing to synchrotron emission from the infrared (IR) to ultra-violet (UV) wavebands. Similarly, we adopted a two-component model to fit the SEDs for the epochs where the optical-UV data are available. In this framework, the inner zone (zone-1) synchrotron+SSC model, characterized by a broken power-law electron distribution, dominates the X-ray to very high energy (VHE) gamma--ray flux, is filled by highly energetic electrons with $\gamma_{min}$~$\sim$~$(0.59-14.5)\times10^{3}$ and $\gamma_{max}$~$\sim$~$(2.52–8.17)\times10^{6}$ (see; Table~\ref{tab:table6}). The outer zone synchrotron+SSC model, with a power-law electron energy distribution, primarily accounts for radio emission but also contributes slightly to the IR/optical/UV bands, and it can explain the ``UV excess". The electron population corresponding to the outer zone is weaker than that in the inner zone region with $\gamma_{min}$$\sim$ $1$ and $\gamma_{max}$$\sim$ $10^{4}$. From Figure~\ref{fig:figure8}, it is argued that there is no interaction between the outer and inner zones. The fitted results correspond to the one-zone synchrotron+SSC model discussed here. Figure~\ref{fig:figure9} shows the SED of 1ES\,2344+514 with the highest flux in X-rays on 2017 July 7 among the observation epochs considered \citep{Giommi_2000,Acciari_2020}, accompanied by elevated $\gamma$-ray flux. 

\begin{figure}[ht]%[h!]
      \centering
      \includegraphics[width = \linewidth]{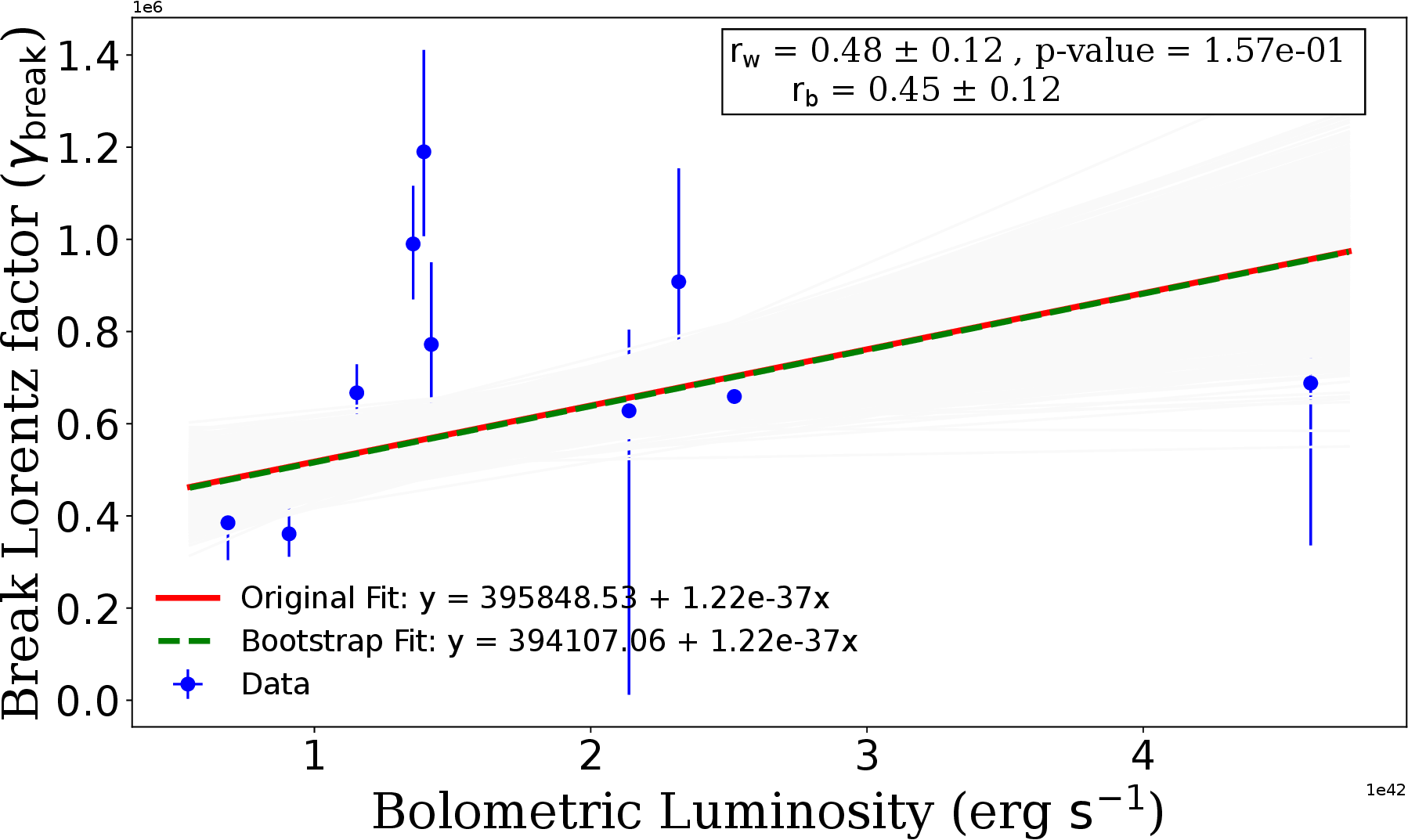}
      \caption{ Variation of the break Lorentz factor ($\gamma_{break}$) with total luminosity fitted with weighted linear regression and bootstrap linear regression. The gray lines are bootstrapped regression lines, the green dashed line is the mean of them, and the red line is the weighted linear regression line.}
      \label{fig:figure13}
\end{figure}

In this work, our results show that the maximum Lorentz factor ($\gamma_{max}$) ranges from $2.52\times10^6$ to $8.17\times10^6$ across all epochs (see Table~\ref{tab:table6}), which is consistent with the previous results of 1ES\,2344+514 \citep{Acciari_2020, Abe_2024}. The synchrotron+SSC model reveals a distinct $\gamma_{break}$ between $3.61\times10^5$ and $11.9 \times 10^5$, about an order of magnitude below $\gamma_{\text{max}}$. Previous work by \citep{Tavecchio_2010} reported that $\gamma_{break}$ lies in the range of $10^4 - 10^5$ during low states for HSP BL Lacs. In addition to the extreme value of $\gamma_{max}$, the low magnetic field between 1.89$\times 10^{-2}$ and 5.40$\times 10^{-2}$ G is found to be essential to describe the SEDs. It is also reported that the magnetic field usually lies between 0.1 and 1 G for HSP BL Lacs in the leptonic model \citep[see][]{Tavecchio_2010}, but for EHSP BL Lacs  the magnetic field is slightly lower $\sim$ $10^{-2}-10^{-3}$ G, aligning with our results.

\begin{figure}[ht]%[h!]
      \centering
      \includegraphics[width = \linewidth]{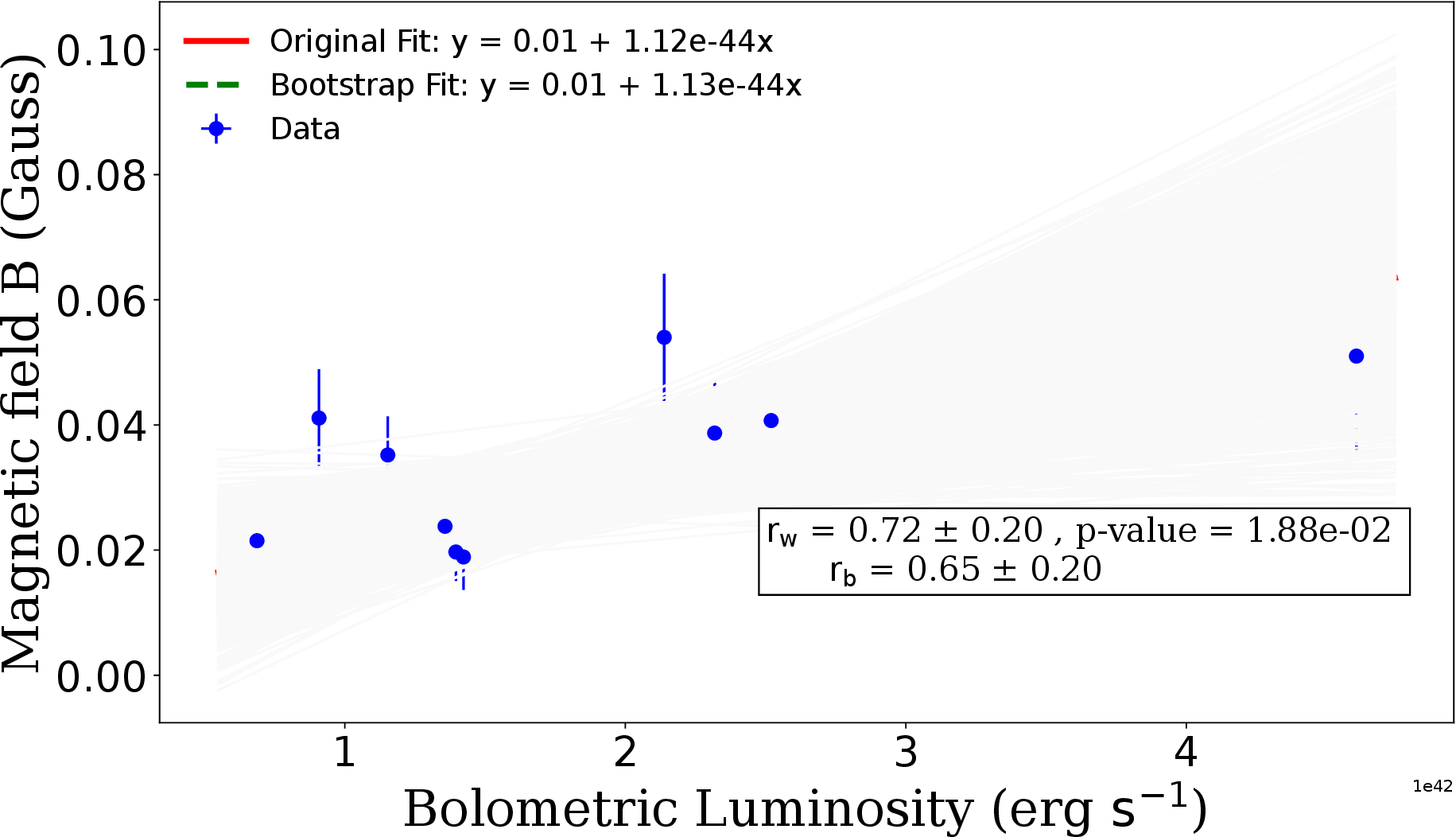}
      \caption{Plot of bolometric luminosity vs magnetic field in the emission region fitted with weighted linear regression and bootstrap linear regression. The gray lines are bootstrapped regression lines, the green dashed line is the mean of them, and the red line is the weighted linear regression line.}
      \label{fig:figure14}
\end{figure}

In the  synchrotron+SSC model, the first index ($p_1$), which defines the slope of the electron energy distribution below $\gamma_{break}$, influences the optical and GeV bands through synchrotron and SSC respectively, while the second index ($p_2$) shapes the X-ray and VHE bands. A very weak correlation is observed between $p_1$ and $p_2$, characterized by a weighted Pearson's correlation coefficient of \( r_w = 0.30 \pm 0.23 \), with a p-value of 0.389, and a bootstrapped correlation coefficient of \( r_b = 0.21 \pm 0.23 \) (see Figure~\ref{fig:figure12}).  Notably, the second index ($p_2$) is much steeper than the expected relation $p_2 = 1 + p_1$, but in a few cases the second index ($p_2$) is less steep than the expected one, thus the origin of the broken power-law spectrum from the cooling break is ruled out. A weak correlation exists between the first index ($p_1$) and the bolometric luminosity, with $r_w = 0.56 \pm 0.13 $, a p-value of 0.09, and $ r_b = 0.42 \pm 0.13 $. Similarly, a weak correlation is found between the bolometric luminosity and the particle break energy ($ \gamma_{\text{break}} $), with $ r_w = 0.48 \pm 0.12 $, a p-value of 0.16, and $ r_b = 0.45 \pm 0.12 $ (see Figure~\ref{fig:figure13}).
Also, a weak correlation is observed between the magnetic field ($B$) and the bolometric luminosity, with $ r_w = 0.72 \pm 0.20 $, a p-value of 0.019, and $ r_b = 0.65 \pm 0.20 $ (see Figure~\ref{fig:figure14}) .

In our analysis, the magnetization parameter $ U_B / U_e $, where $ U_B $ represents the magnetic energy density and $ U_e $ the electron energy density, ranges from approximately $ 6.41 \times 10^{-4} $ to $ 2.92 \times 10^{-2} $. These values align with previous studies of 1ES\,2344+514, which report $ U_B / U_e \approx 10^{-3} $ to $ 10^{-2} $ \citep{Albert_2007, Tavecchio_2010, Acciari_2011, Acciari_2020, Aleksic_2013}. This indicates that the magnetic energy density is significantly lower than the electron energy density, a characteristic commonly observed in BL Lacs during the flaring states. Our findings confirm that the energy equipartition is well below unity, with $ U_B / U_e \ll 1 $, suggesting a particle-dominated emission region rather than a magnetically dominated one, consistent with results from a select group of EHSP BL Lacs \citep{Costamante_2018}, and also for most TeV BL Lacs \citep{Zhang_2012}.

In this work (Table~\ref{tab:table6}), we employed a fixed bulk Doppler factor $\delta = 10$ for 1ES\,2344+514 following \citet{Abe_2024}, due to the absence of simultaneous high-resolution radio observation by VLBI to have a reliable measurement of $\delta$. This value of $\delta$ is lower than $\delta \sim 10–35$ reported by \citet{Zhang_2012} for their 24 TeV BL Lacs. The magnetic field $B \sim (1.89–5.40) \times 10^{-2} \, \text{G}$ is lower than that of Zhang et al.,2012 $0.1–0.6 \, \text{G}$, enabling an extremely high synchrotron peak at $\nu_{\text{sp}} \gtrsim 10^{17} \, \text{Hz}$ versus HSP BL Lac peaks at $10^{15}–10^{16} \, \text{Hz}$. The electron energy density $U_e \sim 3.91 \times 10^{-3}–1.75 \times 10^{-2} \, \text{erg cm}^{-3}$ and magnetization $\eta \sim 6.41 \times 10^{-4}–2.92 \times 10^{-2}$ indicate a jet dominated by particles, more extreme than the values of Zhang et al. 2012. Spectral indices $p_1 \sim 2.08–2.66$ $p_2 \sim 3.26–4.56$ align with Zhang at al. 2012, but a weak correlation between $p_1$ and $p_2$ ($r_w=0.30\pm0.23$), Figure\ref{fig:figure12}) suggests non-cooling spectral breaks. Lorentz factors $\gamma_{\text{min}} \sim (0.59–14.5) \times 10^3$, $\gamma_{\text{break}} \sim (3.61–11.9) \times 10^5$, and $\gamma_{\text{max}} \sim (2.52–8.17) \times 10^6$ exceed the ranges of Zhang et al.,2012, driving hard X-ray and VHE $\gamma$-ray emission (e.g., July 7, 2017 SED, Figure\ref{fig:figure9}). Lower $\delta$, B, $\eta$, and higher $\gamma_{\text{break}}$ produce the hard spectra of EHSP, with a `harder-when-brighter’ X-ray trend (Figure\ref{fig:figure4}) and synchrotron peak shift (Figure~\ref{fig:figure3}), consistent with flaring/quiescent transitions \citep{Tavecchio_2010}.

The synchrotron+SSC modeling of 1ES\,2344+514's broadband SED (2017--2022) reveals a dynamic particle-dominated jet, with electron power ($L_e \approx 10^{43}-10^{44}\,\text{erg}\,\text{s}^{-1}$) exceeding both the radiative power ($L_{\text{rad}} \approx 10^{42}\,\text{erg}\,\text{s}^{-1}$) and the magnetic field power ($L_B \approx 10^{41}\,\text{erg}\,\text{s}^{-1}$). This indicates that the electron power in this BL Lac is sufficient to produce the observed radiation power, estimated from bolometric luminosity. The power carried by the Poynting flux always remains significantly lower than the radiation and electron powers, consistent with the characteristics of EHSP BL Lac \citep{Costamante_2018}. The total jet power ranges from $1.82 \times 10^{43}$ to $9.39 \times 10^{43}\,\text{erg}\,\text{s}^{-1}$ (peaking during the 11-09-2018 flare; see; Table~\ref{tab:table7}), dominated by relativistic electrons ($\langle \gamma \rangle$ ranging from \(3.936 \times 10^3\) to \(4.219 \times 10^4\), with an average \(\langle \gamma \rangle \approx 1.573 \times 10^4\)) under the standard one proton per relativistic electron assumption, while low magnetization (an average $\eta \approx 7.96 \times 10^{-3}$) confirms minimal magnetic field contribution. The radiation power ranges from $6.87 \times 10^{41}$ to $4.61 \times 10^{42}\,\text{erg}\,\text{s}^{-1}$ (peaking during the 09-07-2017 flare; see; Table~\ref{tab:table7}), the source appeared in a high state in X-rays and $\gamma$-rays with peak flux . The soft X-ray spectrum constrains the synchrotron component, which supports the synchrotron+SSC framework \citep{Tavecchio_2010}. The variability in \(N_e\) and \(B\) across observations suggests evolving jet conditions, potentially linked to the accretion process, or turbulence, shocks, or magnetic reconnection events in the jet \citep{Giannios_2013}. Compared to other EHSP BL Lacs, the low magnetization and high jet power of 1ES\,2344+514 align with leptonic models, but do not rule out lepto-hadronic contributions \citep{Mannheim_1993,Cerruti_2015}.

\section{Conclusions}
\label{sec:conl}

In this study, we have presented the multi-epoch spectral study of 1ES\,2344+514 using quasi-simultaneous data. Data in optical, UV, X-ray and $\gamma$-ray bands for various epochs during 2017--2022 from SXT and LAXPC onboard AstroSat, \textit{Swift}-UVOT, \textit{Swift}-XRT, \textit{NuSTAR} and \textit{Fermi}-LAT and have compared our findings with those of the MAGIC collaboration (2024) from the 2019-2021 period. The following are our main conclusions.

\begin{enumerate}
\item During 2017-2022, the source appeared in a high state in X-rays with peak flux seen on 2017 July 9. This is comparable to the highest flux level seen from this source historically.
\item The X-ray spectra are well fitted with both power-law and log-parabola models. A joint fit between SXT and LAXPC and between \textit{Swift}-XRT and \textit{NuSTAR} have been carried out to constrain the location of the synchrotron peak. 
\item A clear shift of synchrotron peak is observed across the observation epochs, suggesting an extreme behaviour of the source. 
\item The source showed spectral evolution such as a `harder-when-brighter' trend in X-rays and a `softer-when-brighter' trend in $\gamma$ - rays.
\item The broadband SEDs were modelled using a one-zone synchrotron+SSC model, with the electron energy distribution characterized by a broken power-law distribution, yielding results consistent with previous modelling of the object. Additionally, the parameter values derived in our study align closely with those reported by \citet{Abe_2024}. 
\item {Our multiwavelength analysis of 1ES 2344+514 from 2017 to 2022, employing a one-zone synchrotron+SSC model, constrains the jet’s physical properties.}
\item The magnetic field inside the emission region is also found to be weakly correlated with the bolometric luminosity.
\item There is no significant correlation observed between $\gamma_{break}$ and  bolometric luminosity. 
\item Jet power calculations, incorporating one proton per relativistic electron, yield $(P_{\text{jet}} \approx 10^{43} – 10^{44} \, \text{erg s}^{-1}$, dominated by relativistic electrons (Table~\ref{tab:table7}). 
\item {The low magnetization (an average $\eta \approx 7.96 \times 10^{-3}$) supports a particle-dominated jet \citep{Costamante_2018}, consistent with leptonic synchrotron+SSC models \citep{Tavecchio_2010}.} 
\item {Variability in electron density and magnetic field suggests dynamic jet conditions, potentially driven by flaring episodes. Although the synchrotron+SSC model adequately describes broadband emission, hadronic contributions cannot be excluded \citep{Böttcher_2013}. Future simultaneous multiwavelength observations, particularly during flares, could further distinguish between leptonic and hadronic processes, refining our understanding of EHSP BL Lacs jet physics.}

\end{enumerate}

\normalem

\section*{Acknowledgements}

This research has used data from the AstroSat mission of the Indian Space Research Organisation (ISRO), archived at the Indian Space Science Data Center (ISSDC). This work has used data from the Soft X-ray Telescope (SXT) developed in a collaboration between the Tata Institute of Fundamental Research (TIFR), Mumbai, India, and the University of Leicester, UK, and the Large X-ray Proportional Counter Array (LAXPC) developed at TIFR, Mumbai, India. The SXT and LAXPC POCs are thanked for verifying and providing the necessary software tools. This research has also made use of the data, software and/or web tools provided by NASAs High Energy Astrophysics Science Archive Research Center (HEASARC), a service of Goddard Space Flight Center (GSFC) and the Smithsonian Astrophysical Observatory. Part of this work is based on archival data, software or online services provided by the ASI (Italian Space Agency) service Data Center (ASDC). The use of XRT Data Analysis Software (XRTDAS) was developed under the responsibility of ASDC, Italy and the \textit{NuSTAR} Data Analysis Software (\textsc{NuSTARDAS}), jointly developed by ASDC, Italy and California Institute of Technology (Caltech), USA, are gratefully acknowledged. This study utilised \textit{Fermi}-LAT data and the Fermitool package, accessed via the \textit{Fermi} Science Support Center (FSSC), which is provided by NASA. We express our gratitude to Axel Arbet-Engels for supplying the MAGIC data. We acknowledge the support of the Department of Atomic Energy, Government of India, under project identification No. RT4002. This program is supported by \textit{Fermi} Guest Investigator grants NNX08AN56G, NNX08AN56G, NNX09AV10G, and NNX15AU81G, and a community-developed python package named \textsc{enrico} to make \textit{Fermi}-LAT data analysis easier and more consistent. We sincerely thank the anonymous reviewer for the thorough reading and constructive feedback.             

%%%%%%%%%%%%%%%%%%%%%%%%%%%%%%%%%%%%%%%%%%%%%%%%%%
\section*{Data Availability}
 
For this work, we have used data from \textit{Fermi}-LAT, \textit{Swift}-XRT, \textit{NuSTAR}, and SXT and LAXPC onboard AstroSat which are available in the public domain. Links are given as follows: 

AstroSat-mission data are available at: ~\url{https://astrobrowse.issdc.gov.in/astro_archive/archive/Home.jsp}

AstroSat-mission data analysis software package is available at: ~\url{http://astrosat-ssc.iucaa.in/data_and_analysis} and \\
~\url{http://astrosat-ssc.iucaa.in/uploads/threadsPageNew_SXT.html}

\textit{Swift}-mission data are available at: ~\url{https://heasarc.gsfc.nasa.gov/docs/archive.html}.

\textit{Swift}-mission data analysis software package is available at: ~\url{https://heasarc.gsfc.nasa.gov/lheasoft/download.html} and \\
~\url{https://swift.gsfc.nasa.gov/analysis/}.

\textit{NuSTAR}-mission data are available at: ~\url{https://heasarc.gsfc.nasa.gov/docs/archive.html}.

\textit{NuSTAR}-mission data analysis software package is available at:~\url{https://heasarc.gsfc.nasa.gov/lheasoft/download.html} and \\ ~\url{https://heasarc.gsfc.nasa.gov/docs/nustar/analysis/}.

\textit{Fermi}-LAT data are available at: ~\url{https://fermi.gsfc.nasa.gov/ssc/data/access/}.

\textit{Fermi}-LAT data analysis software is available at: ~\url{https://fermi.gsfc.nasa.gov/ssc/data/analysis/software/}.

SSDC (ASI) archive: ~\url{https://www.ssdc.asi.it/}.

Software for Broadband SED modelling is available at: ~\url{https://github.com/andreatramacere/jetset}.

\bibliographystyle{raa}
\bibliography{ms2024-0406}

@article{Abdo_2010a,
doi = {10.1088/0004-637X/710/2/1271},
url = {https://dx.doi.org/10.1088/0004-637X/710/2/1271},
journal = {The Astrophysical Journal},
year = {2010a},
month = {jan},
publisher = {The American Astronomical Society},
volume = {710},
number = {2},
pages = {1271},
author = {Abdo, A. A. and Ackermann, M. and Ajello, M. and Atwood, W. B. and Axelsson, M. and Baldini, L. and Ballet, J. and Barbiellini, G. and Bastieri, D. and Bechtol, K. and Bellazzini, R. and Berenji, B. and Blandford, R. D. and Bloom, E. D. and Bonamente, E. and Borgland, A. W. and Bouvier, A. and Bregeon, J. and Brez, A. and Brigida, M. and Bruel, P. and Burnett, T. H. and Buson, S. and Caliandro, G. A. and Cameron, R. A. and Caraveo, P. A. and Carrigan, S. and Casandjian, J. M. and Cavazzuti, E. and Cecchi, C. and Çelik, Ö. and Charles, E. and Chekhtman, A. and Cheung, C. C. and Chiang, J. and Ciprini, S. and Claus, R. and Cohen-Tanugi, J. and Conrad, J. and Cutini, S. and Dermer, C. D. and de Angelis, A. and de Palma, F. and Digel, S. W. and do Couto e Silva, E. and Drell, P. S. and Dubois, R. and Dumora, D. and Farnier, C. and Favuzzi, C. and Fegan, S. J. and Focke, W. B. and Fortin, P. and Frailis, M. and Fukazawa, Y. and Funk, S. and Fusco, P. and Gargano, F. and Gasparrini, D. and Gehrels, N. and Germani, S. and Giebels, B. and Giglietto, N. and Giommi, P. and Giordano, F. and Glanzman, T. and Godfrey, G. and Grenier, I. A. and Grondin, M.-H. and Grove, J. E. and Guillemot, L. and Guiriec, S. and Harding, A. K. and Hartman, R. C. and Hayashida, M. and Hays, E. and Healey, S. E. and Horan, D. and Hughes, R. E. and Jackson, M. S. and Jóhannesson, G. and Johnson, A. S. and Johnson, W. N. and Kamae, T. and Katagiri, H. and Kataoka, J. and Kawai, N. and Kerr, M. and Knödlseder, J. and Kuss, M. and Lande, J. and Latronico, L. and Lemoine-Goumard, M. and Longo, F. and Loparco, F. and Lott, B. and Lovellette, M. N. and Lubrano, P. and Madejski, G. M. and Makeev, A. and Mazziotta, M. N. and McConville, W. and McEnery, J. E. and Meurer, C. and Michelson, P. F. and Mitthumsiri, W. and Mizuno, T. and Moiseev, A. A. and Monte, C. and Monzani, M. E. and Morselli, A. and Moskalenko, I. V. and Murgia, S. and Nolan, P. L. and Norris, J. P. and Nuss, E. and Ohsugi, T. and Omodei, N. and Orlando, E. and Ormes, J. F. and Paneque, D. and Panetta, J. H. and Parent, D. and Pelassa, V. and Pepe, M. and Persic, M. and Pesce-Rollins, M. and Piron, F. and Porter, T. A. and Rainò, S. and Rando, R. and Razzano, M. and Reimer, A. and Reimer, O. and Reposeur, T. and Ritz, S. and Rochester, L. S. and Rodriguez, A. Y. and Romani, R. W. and Roth, M. and Ryde, F. and Sadrozinski, H. F.-W. and Sanchez, D. and Sander, A. and Parkinson, P. M. Saz and Scargle, J. D. and Sgrò, C. and Siskind, E. J. and Smith, D. A. and Smith, P. D. and Spandre, G. and Spinelli, P. and Strickman, M. S. and Suson, D. J. and Tajima, H. and Takahashi, H. and Takahashi, T. and Tanaka, T. and Thayer, J. B. and Thayer, J. G. and Thompson, D. J. and Tibaldo, L. and Torres, D. F. and Tosti, G. and Tramacere, A. and Uchiyama, Y. and Usher, T. L. and Vasileiou, V. and Vilchez, N. and Villata, M. and Vitale, V. and Waite, A. P. and Wang, P. and Winer, B. L. and Wood, K. S. and Ylinen, T. and Ziegler, M.},
title = {SPECTRAL PROPERTIES OF BRIGHT FERMI-DETECTED BLAZARS IN THE GAMMA-RAY BAND},

}

@ARTICLE{Abdo_2010b,
       author = {{Abdo}, A.~A. and {Ackermann}, M. and {Agudo}, I.},
        title = "{The Spectral Energy Distribution of Fermi Bright Blazars}",
      journal = {\apj},
         year = {2010b},
        month = jun,
       volume = {716},
       number = {1},
        pages = {30-70},
          doi = {10.1088/0004-637X/716/1/30},
archivePrefix = {arXiv},
       eprint = {0912.2040},
 primaryClass = {astro-ph.CO},
       adsurl = {https://ui.adsabs.harvard.edu/abs/2010ApJ...716...30A}
}

@article{Abdo_2010c,
    author = {Abdo, A. A. and Ackermann, M. and Ajello, M. and others},
    title = "{Fermi-LAT Observations of the Blazar 3C 279}",
    journal = {The Astrophysical Journal},
    volume = {710},
    pages = {810},
    year = {2010c},
    doi = {10.1088/0004-637X/710/1/810}
}

@ARTICLE{Abdollahi_2020,
       author = {{Abdollahi}, S. and {Acero}, F. and {Ackermann}, M.},
        title = "{Fermi Large Area Telescope Fourth Source Catalog}",
      journal = {\apjs},
         year = 2020,
        month = mar,
       volume = {247},
       number = {1},
          eid = {33},
        pages = {33},
          doi = {10.3847/1538-4365/ab6bcb},
archivePrefix = {arXiv},
       eprint = {1902.10045},
 primaryClass = {astro-ph.HE},
       adsurl = {https://ui.adsabs.harvard.edu/abs/2020ApJS..247...33A}
}

@ARTICLE{Acciari_2011,
       author = {{Acciari}, V.~A. and {Aliu}, E.and {Arlen}, T.},
        title = "{Multiwavelength Observations of the Very High Energy Blazar 1ES 2344+514}",
      journal = {\apj},
         year = 2011,
        month = sep,
       volume = {738},
       number = {2},
          eid = {169},
        pages = {169},
          doi = {10.1088/0004-637X/738/2/169},
archivePrefix = {arXiv},
       eprint = {1106.4594},
 primaryClass = {astro-ph.HE},
       adsurl = {https://ui.adsabs.harvard.edu/abs/2011ApJ...738..169A}
}

@ARTICLE{Acciari_2010,
       author = {{Acciari}, V.~A. and {Aliu}, E. and {Arlen}, T. and {Aune}, T. and {Bautista}, M. and {Beilicke}, M. and {Benbow}, W. and {B{\"o}ttcher}, M. and {Boltuch}, D. and {Bradbury}, S.~M. and {Buckley}, J.~H. and {Bugaev}, V. and {Byrum}, K. and {Cannon}, A. and {Cesarini}, A. and {Ciupik}, L. and {Cui}, W. and {Dickherber}, R. and {Duke}, C. and {Falcone}, A. and {Finley}, J.~P. and {Finnegan}, G. and {Fortson}, L. and {Furniss}, A. and {Galante}, N. and {Gall}, D. and {Gibbs}, K. and {Gillanders}, G.~H. and {Godambe}, S. and {Grube}, J. and {Guenette}, R. and {Gyuk}, G. and {Hanna}, D. and {Holder}, J. and {Hui}, C.~M. and {Humensky}, T.~B. and {Imran}, A. and {Kaaret}, P. and {Karlsson}, N. and {Kertzman}, M. and {Kieda}, D. and {Konopelko}, A. and {Krawczynski}, H. and {Krennrich}, F. and {Lang}, M.~J. and {Lamerato}, A. and {LeBohec}, S. and {Maier}, G. and {McArthur}, S. and {McCann}, A. and {McCutcheon}, M. and {Moriarty}, P. and {Mukherjee}, R. and {Ong}, R.~A. and {Otte}, A.~N. and {Pandel}, D. and {Perkins}, J.~S. and {Petry}, D. and {Pichel}, A. and {Pohl}, M. and {Quinn}, J. and {Ragan}, K. and {Reyes}, L.~C. and {Reynolds}, P.~T. and {Roache}, E. and {Rose}, H.~J. and {Roustazadeh}, P. and {Schroedter}, M. and {Sembroski}, G.~H. and {Senturk}, G. Demet and {Smith}, A.~W. and {Steele}, D. and {Swordy}, S.~P. and {Te{\v{s}}i{\'c}}, G. and {Theiling}, M. and {Thibadeau}, S. and {Varlotta}, A. and {Vassiliev}, V.~V. and {Vincent}, S. and {Wagner}, R.~G. and {Wakely}, S.~P. and {Ward}, J.~E. and {Weekes}, T.~C. and {Weinstein}, A. and {Weisgarber}, T. and {Williams}, D.~A. and {Wissel}, S. and {Wood}, M. and {Zitzer}, B. and {Ackermann}, M. and {Ajello}, M. and {Antolini}, E. and {Baldini}, L. and {Ballet}, J. and {Barbiellini}, G. and {Bastieri}, D. and {Bechtol}, K. and {Bellazzini}, R. and {Berenji}, B. and {Blandford}, R.~D. and {Bloom}, E.~D. and {Bonamente}, E. and {Borgland}, A.~W. and {Bouvier}, A. and {Bregeon}, J. and {Brigida}, M. and {Bruel}, P. and {Buehler}, R. and {Buson}, S. and {Caliandro}, G.~A. and {Cameron}, R.~A. and {Caraveo}, P.~A. and {Carrigan}, S. and {Casandjian}, J.~M. and {Cavazzuti}, E. and {Cecchi}, C. and {{\c{C}}elik}, {\"O}. and {Charles}, E. and {Chekhtman}, A. and {Cheung}, C.~C. and {Chiang}, J. and {Ciprini}, S. and {Claus}, R. and {Cohen-Tanugi}, J. and {Conrad}, J. and {Dermer}, C.~D. and {de Palma}, F. and {Silva}, E. do Couto e. and {Drell}, P.~S. and {Dubois}, R. and {Dumora}, D. and {Farnier}, C. and {Favuzzi}, C. and {Fegan}, S.~J. and {Fortin}, P. and {Frailis}, M. and {Fukazawa}, Y. and {Funk}, S. and {Fusco}, P. and {Gargano}, F. and {Gasparrini}, D. and {Gehrels}, N. and {Germani}, S. and {Giebels}, B. and {Giglietto}, N. and {Giordano}, F. and {Giroletti}, M. and {Glanzman}, T. and {Godfrey}, G. and {Grenier}, I.~A. and {Grove}, J.~E. and {Guiriec}, S. and {Hays}, E. and {Horan}, D. and {Hughes}, R.~E. and {J{\'o}hannesson}, G. and {Johnson}, A.~S. and {Johnson}, W.~N. and {Kamae}, T. and {Katagiri}, H. and {Kataoka}, J. and {Kn{\"o}dlseder}, J. and {Kuss}, M. and {Lande}, J. and {Latronico}, L. and {Lee}, S. -H. and {Llena Garde}, M. and {Longo}, F. and {Loparco}, F. and {Lott}, B. and {Lovellette}, M.~N. and {Lubrano}, P. and {Makeev}, A. and {Mazziotta}, M.~N. and {Michelson}, P.~F. and {Mitthumsiri}, W. and {Mizuno}, T. and {Moiseev}, A.~A. and {Monte}, C. and {Monzani}, M.~E. and {Morselli}, A. and {Moskalenko}, I.~V. and {Murgia}, S. and {Nolan}, P.~L. and {Norris}, J.~P. and {Nuss}, E. and {Ohno}, M. and {Ohsugi}, T. and {Omodei}, N. and {Orlando}, E. and {Ormes}, J.~F. and {Paneque}, D. and {Panetta}, J.~H. and {Pelassa}, V. and {Pepe}, M. and {Pesce-Rollins}, M. and {Piron}, F. and {Porter}, T.~A. and {Rain{\`o}}, S. and {Rando}, R.},
        title = "{The Discovery of {\ensuremath{\gamma}}-Ray Emission from the Blazar RGB J0710+591}",
      journal = {\apjl},
         year = 2010,
        month = may,
       volume = {715},
       number = {1},
        pages = {L49-L55},
          doi = {10.1088/2041-8205/715/1/L49},
archivePrefix = {arXiv},
       eprint = {1005.0041},
 primaryClass = {astro-ph.HE},
       adsurl = {https://ui.adsabs.harvard.edu/abs/2010ApJ...715L..49A}
}

@article{Ackermann_2011,
doi = {10.1088/0004-637X/743/2/171},
url = {https://dx.doi.org/10.1088/0004-637X/743/2/171},
year = {2011},
month = {dec},
publisher = {The American Astronomical Society},
volume = {743},
number = {2},
pages = {171},

author = {{Ackermann}, M. and {Ajello}, M. and {Allafort}, A. and Antolini, E. and Atwood, W. B. and Axelsson, M. and Baldini, L. and Ballet, J. and Barbiellini, G. and Bastieri, D. and Bechtol, K. and Bellazzini, R. and Berenji, B. and Blandford, R. D. and Bloom, E. D. and Bonamente, E. and Borgland, A. W. and Bottacini, E. and Bouvier, A. and Bregeon, J. and Brigida, M. and Bruel, P. and Buehler, R. and Burnett, T. H. and Buson, S. and Caliandro, G. A. and Cameron, R. A. and Caraveo, P. A. and Casandjian, J. M. and Cavazzuti, E. and Cecchi, C. and Charles, E. and Cheung, C. C. and Chiang, J. and Ciprini, S. and Claus, R. and Cohen-Tanugi, J. and Conrad, J. and Costamante, L. and Cutini, S. and de Angelis, A. and de Palma, F. and Dermer, C. D. and Digel, S. W. and do Couto e Silva, E. and Drell, P. S. and Dubois, R. and Escande, L. and Favuzzi, C. and Fegan, S. J. and Ferrara, E. C. and Finke, J. and Focke, W. B. and Fortin, P. and Frailis, M. and Fukazawa, Y. and Funk, S. and Fusco, P. and Gargano, F. and Gasparrini, D. and Gehrels, N. and Germani, S. and Giebels, B. and Giglietto, N. and Giommi, P. and Giordano, F. and Giroletti, M. and Glanzman, T. and Godfrey, G. and Grenier, I. A. and Grove, J. E. and Guiriec, S. and Gustafsson, M. and Hadasch, D. and Hayashida, M. and Hays, E. and Healey, S. E. and Horan, D. and Hou, X. and Hughes, R. E. and Iafrate, G. and Jóhannesson, G. and Johnson, A. S. and Johnson, W. N. and Kamae, T. and Katagiri, H. and Kataoka, J. and Knödlseder, J. and Kuss, M. and Lande, J. and Larsson, S. and Latronico, L. and Longo, F. and Loparco, F. and Lott, B. and Lovellette, M. N. and Lubrano, P. and Madejski, G. M. and Mazziotta, M. N. and McConville, W. and McEnery, J. E. and Michelson, P. F. and Mitthumsiri, W. and Mizuno, T. and Moiseev, A. A. and Monte, C. and Monzani, M. E. and Moretti, E. and Morselli, A. and Moskalenko, I. V. and Murgia, S. and Nakamori, T. and Naumann-Godo, M. and Nolan, P. L. and Norris, J. P. and Nuss, E. and Ohno, M. and Ohsugi, T. and Okumura, A. and Omodei, N. and Orienti, M. and Orlando, E. and Ormes, J. F. and Ozaki, M. and Paneque, D. and Parent, D. and Pesce-Rollins, M. and Pierbattista, M. and Piranomonte, S. and Piron, F. and Pivato, G. and Porter, T. A. and Rainò, S. and Rando, R. and Razzano, M. and Razzaque, S. and Reimer, A. and Reimer, O. and Ritz, S. and Rochester, L. S. and Romani, R. W. and Roth, M. and Sanchez, D. A. and Sbarra, C. and Scargle, J. D. and Schalk, T. L. and Sgrò, C. and Shaw, M. S. and Siskind, E. J. and Spandre, G. and Spinelli, P. and Strong, A. W. and Suson, D. J. and Tajima, H. and Takahashi, H. and Takahashi, T. and Tanaka, T. and Thayer, J. G. and Thayer, J. B. and Thompson, D. J. and Tibaldo, L. and Tinivella, M. and Torres, D. F. and Tosti, G. and Troja, E. and Uchiyama, Y. and Vandenbroucke, J. and Vasileiou, V. and Vianello, G. and Vitale, V. and Waite, A. P. and Wallace, E. and Wang, P. and Winer, B. L. and Wood, D. L. and Wood, K. S. and Zimmer, S.},
title = {THE SECOND CATALOG OF ACTIVE GALACTIC NUCLEI DETECTED BY THE FERMI LARGE AREA TELESCOPE},
journal = {The Astrophysical Journal}
}

@ARTICLE{Ackermann_2012,
       author = {{Ackermann}, M. and {Ajello}, M. and {Albert}, A.},
        title = "{The Fermi Large Area Telescope on Orbit: Event Classification, Instrument Response Functions, and Calibration}",
      journal = {\apjs},
         year = 2012,
        month = nov,
       volume = {203},
       number = {1},
          eid = {4},
        pages = {4},
          doi = {10.1088/0067-0049/203/1/4},
archivePrefix = {arXiv},
       eprint = {1206.1896},
 primaryClass = {astro-ph.IM},
       adsurl = {https://ui.adsabs.harvard.edu/abs/2012ApJS..203....4A}
}

@ARTICLE{Ackermann_2015,
       author = {{Ackermann}, M. and {Ajello}, M. and {Atwood}, W.~B. and {Baldini}, L. and {Ballet}, J. and {Barbiellini}, G. and {Bastieri}, D. and {Becerra Gonzalez}, J. and {Bellazzini}, R. and {Bissaldi}, E. and {Blandford}, R.~D. and {Bloom}, E.~D. and {Bonino}, R. and {Bottacini}, E. and {Brandt}, T.~J. and {Bregeon}, J. and {Britto}, R.~J. and {Bruel}, P. and {Buehler}, R. and {Buson}, S. and {Caliandro}, G.~A. and {Cameron}, R.~A. and {Caragiulo}, M. and {Caraveo}, P.~A. and {Carpenter}, B. and {Casandjian}, J.~M. and {Cavazzuti}, E. and {Cecchi}, C. and {Charles}, E. and {Chekhtman}, A. and {Cheung}, C.~C. and {Chiang}, J. and {Chiaro}, G. and {Ciprini}, S. and {Claus}, R. and {Cohen-Tanugi}, J. and {Cominsky}, L.~R. and {Conrad}, J. and {Cutini}, S. and {D'Abrusco}, R. and {D'Ammando}, F. and {de Angelis}, A. and {Desiante}, R. and {Digel}, S.~W. and {Di Venere}, L. and {Drell}, P.~S. and {Favuzzi}, C. and {Fegan}, S.~J. and {Ferrara}, E.~C. and {Finke}, J. and {Focke}, W.~B. and {Franckowiak}, A. and {Fuhrmann}, L. and {Fukazawa}, Y. and {Furniss}, A.~K. and {Fusco}, P. and {Gargano}, F. and {Gasparrini}, D. and {Giglietto}, N. and {Giommi}, P. and {Giordano}, F. and {Giroletti}, M. and {Glanzman}, T. and {Godfrey}, G. and {Grenier}, I.~A. and {Grove}, J.~E. and {Guiriec}, S. and {Hewitt}, J.~W. and {Hill}, A.~B. and {Horan}, D. and {Itoh}, R. and {J{\'o}hannesson}, G. and {Johnson}, A.~S. and {Johnson}, W.~N. and {Kataoka}, J. and {Kawano}, T. and {Krauss}, F. and {Kuss}, M. and {La Mura}, G. and {Larsson}, S. and {Latronico}, L. and {Leto}, C. and {Li}, J. and {Li}, L. and {Longo}, F. and {Loparco}, F. and {Lott}, B. and {Lovellette}, M.~N. and {Lubrano}, P. and {Madejski}, G.~M. and {Mayer}, M. and {Mazziotta}, M.~N. and {McEnery}, J.~E. and {Michelson}, P.~F. and {Mizuno}, T. and {Moiseev}, A.~A. and {Monzani}, M.~E. and {Morselli}, A. and {Moskalenko}, I.~V. and {Murgia}, S. and {Nuss}, E. and {Ohno}, M. and {Ohsugi}, T. and {Ojha}, R. and {Omodei}, N. and {Orienti}, M. and {Orlando}, E. and {Paggi}, A. and {Paneque}, D. and {Perkins}, J.~S. and {Pesce-Rollins}, M. and {Piron}, F. and {Pivato}, G. and {Porter}, T.~A. and {Rain{\`o}}, S. and {Rando}, R. and {Razzano}, M. and {Razzaque}, S. and {Reimer}, A. and {Reimer}, O. and {Romani}, R.~W. and {Salvetti}, D. and {Schaal}, M. and {Schinzel}, F.~K. and {Schulz}, A. and {Sgr{\`o}}, C. and {Siskind}, E.~J. and {Sokolovsky}, K.~V. and {Spada}, F. and {Spandre}, G. and {Spinelli}, P. and {Stawarz}, L. and {Suson}, D.~J. and {Takahashi}, H. and {Takahashi}, T. and {Tanaka}, Y. and {Thayer}, J.~G. and {Thayer}, J.~B. and {Tibaldo}, L. and {Torres}, D.~F. and {Torresi}, E. and {Tosti}, G. and {Troja}, E. and {Uchiyama}, Y. and {Vianello}, G. and {Winer}, B.~L. and {Wood}, K.~S. and {Zimmer}, S.},
        title = "{The Third Catalog of Active Galactic Nuclei Detected by the Fermi Large Area Telescope}",
      journal = {\apj},
         year = 2015,
        month = sep,
       volume = {810},
       number = {1},
          eid = {14},
        pages = {14},
          doi = {10.1088/0004-637X/810/1/14},
archivePrefix = {arXiv},
       eprint = {1501.06054},
 primaryClass = {astro-ph.HE},
       adsurl = {https://ui.adsabs.harvard.edu/abs/2015ApJ...810...14A}
}

@ARTICLE{Agrawal_2017,
       author = {{Agrawal}, P.~C. and {Yadav}, J.~S. and {Antia}, H.~M.},
        title = "{Large Area X-Ray Proportional Counter (LAXPC) Instrument on AstroSat and Some Preliminary Results from its Performance in the Orbit}",
      journal = {Journal of Astrophysics and Astronomy},
         year = 2017,
        month = jun,
       volume = {38},
       number = {2},
          eid = {30},
        pages = {30},
          doi = {10.1007/s12036-017-9451-z},
archivePrefix = {arXiv},
       eprint = {1705.06446},
 primaryClass = {astro-ph.IM},
       adsurl = {https://ui.adsabs.harvard.edu/abs/2017JApA...38...30A}
}

@ARTICLE{Aharonian_2007,
       author = {{Aharonian}, F. and {Akhperjanian}, A.~G. and {Bazer-Bachi}, A.~R. and {Beilicke}, M. and {Benbow}, W. and {Berge}, D. and {Bernl{\"o}hr}, K. and {Boisson}, C. and {Bolz}, O. and {Borrel}, V. and {Braun}, I. and {Brion}, E. and {Brown}, A.~M. and {B{\"u}hler}, R. and {B{\"u}sching}, I. and {Boutelier}, T. and {Carrigan}, S. and {Chadwick}, P.~M. and {Chounet}, L. -M. and {Coignet}, G. and {Cornils}, R. and {Costamante}, L. and {Degrange}, B. and {Dickinson}, H.~J. and {Djannati-Ata{\"\i}}, A. and {O'C. Drury}, L. and {Dubus}, G. and {Egberts}, K. and {Emmanoulopoulos}, D. and {Espigat}, P. and {Farnier}, C. and {Feinstein}, F. and {Ferrero}, E. and {Fiasson}, A. and {Fontaine}, G. and {Funk}, Seb. and {Funk}, S. and {F{\"u}{\ss}ling}, M. and {Gallant}, Y.~A. and {Giebels}, B. and {Glicenstein}, J.~F. and {Gl{\"u}ck}, B. and {Goret}, P. and {Hadjichristidis}, C. and {Hauser}, D. and {Hauser}, M. and {Heinzelmann}, G. and {Henri}, G. and {Hermann}, G. and {Hinton}, J.~A. and {Hoffmann}, A. and {Hofmann}, W. and {Holleran}, M. and {Hoppe}, S. and {Horns}, D. and {Jacholkowska}, A. and {de Jager}, O.~C. and {Kendziorra}, E. and {Kerschhaggl}, M. and {Kh{\'e}lifi}, B. and {Komin}, Nu. and {Kosack}, K. and {Lamanna}, G. and {Latham}, I.~J. and {Le Gallou}, R. and {Lemi{\`e}re}, A. and {Lemoine-Goumard}, M. and {Lohse}, T. and {Martin}, J.~M. and {Martineau-Huynh}, O. and {Marcowith}, A. and {Masterson}, C. and {Maurin}, G. and {McComb}, T.~J.~L. and {Moulin}, E. and {de Naurois}, M. and {Nedbal}, D. and {Nolan}, S.~J. and {Noutsos}, A. and {Olive}, J. -P. and {Orford}, K.~J. and {Osborne}, J.~L. and {Panter}, M. and {Pelletier}, G. and {Petrucci}, P. -O. and {Pita}, S. and {P{\"u}hlhofer}, G. and {Punch}, M. and {Ranchon}, S. and {Raubenheimer}, B.~C. and {Raue}, M. and {Rayner}, S.~M. and {Ripken}, J. and {Rob}, L. and {Rolland}, L. and {Rosier-Lees}, S. and {Rowell}, G. and {Sahakian}, V. and {Santangelo}, A. and {Saug{\'e}}, L. and {Schlenker}, S. and {Schlickeiser}, R. and {Schr{\"o}der}, R. and {Schwanke}, U. and {Schwarzburg}, S. and {Schwemmer}, S. and {Shalchi}, A. and {Sol}, H. and {Spangler}, D. and {Spanier}, F. and {Steenkamp}, R. and {Stegmann}, C. and {Superina}, G. and {Tam}, P.~H. and {Tavernet}, J. -P. and {Terrier}, R. and {Tluczykont}, M. and {van Eldik}, C. and {Vasileiadis}, G. and {Venter}, C. and {Vialle}, J.~P. and {Vincent}, P. and {V{\"o}lk}, H.~J. and {Wagner}, S.~J. and {Ward}, M.},
        title = "{Detection of VHE gamma-ray emission from the distant blazar 1ES{\,}1101-232 with HESS and broadband characterisation}",
      journal = {\aap},
         year = 2007,
        month = aug,
       volume = {470},
       number = {2},
        pages = {475-489},
          doi = {10.1051/0004-6361:20077057},
archivePrefix = {arXiv},
       eprint = {0705.2946},
 primaryClass = {astro-ph},
       adsurl = {https://ui.adsabs.harvard.edu/abs/2007A&A...470..475A}
}

@article{Aharonian_2000,
title = {TeV gamma rays from BL Lac objects due to synchrotron radiation of extremely high energy protons},
journal = {New Astronomy},
volume = {5},
number = {7},
pages = {377-395},
year = {2000},
issn = {1384-1076},
doi = {https://doi.org/10.1016/S1384-1076(00)00039-7},
author = {F.A. Aharonian}
}

@article{Albert_2007,
doi = {10.1086/518431},
url = {https://dx.doi.org/10.1086/518431},
year = {2007},
month = {jun},
publisher = {},
volume = {662},
number = {2},
pages = {892},
author = {J. Albert and E. Aliu and H. Anderhub},
title = {Observation of Very High Energy $\gamma$-Rays from the AGN 1ES 2344+514 in a Low Emission State with the MAGIC Telescope},
journal = {The Astrophysical Journal}
}

@ARTICLE{Aleksic_2013,
       author = {{Aleksi{\'c}}, J. and {Antonelli}, L.~A. and {Antoranz}, P.},
        title = "{The simultaneous low state spectral energy distribution of 1ES 2344+514 from radio to very high energies}",
      journal = {\aap},
         year = 2013,
        month = aug,
       volume = {556},
          eid = {A67},
        pages = {A67},
          doi = {10.1051/0004-6361/201220714},
archivePrefix = {arXiv},
       eprint = {1211.2608},
 primaryClass = {astro-ph.HE},
       adsurl = {https://ui.adsabs.harvard.edu/abs/2013A&A...556A..67A}
}

@article{Allen_2017,
    author = {Allen, C. and Archambault, S. and Archer, A.},
    title = "{Very-High-Energy $\gamma$-Ray Observations of the Blazar 1ES 2344+514 with VERITAS}",
    journal = {Monthly Notices of the Royal Astronomical Society},
    volume = {471},
    number = {2},
    pages = {2117-2123},
    year = {2017},
    month = {07},
    issn = {0035-8711},
    doi = {10.1093/mnras/stx1756},
    url = {https://doi.org/10.1093/mnras/stx1756},
    }

@ARTICLE{Antia_2017,
       author = {{Antia}, H.~M. and {Yadav}, J.~S. and {Agrawal}, P.~C.},
        title = "{Calibration of the Large Area X-Ray Proportional Counter (LAXPC) Instrument on board AstroSat}",
      journal = {\apjs},
         year = 2017,
        month = jul,
       volume = {231},
       number = {1},
          eid = {10},
        pages = {10},
          doi = {10.3847/1538-4365/aa7a0e},
archivePrefix = {arXiv},
       eprint = {1702.08624},
 primaryClass = {astro-ph.IM},
       adsurl = {https://ui.adsabs.harvard.edu/abs/2017ApJS..231...10A}
}

@INPROCEEDINGS{Arnaud_1996,
       author = {{Arnaud}, K.~A.},
        title = "{XSPEC: The First Ten Years}",
    booktitle = {Astronomical Data Analysis Software and Systems V},
         year = 1996,
       editor = {{Jacoby}, George H. and {Barnes}, Jeannette},
       series = {Astronomical Society of the Pacific Conference Series},
       volume = {101},
        month = jan,
        pages = {17},
       adsurl = {https://ui.adsabs.harvard.edu/abs/1996ASPC..101...17A}
}

@ARTICLE{Atwood_2009,
       author = {{Atwood}, W.~B. and {Abdo}, A.~A. and {Ackermann}, M.},
        title = "{The Large Area Telescope on the Fermi Gamma-Ray Space Telescope Mission}",
      journal = {\apj},
         year = 2009,
        month = jun,
       volume = {697},
       number = {2},
        pages = {1071-1102},
          doi = {10.1088/0004-637X/697/2/1071},
archivePrefix = {arXiv},
       eprint = {0902.1089},
 primaryClass = {astro-ph.IM},
       adsurl = {https://ui.adsabs.harvard.edu/abs/2009ApJ...697.1071A}
}

@ARTICLE{Barth_2003,
       author = {{Barth}, Aaron J. and {Ho}, Luis C. and {Sargent}, Wallace L.~W.},
        title = "{The Black Hole Masses and Host Galaxies of BL Lacertae Objects}",
      journal = {\apj},
         year = 2003,
        month = jan,
       volume = {583},
       number = {1},
        pages = {134-144},
          doi = {10.1086/345083},
archivePrefix = {arXiv},
       eprint = {astro-ph/0209562},
 primaryClass = {astro-ph},
       adsurl = {https://ui.adsabs.harvard.edu/abs/2003ApJ...583..134B}
}

@article{Blandford_1979,
       author = {{Blandford}, R.~D. and {K{\"o}nigl}, A.},
      title = "{Relativistic jets as compact radio sources.}",
      journal = {\apj},
         year = {1979},
        month = {aug},
       volume = {232},
        pages = {34-48},
          doi = {10.1086/157262},
       adsurl = {https://ui.adsabs.harvard.edu/abs/1979ApJ...232...34B}

}

@Article{Böttcher_2007,
author={B{\"o}ttcher, Markus},
title={Modeling the emission processes in blazars},
journal={Astrophysics and Space Science},
year={2007},
month={Jun},
day={01},
volume={309},
number={1},
pages={95-104},
issn={1572-946X},
doi={10.1007/s10509-007-9404-0},
url={https://doi.org/10.1007/s10509-007-9404-0}
}

@article{Böttcher_2013,
   author = {{B{\"o}ttcher}, M. and {Reimer}, A. and {Sweeney}, K. and {Prakash}, A.},
    title = "{Leptonic and Hadronic Modeling of Fermi-detected Blazars}",
  journal = {\apj},
   volume = {768},
    pages = {54},
      year = {2013},
     month = may,
      doi = {10.1088/0004-637X/768/1/54}
}

@inproceedings{Burrows_2004,
author = {David N. Burrows and Joanne E. Hill and John A. Nousek},
title = {{The Swift X-Ray Telescope}},
volume = {5165},
booktitle = {X-Ray and Gamma-Ray Instrumentation for Astronomy XIII},
editor = {Kathryn A. Flanagan and Oswald H. W. Siegmund},
organization = {International Society for Optics and Photonics},
publisher = {SPIE},
pages = {201 -- 216},
year = {2004},
doi = {10.1117/12.504868},
URL = {https://doi.org/10.1117/12.504868}
}

@ARTICLE{Burrows_2005,
       author = {{Burrows}, David N. and {Hill}, J.~E. and {Nousek}, J.~A.},
        title = "{The Swift X-Ray Telescope}",
      journal = {\ssr},
         year = 2005,
        month = oct,
       volume = {120},
       number = {3-4},
        pages = {165-195},
          doi = {10.1007/s11214-005-5097-2},
archivePrefix = {arXiv},
       eprint = {astro-ph/0508071},
 primaryClass = {astro-ph},
       adsurl = {https://ui.adsabs.harvard.edu/abs/2005SSRv..120..165B}
}

@book{carroll_1988,
  author = {Carroll, Raymond J. and Ruppert, David},
  title = {Transformation and Weighting in Regression},
  year = {1988},
  publisher = {Chapman and Hall/CRC},
  address = {New York},
  isbn = {978-0412014215},
  series = {Monographs on Statistics and Applied Probability},
  volume = {30}
}

@ARTICLE{Catanese_1998,
       author = {{Catanese}, M. and {Akerlof}, C.~W. and {Badran}, H.~M.},
        title = "{Discovery of Gamma-Ray Emission above 350 GeV from the BL Lacertae Object 1ES 2344+514}",
      journal = {\apj},
         year = 1998,
        month = jul,
       volume = {501},
       number = {2},
        pages = {616-623},
          doi = {10.1086/305857},
archivePrefix = {arXiv},
       eprint = {astro-ph/9712325},
 primaryClass = {astro-ph},
       adsurl = {https://ui.adsabs.harvard.edu/abs/1998ApJ...501..616C}
}

@article{Celotti_2008,
    author = {Celotti, Annalisa and Ghisellini, Gabriele},
    title = {The power of blazar jets},
    journal = {Monthly Notices of the Royal Astronomical Society},
    volume = {385},
    number = {1},
    pages = {283-300},
    year = {2008},
    month = {02},
    issn = {0035-8711},
    doi = {10.1111/j.1365-2966.2007.12758.x},
    url = {https://doi.org/10.1111/j.1365-2966.2007.12758.x},
    eprint = {https://academic.oup.com/mnras/article-pdf/385/1/283/3445987/mnras0385-0283.pdf},
}

@article{Cerruti_2015,
    author = {Cerruti, M. and Zech, A. and Boisson, C. and Inoue, S.},
    title = {A hadronic origin for ultra-high-frequency-peaked BL Lac objects},
    journal = {Monthly Notices of the Royal Astronomical Society},
    volume = {448},
    number = {1},
    pages = {910-927},
    year = {2015},
    month = {02},
    issn = {0035-8711},
    doi = {10.1093/mnras/stu2691},
    url = {https://doi.org/10.1093/mnras/stu2691},
    eprint = {https://academic.oup.com/mnras/article-pdf/448/1/910/9379400/stu2691.pdf},
}

@ARTICLE{Chau_2018,
       author = {{Chaudhury}, K. and {Chitnis}, V.~R. and {Rao}, A.~R.},
        title = "{Long-term X-ray variability characteristics of the narrow-line Seyfert 1 galaxy RE J1034+396}",
      journal = {\mnras},
         year = 2018,
        month = aug,
       volume = {478},
       number = {4},
        pages = {4830-4836},
          doi = {10.1093/mnras/sty1366},
archivePrefix = {arXiv},
       eprint = {1807.06460},
 primaryClass = {astro-ph.HE},
       adsurl = {https://ui.adsabs.harvard.edu/abs/2018MNRAS.478.4830C}
}

@book{cheng_1999,
  author = {Cheng, Ching-Ling and Van Ness, John W.},
  title = {Statistical Regression with Measurement Error},
  year = {1999},
  publisher = {Arnold},
  address = {London},
  isbn = {978-0340614617},
  series = {Kendall's Advanced Theory of Statistics}
}

@ARTICLE{Costamante_2001,
       author = {{Costamante}, L. and {Ghisellini}, G. and {Giommi}, P.},
        title = "{Extreme synchrotron BL Lac objects. Stretching the blazar sequence}",
      journal = {\aap},
         year = 2001,
        month = may,
       volume = {371},
        pages = {512-526},
          doi = {10.1051/0004-6361:20010412},
archivePrefix = {arXiv},
       eprint = {astro-ph/0103343},
 primaryClass = {astro-ph},
       adsurl = {https://ui.adsabs.harvard.edu/abs/2001A&A...371..512C}
}

@article{Costamante_2018,
    author = {Costamante, L and Bonnoli, G and Tavecchio, F},
    title = "{The NuSTAR view on hard-TeV BL Lacs}",
    journal = {Monthly Notices of the Royal Astronomical Society},
    volume = {477},
    number = {3},
    pages = {4257-4268},
    year = {2018},
    month = {05},
    issn = {0035-8711},
    doi = {10.1093/mnras/sty857},
    url = {https://doi.org/10.1093/mnras/sty857},
    
}

@ARTICLE{Dermer_1993,
       author = {{Dermer}, Charles D. and {Schlickeiser}, Reinhard},
        title = "{Model for the High-Energy Emission from Blazars}",
      journal = {\apj},
         year = 1993,
        month = oct,
       volume = {416},
        pages = {458},
          doi = {10.1086/173251},
       adsurl = {https://ui.adsabs.harvard.edu/abs/1993ApJ...416..458D}
}

@article{efron_1979,
  author = {Efron, Bradley},
  title = {Bootstrap Methods: Another Look at the Jackknife},
  journal = {The Annals of Statistics},
  volume = {7},
  number = {1},
  year = {1979},
  pages = {1--26},
  doi = {10.1214/aos/1176344552},
  publisher = {Institute of Mathematical Statistics},
  issn = {0090-5364}
}

@ARTICLE{Elvis_1992,
       author = {{Elvis}, Martin and {Plummer}, David and {Schachter}, Jonathan and {Fabbiano}, G.},
        title = "{The Einstein Slew Survey}",
      journal = {\apjs},
         year = 1992,
        month = may,
       volume = {80},
        pages = {257},
          doi = {10.1086/191665},
       adsurl = {https://ui.adsabs.harvard.edu/abs/1992ApJS...80..257E}
}

@article{Fan_2006,
doi = {10.1086/504864},
url = {https://dx.doi.org/10.1086/504864},
year = {2006},
month = {jul},
publisher = {},
volume = {646},
number = {1},
pages = {8},
author = {Zhonghui Fan and Xinwu Cao and Minfeng Gu},
title = {A Test of External Compton Models for Gamma-Ray Active Galactic Nuclei},
journal = {The Astrophysical Journal},
}

@ARTICLE{Franceschini_2017,
       author = {{Franceschini}, Alberto and {Rodighiero}, Giulia},
        title = "{The extragalactic background light revisited and the cosmic photon-photon opacity}",
      journal = {\aap},
         year = 2017,
        month = jul,
       volume = {603},
          eid = {A34},
        pages = {A34},
          doi = {10.1051/0004-6361/201629684},
archivePrefix = {arXiv},
       eprint = {1705.10256},
 primaryClass = {astro-ph.HE},
       adsurl = {https://ui.adsabs.harvard.edu/abs/2017A&A...603A..34F}
}

@book{fuller_1987,
  author = {Fuller, Wayne A.},
  title = {Measurement Error Models},
  year = {1987},
  publisher = {Wiley},
  address = {New York},
  isbn = {978-0471861874},
  series = {Wiley Series in Probability and Statistics}
}

@article{Ghisellini_2001,
    author = {Ghisellini, G. and Celotti, A.},
    title = {Relativistic large-scale jets and minimum power requirements},
    journal = {Monthly Notices of the Royal Astronomical Society},
    volume = {327},
    number = {3},
    pages = {739-743},
    year = {2001},
    month = {11},
    issn = {0035-8711},
    doi = {10.1046/j.1365-8711.2001.04700.x},
    url = {https://doi.org/10.1046/j.1365-8711.2001.04700.x},
    eprint = {https://academic.oup.com/mnras/article-pdf/327/3/739/2872684/327-3-739.pdf},
}

@ARTICLE{Ghisellini_2009,
       author = {{Ghisellini}, G. and {Maraschi}, L. and {Tavecchio}, F.},
        title = "{The Fermi blazars' divide}",
      journal = {\mnras},
         year = 2009,
        month = jun,
       volume = {396},
       number = {1},
        pages = {L105-L109},
          doi = {10.1111/j.1745-3933.2009.00673.x},
archivePrefix = {arXiv},
       eprint = {0903.2043},
 primaryClass = {astro-ph.CO},
       adsurl = {https://ui.adsabs.harvard.edu/abs/2009MNRAS.396L.105G}
}

@article{Ghisellini_2010,
   author = {{Ghisellini}, G. and {Tavecchio}, F. and {Foschini}, L. and 
             {Ghirlanda}, G. and {Maraschi}, L. and {Celotti}, A.},
    title = "{The transition between BL Lac objects and flat-spectrum radio quasars}",
  journal = {\mnras},
   volume = {402},
    pages = {497-518},
      year = {2010},
     month = feb,
      doi = {10.1111/j.1365-2966.2009.15898.x}
}

@article{Giannios_2013,
    author = {Giannios, Dimitrios},
    title = {Reconnection-driven plasmoids in blazars: fast flares on a slow envelope},
    journal = {Monthly Notices of the Royal Astronomical Society},
    volume = {431},
    number = {1},
    pages = {355-363},
    year = {2013},
    month = {02},
    issn = {0035-8711},
    doi = {10.1093/mnras/stt167},
    url = {https://doi.org/10.1093/mnras/stt167},
    eprint = {https://academic.oup.com/mnras/article-pdf/431/1/355/18241397/stt167.pdf},
}

@article{Giommi_2000,
    author = {Giommi, P. and Padovani, P. and Perlman, E.},
    title = "{Detection of exceptional X-ray spectral variability in the TeV BL Lac 1ES 2344+514}",
    journal = {Monthly Notices of the Royal Astronomical Society},
    volume = {317},
    number = {4},
    pages = {743-749},
    year = {2000},
    month = {10},
    issn = {0035-8711},
    doi = {10.1046/j.1365-8711.2000.03353.x},
    url = {https://doi.org/10.1046/j.1365-8711.2000.03353.x},
    
}

@ARTICLE{Godambe_2007,
       author = {{Godambe}, S.~V. and {Rannot}, R.~C. and {Baliyan}, K.~S.},
        title = "{Very high energy {\ensuremath{\gamma}}-ray and near infrared observations of 1ES2344+514 during 2004 05}",
      journal = {Journal of Physics G Nuclear Physics},
         year = 2007,
        month = jul,
       volume = {34},
       number = {7},
        pages = {1683-1695},
          doi = {10.1088/0954-3899/34/7/009},
archivePrefix = {arXiv},
       eprint = {0704.3533},
 primaryClass = {astro-ph},
       adsurl = {https://ui.adsabs.harvard.edu/abs/2007JPhG...34.1683G}
}

@ARTICLE{Goswami_2024,
       author = {{Goswami}, P. and {Zacharias}, M. and {Zech}, A. and {Chandra}, S. and {Boettcher}, M. and {Sushch}, I.},
        title = "{The variety of extreme blazars in the AstroSat view}",
      journal = {\aap},
         year = 2024,
        month = feb,
       volume = {682},
          eid = {A134},
        pages = {A134},
          doi = {10.1051/0004-6361/202348121},
archivePrefix = {arXiv},
       eprint = {2311.12695},
 primaryClass = {astro-ph.HE},
       adsurl = {https://ui.adsabs.harvard.edu/abs/2024A&A...682A.134G}
}

@ARTICLE{Harrison_2013,
       author = {{Harrison}, Fiona A. and {Craig}, William W. and {Christensen}, Finn E.},
        title = "{The Nuclear Spectroscopic Telescope Array (NuSTAR) High-energy X-Ray Mission}",
      journal = {\apj},
         year = 2013,
        month = jun,
       volume = {770},
       number = {2},
          eid = {103},
        pages = {103},
          doi = {10.1088/0004-637X/770/2/103},
archivePrefix = {arXiv},
       eprint = {1301.7307},
 primaryClass = {astro-ph.IM},
       adsurl = {https://ui.adsabs.harvard.edu/abs/2013ApJ...770..103H}
}

@ARTICLE{HI4PI_2016,
       author = {{HI4PI Collaboration} and {Ben Bekhti}, N. and {Fl{\"o}er}, L.},
        title = "{HI4PI: A full-sky H I survey based on EBHIS and GASS}",
      journal = {\aap},
         year = 2016,
        month = oct,
       volume = {594},
          eid = {A116},
        pages = {A116},
          doi = {10.1051/0004-6361/201629178},
archivePrefix = {arXiv},
       eprint = {1610.06175},
 primaryClass = {astro-ph.GA},
       adsurl = {https://ui.adsabs.harvard.edu/abs/2016A&A...594A.116H}
}

@article{ Hovatta_2009,
	author = {{Hovatta, T.} and {Valtaoja, E.} and {Tornikoski, M.} and {Lähteenmäki, A.}},
	title = {Doppler factors, Lorentz factors and viewing angles  for quasars, BL Lacertae objects and radio galaxies},
	DOI= "10.1051/0004-6361:200811150",
	url= "https://doi.org/10.1051/0004-6361:200811150",
	journal = {A\&A},
	year = 2009,
	volume = 494,
	number = 2,
	pages = "527-537",
}

@ARTICLE{Kapanadze_2017,
       author = {{Kapanadze}, S. and {Kapanadze}, B. and {Romano}, P.},
        title = "{The swift observations of BL Lacertae object 1ES 2344+514}",
      journal = {\apss},
         year = 2017,
        month = oct,
       volume = {362},
       number = {10},
          eid = {196},
        pages = {196},
          doi = {10.1007/s10509-017-3170-4},
       adsurl = {https://ui.adsabs.harvard.edu/abs/2017Ap&SS.362..196K}
}

@article{Krawczynski_2004,
    author = {Krawczynski, H. and Hughes, S. B. and Horan, D. and others},
    title = "{Multiwavelength Observations of Strong Flares from the TeV Blazar 1ES 1959+650}",
    journal = {The Astrophysical Journal},
    volume = {601},
    pages = {151},
    year = {2004},
    doi = {10.1086/380392}
}

@ARTICLE{Acciari_2020,
       author = {{MAGIC Collaboration} and {Acciari}, V.~A. and {Ansoldi}, S. and {Antonelli}, L.~A.},
        title = "{An intermittent extreme BL Lac: MWL study of 1ES 2344+514 in an enhanced state}",
      journal = {\mnras},
         year = 2020,
        month = aug,
       volume = {496},
       number = {3},
        pages = {3912-3928},
          doi = {10.1093/mnras/staa1702},
archivePrefix = {arXiv},
       eprint = {2006.06796},
 primaryClass = {astro-ph.HE},
       adsurl = {https://ui.adsabs.harvard.edu/abs/2020MNRAS.496.3912M}
}

@article{ Abe_2024,
	author = {{MAGIC Collaboration} and {Abe, H.} and {Abe, S.} and {Acciari, V. A.}},
	title = {Multi-year characterisation of the broad-band emission from the intermittent extreme BL Lac 1ES 2344+514},
	DOI= "10.1051/0004-6361/202347845",
	url= "https://doi.org/10.1051/0004-6361/202347845",
	journal = {\aap},
	year = 2024,
	volume = 682,
	pages = "A114",
}

@ARTICLE{Mannheim_1993,
       author = {{Mannheim}, K.},
        title = "{The proton blazar.}",
      journal = {\aap},
         year = 1993,
        month = mar,
       volume = {269},
        pages = {67-76},
          doi = {10.48550/arXiv.astro-ph/9302006},
archivePrefix = {arXiv},
       eprint = {astro-ph/9302006},
 primaryClass = {astro-ph},
       adsurl = {https://ui.adsabs.harvard.edu/abs/1993A&A...269...67M}
}

@ARTICLE{Maraschi_1992,
       author = {{Maraschi}, L. and {Ghisellini}, G. and {Celotti}, A.},
        title = "{A Jet Model for the Gamma-Ray--emitting Blazar 3C 279}",
      journal = {\apjl},
         year = 1992,
        month = sep,
       volume = {397},
        pages = {L5},
          doi = {10.1086/186531},
       adsurl = {https://ui.adsabs.harvard.edu/abs/1992ApJ...397L...5M}
}

@ARTICLE{ Massaro_2004,
       author = {{Massaro}, E. and {Perri}, M. and {Giommi}, P. and {Nesci}, R. and {Verrecchia}, F.},
        title = "{Log-parabolic spectra and particle acceleration in blazars.  II. The BeppoSAX wide band X-ray spectra of Mkn 501}",
      journal = {\aap},
         year = 2004,
        month = jul,
       volume = {422},
        pages = {103-111},
          doi = {10.1051/0004-6361:20047148},
archivePrefix = {arXiv},
       eprint = {astro-ph/0405152},
 primaryClass = {astro-ph},
       adsurl = {https://ui.adsabs.harvard.edu/abs/2004A&A...422..103M}
}

@ARTICLE{Massaro_2006,
       author = {{Massaro}, E. and {Tramacere}, A. and {Perri}, M.},
        title = "{Log-parabolic spectra and particle acceleration in blazars. III. SSC emission in the TeV band from Mkn501}",
      journal = {\aap},
         year = 2006,
        month = mar,
       volume = {448},
       number = {3},
        pages = {861-871},
          doi = {10.1051/0004-6361:20053644},
archivePrefix = {arXiv},
       eprint = {astro-ph/0511673},
 primaryClass = {astro-ph},
       adsurl = {https://ui.adsabs.harvard.edu/abs/2006A&A...448..861M}
}

@ARTICLE{Massaro_2008,
       author = {{Massaro}, F. and {Tramacere}, A. and {Cavaliere}, A.},
        title = "{X-ray spectral evolution of TeV BL Lacertae objects: eleven years of observations with BeppoSAX, XMM-Newton and Swift satellites}",
      journal = {\aap},
         year = 2008,
        month = feb,
       volume = {478},
       number = {2},
        pages = {395-401},
          doi = {10.1051/0004-6361:20078639},
archivePrefix = {arXiv},
       eprint = {0712.2116},
 primaryClass = {astro-ph},
       adsurl = {https://ui.adsabs.harvard.edu/abs/2008A&A...478..395M}
}

@ARTICLE{Mucker_2001,
       author = {{M{\"u}cke}, A. and {Protheroe}, R.~J.},
        title = "{A proton synchrotron blazar model for flaring in Markarian 501}",
      journal = {Astroparticle Physics},
         year = 2001,
        month = mar,
       volume = {15},
       number = {1},
        pages = {121-136},
          doi = {10.1016/S0927-6505(00)00141-9},
archivePrefix = {arXiv},
       eprint = {astro-ph/0004052},
 primaryClass = {astro-ph},
       adsurl = {https://ui.adsabs.harvard.edu/abs/2001APh....15..121M}
}

@ARTICLE{Nilsson_2018,
       author = {{Nilsson}, K. and {Lindfors}, E. and {Takalo}, L.~O.},
        title = "{Long-term optical monitoring of TeV emitting blazars. I. Data analysis}",
      journal = {\aap},
         year = 2018,
        month = dec,
       volume = {620},
          eid = {A185},
        pages = {A185},
          doi = {10.1051/0004-6361/201833621},
archivePrefix = {arXiv},
       eprint = {1810.01751},
 primaryClass = {astro-ph.HE},
       adsurl = {https://ui.adsabs.harvard.edu/abs/2018A&A...620A.185N}
}

@article{Padovani_2017,
   author = {{Padovani}, P. and {Alexander}, D.~M. and {Assef}, R.~J. and 
            {De Marco}, B. and {Giommi}, P. and {Hickox}, R.~C. and 
            {Richards}, G.~T. and {Smol{\v{c}}i{\'c}}, V. and {Hatziminaoglou}, E. and 
            {Mainieri}, V. and {Salvato}, M.},
    title = "{Active Galactic Nuclei: What's in a Name?}",
  journal = {\aapr},
   volume = {25},
    pages = {2},
      year = {2017},
     month = jun,
      doi = {10.1007/s00159-017-0102-9},
   adsurl = {https://ui.adsabs.harvard.edu/abs/2017A%26ARv..25....2P}
}

@ARTICLE{Perlman_1996,
       author = {{Perlman}, Eric S. and {Stocke}, John T. and {Schachter}, Jonathan F.},
        title = "{The Einstein Slew Survey Sample of BL Lacertae Objects}",
      journal = {\apjs},
         year = 1996,
        month = jun,
       volume = {104},
        pages = {251},
          doi = {10.1086/192300},
       adsurl = {https://ui.adsabs.harvard.edu/abs/1996ApJS..104..251P}
}

@ARTICLE{Poole2008,
       author = {{Poole}, T.~S. and {Breeveld}, A.~A. and {Page}, M.~J. and {Landsman}, W. and {Holland}, S.~T. and {Roming}, P. and {Kuin}, N.~P.~M. and {Brown}, P.~J. and {Gronwall}, C. and {Hunsberger}, S. and {Koch}, S. and {Mason}, K.~O. and {Schady}, P. and {vanden Berk}, D. and {Blustin}, A.~J. and {Boyd}, P. and {Broos}, P. and {Carter}, M. and {Chester}, M.~M. and {Cucchiara}, A. and {Hancock}, B. and {Huckle}, H. and {Immler}, S. and {Ivanushkina}, M. and {Kennedy}, T. and {Marshall}, F. and {Morgan}, A. and {Pandey}, S.~B. and {de Pasquale}, M. and {Smith}, P.~J. and {Still}, M.},
        title = "{Photometric calibration of the Swift ultraviolet/optical telescope}",
      journal = {\mnras},
         year = 2008,
        month = jan,
       volume = {383},
       number = {2},
        pages = {627-645},
          doi = {10.1111/j.1365-2966.2007.12563.x},
archivePrefix = {arXiv},
       eprint = {0708.2259},
 primaryClass = {astro-ph},
       adsurl = {https://ui.adsabs.harvard.edu/abs/2008MNRAS.383..627P}
}

@ARTICLE{Ramadevi_2018,
       author = {{Ramadevi}, M.~C. and {Ravishankar}, B.~T. and {Sarwade}, Abhilash R.},
        title = "{Study of X-ray transients with Scanning Sky Monitor (SSM) onboard AstroSat}",
      journal = {Journal of Astrophysics and Astronomy},
         year = 2018,
        month = feb,
       volume = {39},
       number = {1},
          eid = {11},
        pages = {11},
          doi = {10.1007/s12036-017-9506-1},
       adsurl = {https://ui.adsabs.harvard.edu/abs/2018JApA...39...11R}
}

@ARTICLE{Rao_2016,
       author = {{Rao}, A.~R. and {Singh}, K.~P. and {Bhattacharya}, D.},
        title = "{AstroSat - a multi-wavelength astronomy satellite}",
      journal = {arXiv e-prints},
         year = 2016,
        month = aug,
          eid = {arXiv:1608.06051},
        pages = {arXiv:1608.06051},
          doi = {10.48550/arXiv.1608.06051},
archivePrefix = {arXiv},
       eprint = {1608.06051},
 primaryClass = {astro-ph.IM},
       adsurl = {https://ui.adsabs.harvard.edu/abs/2016arXiv160806051R}
}

@ARTICLE{Rao_2017,
       author = {{Rao}, A.~R. and {Bhattacharya}, D. and {Bhalerao}, V.~B. and {Vadawale}, S.~V. and {Sreekumar}, S.},
        title = "{Cadmium-Zinc-Telluride Imager on-board AstroSat: a multi-faceted hard X-ray instrument}",
      journal = {Current Science},
         year = 2017,
        month = aug,
       volume = {113},
       number = {4},
        pages = {595},
          doi = {10.18520/cs/v113/i04/595-598},
archivePrefix = {arXiv},
       eprint = {1710.10773},
 primaryClass = {astro-ph.IM},
       adsurl = {https://ui.adsabs.harvard.edu/abs/2017CSci..113..595R}
}

@INPROCEEDINGS{Romano_2005,
       author = {{Romano}, P. and {Cusumano}, G. and {Campana}, S.},
        title = "{In-flight calibration of the SWIFT XRT effective area}",
    booktitle = {UV, X-Ray, and Gamma-Ray Space Instrumentation for Astronomy XIV},
         year = 2005,
       editor = {{Siegmund}, Oswald H.~W.},
       series = {Society of Photo-Optical Instrumentation Engineers (SPIE) Conference Series},
       volume = {5898},
        month = aug,
        pages = {369-376},
          doi = {10.1117/12.616974},
       adsurl = {https://ui.adsabs.harvard.edu/abs/2005SPIE.5898..369R}
}

@article{Roming_2005,
  title={The Swift ultra-violet/optical telescope},
  author={Roming, Peter WA and Kennedy, Thomas E and Mason, Keith O and Nousek, John A and Ahr, Lindy and Bingham, Richard E and Broos, Patrick S and Carter, Mary J and Hancock, Barry K and Huckle, Howard E and others},
  journal={Space Science Reviews},
  volume={120},
  pages={95--142},
  year={2005},
  publisher={Springer}
}

@INPROCEEDINGS{Sanchez_2013,
       author = {{Sanchez}, D.~A. and {Deil}, C.},
        title = "{Enrico : A Python Package to Simplify Fermi-LAT Analysis}",
    booktitle = {International Cosmic Ray Conference},
         year = 2013,
       series = {International Cosmic Ray Conference},
       volume = {33},
        month = jan,
        pages = {2784},
          doi = {10.48550/arXiv.1307.4534},
archivePrefix = {arXiv},
       eprint = {1307.4534},
 primaryClass = {astro-ph.IM},
       adsurl = {https://ui.adsabs.harvard.edu/abs/2013ICRC...33.2784S}
}

@article{Schlegel1998,
  title={Maps of dust infrared emission for use in estimation of reddening and cosmic microwave background radiation foregrounds},
  author={Schlegel, David J and Finkbeiner, Douglas P and Davis, Marc},
  journal={The Astrophysical Journal},
  volume={500},
  number={2},
  pages={525},
  year={1998},
  publisher={IOP Publishing}
}

@article{Schroedter_2005,
doi = {10.1086/496968},
url = {https://dx.doi.org/10.1086/496968},
year = {2005},
month = {dec},
publisher = {},
volume = {634},
number = {2},
pages = {947},
author = {M. Schroedter and H. M. Badran and J. H. Buckley},
title = {A Very High Energy Gamma-Ray Spectrum of 1ES 2344+514},
journal = {The Astrophysical Journal},
}

@article{Sikora_1994,
   author = {{Sikora}, M. and {Begelman}, M.~C. and {Rees}, M.~J.},
    title = "{Comptonization of Diffuse Ambient Radiation by a Relativistic Jet: The Source of Gamma Rays from Blazars?}",
  journal = {\apj},
   volume = {421},
    pages = {153-162},
      year = {1994},
     month = jan,
      doi = {10.1086/173633},
   adsurl = {https://ui.adsabs.harvard.edu/abs/1994ApJ...421..153S}
}

@INPROCEEDINGS{Singh_2014,
       author = {{Singh}, Kulinder Pal and {Tandon}, S.~N. and {Agrawal}, P.~C.},
        title = "{ASTROSAT mission}",
    booktitle = {Space Telescopes and Instrumentation 2014: Ultraviolet to Gamma Ray},
         year = 2014,
       editor = {{Takahashi}, Tadayuki and {den Herder}, Jan-Willem A. and {Bautz}, Mark},
       series = {Society of Photo-Optical Instrumentation Engineers (SPIE) Conference Series},
       volume = {9144},
        month = jul,
          eid = {91441S},
        pages = {91441S},
          doi = {10.1117/12.2062667},
       adsurl = {https://ui.adsabs.harvard.edu/abs/2014SPIE.9144E..1SS}
}

@INPROCEEDINGS{Singh_2016,
       author = {{Singh}, Kulinder Pal and {Stewart}, Gordon C. and {Chandra}, Sunil},
        title = "{In-orbit performance of SXT aboard AstroSat}",
    booktitle = {Space Telescopes and Instrumentation 2016: Ultraviolet to Gamma Ray},
         year = 2016,
       editor = {{den Herder}, Jan-Willem A. and {Takahashi}, Tadayuki and {Bautz}, Marshall},
       series = {Society of Photo-Optical Instrumentation Engineers (SPIE) Conference Series},
       volume = {9905},
        month = jul,
          eid = {99051E},
        pages = {99051E},
          doi = {10.1117/12.2235309},
       adsurl = {https://ui.adsabs.harvard.edu/abs/2016SPIE.9905E..1ES}
}

@ARTICLE{Signh_2017,
       author = {{Singh}, K.~P. and {Stewart}, G.~C. and {Westergaard}, N.~J.},
        title = "{Soft X-ray Focusing Telescope Aboard AstroSat: Design, Characteristics and Performance}",
      journal = {Journal of Astrophysics and Astronomy},
         year = 2017,
        month = jun,
       volume = {38},
       number = {2},
          eid = {29},
        pages = {29},
          doi = {10.1007/s12036-017-9448-7},
       adsurl = {https://ui.adsabs.harvard.edu/abs/2017JApA...38...29S}
}

@ARTICLE{Tandon_2017,
       author = {{Tandon}, S.~N. and {Subramaniam}, Annapurni and {Girish}, V.},
        title = "{In-orbit Calibrations of the Ultraviolet Imaging Telescope}",
      journal = {\aj},
         year = 2017,
        month = sep,
       volume = {154},
       number = {3},
          eid = {128},
        pages = {128},
          doi = {10.3847/1538-3881/aa8451},
archivePrefix = {arXiv},
       eprint = {1705.03715},
 primaryClass = {astro-ph.IM},
       adsurl = {https://ui.adsabs.harvard.edu/abs/2017AJ....154..128T}
}

@ARTICLE{Tandon_2017a,
       author = {{Tandon}, S.~N. and {Hutchings}, J.~B. and {Ghosh}, S.~K.},
        title = "{In-orbit Performance of UVIT and First Results}",
      journal = {Journal of Astrophysics and Astronomy},
         year = 2017,
        month = jun,
       volume = {38},
       number = {2},
          eid = {28},
        pages = {28},
          doi = {10.1007/s12036-017-9445-x},
archivePrefix = {arXiv},
       eprint = {1612.00612},
 primaryClass = {astro-ph.IM},
       adsurl = {https://ui.adsabs.harvard.edu/abs/2017JApA...38...28T}
}

@article{Tavecchio_2010,
    author = {Tavecchio, F. and Ghisellini, G. and Ghirlanda, G.},
    title = "{TeV BL Lac objects at the dawn of the Fermi era}",
    journal = {Monthly Notices of the Royal Astronomical Society},
    volume = {401},
    number = {3},
    pages = {1570-1586},
    year = {2010},
    month = {01},
    issn = {0035-8711},
    doi = {10.1111/j.1365-2966.2009.15784.x},
    url = {https://doi.org/10.1111/j.1365-2966.2009.15784.x},
 
}

@ARTICLE{Tavecchio_2016,
       author = {{Tavecchio}, F. and {Ghisellini}, G.},
        title = "{On the magnetization of BL Lac jets}",
      journal = {\mnras},
         year = 2016,
        month = mar,
       volume = {456},
       number = {3},
        pages = {2374-2382},
          doi = {10.1093/mnras/stv2790},
archivePrefix = {arXiv},
       eprint = {1509.08710},
 primaryClass = {astro-ph.HE},
       adsurl = {https://ui.adsabs.harvard.edu/abs/2016MNRAS.456.2374T}
}

@ARTICLE{Tramacere_2009,
       author = {{Tramacere}, A. and {Giommi}, P. and {Perri}, M. and {Verrecchia}, F. and {Tosti}, G.},
        title = "{Swift observations of the very intense flaring activity of Mrk 421 during 2006. I. Phenomenological picture of electron acceleration and predictions for MeV/GeV emission}",
      journal = {\aap},
         year = 2009,
        month = jul,
       volume = {501},
       number = {3},
        pages = {879-898},
          doi = {10.1051/0004-6361/200810865},
archivePrefix = {arXiv},
       eprint = {0901.4124},
 primaryClass = {astro-ph.HE},
       adsurl = {https://ui.adsabs.harvard.edu/abs/2009A&A...501..879T}
}

@ARTICLE{Tramacere_2011,
       author = {{Tramacere}, A. and {Massaro}, E. and {Taylor}, A.~M.},
        title = "{Stochastic Acceleration and the Evolution of Spectral Distributions in Synchro-Self-Compton Sources: A Self-consistent Modeling of Blazars' Flares}",
      journal = {\apj},
         year = 2011,
        month = oct,
       volume = {739},
       number = {2},
          eid = {66},
        pages = {66},
          doi = {10.1088/0004-637X/739/2/66},
archivePrefix = {arXiv},
       eprint = {1107.1879},
 primaryClass = {astro-ph.HE},
       adsurl = {https://ui.adsabs.harvard.edu/abs/2011ApJ...739...66T}
}

@MISC{Tramacere_2020,
       author = {{Tramacere}, Andrea},
        title = "{JetSeT: Numerical modeling and SED fitting tool for relativistic jets}",
 howpublished = {Astrophysics Source Code Library, record ascl:2009.001},
         year = 2020,
        month = sep,
          eid = {ascl:2009.001},
        pages = {ascl:2009.001},
archivePrefix = {ascl},
       eprint = {2009.001},
       adsurl = {https://ui.adsabs.harvard.edu/abs/2020ascl.soft09001T}
}

@ARTICLE{Tramacere_2007,
       author = {{Tramacere}, A. and {Giommi}, P. and {Massaro}, E.},
        title = "{SWIFT observations of TeV BL Lacertae objects}",
      journal = {\aap},
         year = 2007,
        month = may,
       volume = {467},
       number = {2},
        pages = {501-508},
          doi = {10.1051/0004-6361:20066226},
archivePrefix = {arXiv},
       eprint = {astro-ph/0611276},
 primaryClass = {astro-ph},
       adsurl = {https://ui.adsabs.harvard.edu/abs/2007A&A...467..501T}
}

@article{Urry_1995,
doi = {10.1086/133630},
url = {https://dx.doi.org/10.1086/133630},
year = {1995},
month = {sep},
publisher = {The Astronomical Society of the Pacific},
volume = {107},
number = {715},
pages = {803},
author = {C. Megan Urry and Paolo Padovani},
title = {UNIFIED SCHEMES FOR RADIO-LOUD ACTIVE GALACTIC NUCLEI},
journal = {Publications of the Astronomical Society of the Pacific},
}

@ARTICLE{Wilms_2000,
       author = {{Wilms}, J. and {Allen}, A. and {McCray}, R.},
        title = "{On the Absorption of X-Rays in the Interstellar Medium}",
      journal = {\apj},
         year = 2000,
        month = oct,
       volume = {542},
       number = {2},
        pages = {914-924},
          doi = {10.1086/317016},
archivePrefix = {arXiv},
       eprint = {astro-ph/0008425},
 primaryClass = {astro-ph},
       adsurl = {https://ui.adsabs.harvard.edu/abs/2000ApJ...542..914W}
}

@INPROCEEDINGS{Yadav_2016,
       author = {{Yadav}, J.~S. and {Agrawal}, P.~C. and {Antia}, H.~M.},
        title = "{Large Area X-ray Proportional Counter (LAXPC) instrument onboard ASTROSAT}",
    booktitle = {Space Telescopes and Instrumentation 2016: Ultraviolet to Gamma Ray},
         year = 2016,
       editor = {{den Herder}, Jan-Willem A. and {Takahashi}, Tadayuki and {Bautz}, Marshall},
       series = {Society of Photo-Optical Instrumentation Engineers (SPIE) Conference Series},
       volume = {9905},
        month = jul,
          eid = {99051D},
        pages = {99051D},
          doi = {10.1117/12.2231857},
       adsurl = {https://ui.adsabs.harvard.edu/abs/2016SPIE.9905E..1DY}
}

@ARTICLE{Zhang_2012,
       author = {{Zhang}, Jin and {Liang}, En-Wei and {Zhang}, Shuang-Nan and {Bai}, J.~M.},
        title = "{Radiation Mechanisms and Physical Properties of GeV-TeV BL Lac Objects}",
      journal = {\apj},
         year = 2012,
        month = jun,
       volume = {752},
       number = {2},
          eid = {157},
        pages = {157},
          doi = {10.1088/0004-637X/752/2/157},
archivePrefix = {arXiv},
       eprint = {1108.0607},
 primaryClass = {astro-ph.HE},
       adsurl = {https://ui.adsabs.harvard.edu/abs/2012ApJ...752..157Z}
}

@ARTICLE{Zhang_2014,
       author = {{Zhang}, Jin and {Sun}, Xiao-Na and {Liang}, En-Wei and {Lu}, Rui-Jing and {Lu}, Ye and {Zhang}, Shuang-Nan},
        title = "{Relativistic Jet Properties of GeV-TeV Blazars and Possible Implications for the Jet Formation, Composition, and Cavity Kinematics}",
      journal = {\apj},
         year = 2014,
        month = jun,
       volume = {788},
       number = {2},
          eid = {104},
        pages = {104},
          doi = {10.1088/0004-637X/788/2/104},
       adsurl = {https://ui.adsabs.harvard.edu/abs/2014ApJ...788..104Z}
}

\end{document}